\documentclass[sigplan,10pt]{acmart}
\renewcommand\footnotetextcopyrightpermission[1]{}

\usepackage{appendix}
\usepackage{listings}
\usepackage{comment}
\usepackage{paralist}
\usepackage{amsmath}
\usepackage{xspace}
\usepackage{array}
\usepackage{tikz}
\usepackage{enumitem}
\usepackage[normalem]{ulem}
\usetikzlibrary{calc}
\usetikzlibrary{shapes}
\usetikzlibrary{decorations.pathreplacing, decorations.markings}
\usetikzlibrary{shadows}
\usetikzlibrary{fit, backgrounds}
\usetikzlibrary{arrows.meta,chains,positioning}
\usetikzlibrary{arrows}
\pgfdeclarelayer{lp}
\pgfdeclarelayer{lr}
\pgfdeclarelayer{lr2}
\pgfdeclarelayer{bg}
\pgfsetlayers{bg,lr,lr2,lp,main}
\graphicspath{{./figures}}
\usepackage{filecontents}
\usepackage{subcaption}
\usepackage{colortbl}

\usepackage{algorithmicx}
\usepackage{algorithm}
\usepackage[noend]{algpseudocode}
\algnewcommand\algorithmicswitch{\textbf{switch}}
\algnewcommand\algorithmiccase{\textbf{case}}
\algnewcommand\algorithmicassert{\texttt{assert}}
\algnewcommand\Assert[1]{\State \algorithmicassert(#1)}%
\algdef{SE}[SWITCH]{Switch}{EndSwitch}[1]{\algorithmicswitch\ #1\ \algorithmicdo}{\algorithmicend\ \algorithmicswitch}%
\algdef{SE}[CASE]{Case}{EndCase}[1]{\algorithmiccase\ #1}{\algorithmicend\ \algorithmiccase}%
\algtext*{EndSwitch}%
\algtext*{EndCase}%⇧
\algnewcommand\algorithmiccontinue{\textbf{continue}}
\algnewcommand\algorithmicbreak{\textbf{break}}
\algnewcommand\Continue{\algorithmiccontinue}
\algnewcommand\Break{\algorithmicbreak}

\algnewcommand{\LineComment}[1]{\State {\color{gray}\textrm{// #1}}}
\algnewcommand{\InlineComment}[1]{{\hspace{0.5em}\color{gray}\textrm{// #1}}}
\algnewcommand{\SectionComment}[2]{\State {\color{#2}\textrm{// #1}}}

\newcommand{\key}{\mathit{key}}

\newcommand{\node}{u}
\newcommand{\dataNode}{d}

\newcommand{\nnull}{\bot}

\newif\ifpdfbars

\pdfbarsfalse

\newcommand{\bptree}{B\textsuperscript{+}-tree\xspace}

\newcommand{\DB}{CedrusDB\xspace}

\newcommand{\remove}[1]{\textcolor{red}{\sout{#1}}}
\begin{document}
%-------------------------------------------------------------------------------

%don't want date printed
\date{}

\title{CedrusDB: Persistent Key-Value Store with Memory-Mapped Lazy-Trie}

\ifdefined\submission
\author{Anonymous Submission 504}
\else
\author{Maofan~Yin, Hongbo~Zhang, Robbert~van~Renesse, Emin~G\"un~Sirer}
\affiliation{
\institution{Cornell University}
\country{}}
\fi

\begin{abstract}
%\begin{comment}
%    Most of today's persistent key-value stores are based either on a
%    write-optimized Log-Structured Merge tree (LSM) or a read-optimized \bptree,
%    both of which assume that there is not enough memory to cache the entire
%    data set.
%\end{comment}
    As a result of RAM becoming cheaper, there has been a trend in
    key-value store design towards maintaining a fast in-memory index
    (such as a hash table)
    while logging user operations to disk, allowing high performance
    under failure-free conditions while still being able to recover
    from failures.  This design, however, comes at the cost of long
    recovery times or expensive checkpoint operations.
%
    %it is time to revisit the
    %design of persistent key-value storage systems.
    %This paper introduces a new design called ``lazy-trie''
    This paper presents a new in-memory index that is also storage-friendly.
    A ``lazy-trie'' is a variant of the hash-trie data structure that
    achieves near-optimal height, has practical storage overhead,
    and can be maintained on-disk with standard write-ahead logging.
    %but does not support range queries.

    % Unlike other balanced data structures, it achieves this without the need for
    % expensive reorganization of subtrees, allowing more efficient concurrent access.

    We implemented \DB,
    persistent key-value store based on a lazy-trie. % that supports point lookup and updates.
    The lazy-trie is kept on disk while made available in memory using
    standard memory-mapping.  The lazy-trie
    organization in virtual memory allows \DB to better leverage
    concurrent processing than other on-disk index schemes (LSMs, \bptree{}s).
    \DB achieves comparable or superior performance to recent
    log-based in-memory key-value stores in
    mixed workloads while being able to recover quickly from failures.
\end{abstract}

\settopmatter{printfolios=true}
\settopmatter{printacmref=false}
\maketitle
\pagestyle{plain}

%-------------------------------------------------------------------------------
\section{Introduction}
%-------------------------------------------------------------------------------

Persistent key-value stores have become an indispensable part of
applications such as web servers~\cite{BadamPPP09}, cloud
storage~\cite{dynamo07, alibaba2019}, machine learning~\cite{mlsurvey},
mobile apps~\cite{snappydb}, and
% even
blockchain infrastructures~\cite{mlsm, WangDLXZCCOR18}.
There are two
% main
on-disk data structures that are often used for a
general-purpose persistent key-value store: \bptree{}s
% (or \btree{}s)
and Log-Structured Merge Trees (LSMs). \bptree{}s have stable,
predictable degradation as the data set grows large and supports
fast read access, but they incur non-sequential small writes that
sometimes get amplified in order to maintain balance.
LSMs, on the other hand, are optimized for write-intensive
workloads. They usually have high write-throughput as most of the
writes are sequential and the resulting storage space is compact.
An LSM does not have an explicit
tree structure that organizes nodes as in \bptree{}s. Instead, the
merge tree serves as a conceptual hierarchy that directs how to
merge-sort user data. The amount of merge-sorted data gets amplified
as the level goes deeper and thus may cause severe write amplification
during log compaction.  Compared to \bptree{}s, LSMs also have
higher read amplification.

Both designs are intended for scenarios where the data set
cannot fit in memory and the underlying
secondary storage is much slower than memory.  But servers and consumer
devices have increasingly larger memories.  Today, servers may have
tens to thousands of gigabytes of memory with a secondary storage
using flash or non-volatile memory technology.  As a result, more
applications can have all or at least most of their data set fully reside
in memory, and
this has ignited interest in exploring new data structures that
better utilize the characteristics of the abundant memory~\cite{lmdb, alibaba2019} or even non-volatile main memory~\cite{MarmolGA16, recipe19}.  At the same time, solid state devices (SSDs) may also require
changes in data structures on secondary storage to take advantage
of the much faster access times~\cite{silt2011, ShenCJS16}.

Some recent key-value store designs have eschewed on-disk indexes
altogether, favoring a fast in-memory index and an unordered solution
for persistence such as logging updates sequentially or slab
allocation~\cite{masstree2012, nibble17, faster2018, kvell}.
While such designs can perform very well in the failure-free case,
crash recovery involves reading large sections
if not all of the disk, leading to lengthy recovery times.
Checkpointing can improve recovery time but comes at a considerable
overhead during normal operation (\S\ref{sec:eval-crash}).
As an example, FASTER~\cite{faster2018} is a recent work that keeps an
in-memory hash table and logs all writes to the disk.
% It can be used as a building block
% in the containing application or system to provide persistent,
% unordered mappings. 
While the design utilizes the
disk bandwidth well for intensive in-place updates,
% mixed operations are less efficient and
a user has to invoke checkpoints manually to
make the store persistent, where both checkpointing and recovery take substantial
time due to the lack of an on-disk index.

% We consider the design of a persistent key-value store
% that uses SSDs for persistence and where the volatile main memory is
% capable of storing most of the data set for high access demand.
As a result, applications that can fit their data in memory but also
need persistence currently have to choose between stores that maintain
relatively slow on-disk indexes and stores that have high failure
recovery times.
We therefore explore the design of key-value stores with the same
persistence model as \bptree{}- or LSM-based approaches
but whose performance is competitive with a log-based approach to persistence.

To this end, we propose \emph{lazy-trie}, a storage-friendly
data structure.
All nodes in a lazy-trie have the same
number of children slots, simplifying maintenance.
To bound the depth of the trie and
probabilistically balance the load, user keys are hashed
to index into the trie.
To further reduce the depth of the trie
and improve utilization, the lazy-trie uses a path compression technique
similar to radix trees~\cite{radixtree}.
% In addition, we introduce a path compression
% technique similar to the one used by ART. Unlike ART, which uses
% trie nodes with four sizes,
Finally, some small subtrees are collapsed into linked lists
at leaves, greatly reducing storage overhead and read/write amplification.

We use the lazy-trie data structure to implement a memory-mapped, persistent
key-value store, \emph{\DB}.
It is able to achieve near-optimal dynamic tree height with practical storage
overhead.
% \ted{\DB outperforms carefully
% engineered key-value stores such as FASTER, RocksDB and LMDB in most of %nearly all
% read-intensive workloads and some write-intensive workloads.}
%
Like LMDB~\cite{lmdb}, the implementation of \DB uses memory-mapping, but,
unlike LMDB, \DB does not require that all of the data set fit in
available memory---\DB implements its own page replacement and does
not rely on the operating system kernel to do so.  The lazy-trie organization
in virtual memory allows \DB to better leverage concurrency.

A shortcoming that \DB shares with FASTER is that hashing makes support
for range queries difficult.
Fortunately, not all applications require support for range queries and
there are various proprietary and open source key-value stores that do not support
them~\cite{dynamo07,voldemort,faster2018,orientdb,fastercloud}.
Many applications use a key-value store to persist
user data by keys. For example, a blockchain application stores data
using keys that are already hashes.
The Shadowfax distributed key-value store based on FASTER which is unordered and
persistent like \DB, serves 130 Mops/s/VM in the Microsoft Azure
cloud~\cite{fastercloud}.
There also exist practical techniques for supporting range queries on top
of key-value stores that do not~\cite{rangequery}.
Another limitation of \DB is that it does not perform as well as some of its
competitors for write-only workloads or if the working set size is much
larger than available memory.

This paper makes the following contributions:
\begin{itemize}[leftmargin=*,itemsep=1pt,topsep=1pt]
%\begin{compactitem}
    \item The lazy-trie data structure, which dynamically
        grows with near-optimal tree height, has a practical
        storage footprint, and allows for efficient concurrent access.
    \item The design and implementation of \DB, a high-perfor\-mance
    key-value store that uses the lazy-trie.
    \item An evaluation of \DB for mixed workloads, comparing to the
        most important competitors.
\begin{comment}
        results show that, compared to its most important
    competitors, \DB provides comparable or superior performance for mixed workloads and
        is able to better leverage multi-core computing.
\end{comment}
%\end{compactitem}
\end{itemize}

\begin{comment}
Section~\ref{sec:design} introduces the lazy-trie data structure and its
unique properties.
Section~\ref{sec:implementation} discusses the storage model, design decisions,
and optimizations that went into building \DB.
Section~\ref{sec:evaluation} presents our evaluation of \DB.
Section~\ref{sec:related-work} discusses related work,
and Section~\ref{sec:conclusion} concludes.
\end{comment}

%-------------------------------------------------------------------------------
\section{The Lazy-Trie Data Structure}
\label{sec:design}
%-------------------------------------------------------------------------------

The design of \DB is inspired by memory-mapped key-value stores like LMDB~\cite{lmdb}. Instead of using \bptree variants or other tree
structures that require complex operations to move nodes across sibling
subtrees for a logarithmic tree height, we propose \emph{lazy-trie}, a
trie-tree structure tailored for persistent storage.
In this section, we describe the lazy-trie by progressively adding
its crucial elements, along with a discussion of its properties and their implications for storage.

\begin{figure}
\begin{center}
{%
\makeatletter
\newcommand*{\Strut}[1][1em]{\vrule\@width\z@\@height#1\@depth\z@\relax}
\makeatother
\definecolor{lightgray}{HTML}{dddddd}
\definecolor{medgray}{HTML}{cccccc}
\definecolor{medgray2}{HTML}{bbbbbb}
\definecolor{darkgray}{HTML}{aaaaaa}
\definecolor{darkgray2}{HTML}{666666}
\definecolor{lightgp}{HTML}{ddddee}
\definecolor{lightyellow}{HTML}{ffa529}
%\begin{tikzpicture}[x=1.12cm, scale=0.8, every node/.append style={transform shape}]
\begin{tikzpicture}[x=1.12cm, scale=0.7, every node/.append style={transform shape}]
    \tikzstyle{data}=[inner sep=+0pt, drop shadow={shadow xshift=0.3ex,shadow yshift=-0.3ex}]
    \begin{scope}[color=darkgray2]
        \begin{scope}[rotate around={90:(-6, -3)}]
        \node[draw] (key) at (-6, -3) {\texttt{$\cdots$01001001101010101101101001110001010100}};
        \node[draw] (hkey) at (-6, -4.5) {\texttt{$\cdots$00000010 10100001 01000010}};
        \node[anchor=south] at (key.north) {user key};
        \path[->, line width=0.2ex] (key) edge[out=south, in=north] node[sloped,above] {$h_k(\cdot)$} (hkey);
        \draw[decorate,decoration={brace,amplitude=2ex, mirror}, yshift=0.5ex, xshift=14.5ex] (-7, -5) -- (-5.75, -5) node [midway] (c1) {\footnotesize\texttt{0x42}};
        \draw[decorate,decoration={brace,amplitude=2ex, mirror}, yshift=0.5ex, xshift=4ex] (-7, -5) -- (-5.75, -5) node [midway] (c2) {\footnotesize\texttt{0xa1}};
        \draw[decorate,decoration={brace,amplitude=2ex, mirror}, yshift=0.5ex, xshift=-6ex] (-7, -5) -- (-5.75, -5) node [midway] (c3) {\footnotesize\texttt{0x02}};
        \end{scope}
    \end{scope}
    \begin{scope}[all/.style={draw, minimum height=0.5cm, minimum width=0.5cm},line width=0.08ex]
        \setlength{\tabcolsep}{0.6ex}
        \newcommand{\tNodeChild}[1]{\rotatebox[origin=c]{90}{\footnotesize\texttt{\space#1\space}}}
        \newcommand{\mydots}{...}
        \node[all, data, draw,fill=white] (u0) at (0, 0) {\begin{tabular}{c|c|c|c|c|c}\multicolumn{6}{c}{$u_0$}\\\hline\tNodeChild{00}&\tNodeChild{01}&\mydots&\tNodeChild{42}&\tNodeChild{43}&\mydots\end{tabular}};
        \node[all, data, draw,fill=white] (u1) at (-1, -2) {\begin{tabular}{c|c|c|c|c|c}\multicolumn{6}{c}{$u_1$}\\\hline\tNodeChild{00}&\mydots&\tNodeChild{a1}&\mydots&\tNodeChild{e4}&\mydots\end{tabular}};
        \node[all, data, draw,fill=white] (u2) at (1.5, -2) {\begin{tabular}{c|c|c|c|c}\multicolumn{5}{c}{$u_2$}\\\hline\tNodeChild{00}&\tNodeChild{01}&\tNodeChild{02}&\tNodeChild{03}&\mydots\end{tabular}};
        \node[all, data, draw,fill=white] (u3) at (-2.1, -4) {\begin{tabular}{c|c|c|c|c}\multicolumn{5}{c}{$u_3$}\\\hline\tNodeChild{00}&\tNodeChild{01}&\tNodeChild{02}&\tNodeChild{03}&\mydots\end{tabular}};
        \node[all, data, draw,fill=white] (u4) at (0.1, -4) {\begin{tabular}{c|c|c|c|c}\multicolumn{5}{c}{$u_4$}\\\hline\tNodeChild{00}&\tNodeChild{01}&\tNodeChild{02}&\tNodeChild{03}&\mydots\end{tabular}};
        \node[all, data, draw,fill=lightyellow!20] (u5) at (-2.1, -6.5) {\begin{tabular}{c|c}\multicolumn{2}{c}{$u_5$}\\\hline$(\key_1, \mathit{value}_1)$ & $\mathit{next}$\end{tabular}};
        \node[all, data, draw,fill=lightyellow!20] (u6) at (1.5, -6.5) {\begin{tabular}{c|c}\multicolumn{2}{c}{$u_6$}\\\hline$(\key_2, \mathit{value}_2)$ & $\nnull$\end{tabular}};
        \node[fill=none] (x) at (-2.1, -5.3) {\rotatebox[origin=c]{-90}{$\cdots$}};

        \begin{scope}[line width=0.1ex]
        %\path[->] (u0) ++ (0.2, -0.5) --
        \draw ($(u0) + (-0.85, -0.52)$) -- ($(u0) + (-1, -0.8)$);
        \draw ($(u0) + (-0.85, -0.52) + (0.3, 0)$) -- ($(u0) + (-1, -0.8) + (0.35, 0)$);
        \draw ($(u0) + (-0.75, -0.52) + (0.55, 0)$) -- ($(u0) + (-1, -0.8) + (0.75, 0)$);
        \draw ($(u0) + (0.85, -0.52)$) -- ($(u0) + (1, -0.8)$);

        \draw ($(u1) + (-0.95, -0.52)$) -- ($(u1) + (-1.2, -0.8)$);
        \draw ($(u1) + (-0.55, -0.52)$) -- ($(u1) + (-1.2, -0.8) + (0.45, 0)$);
        \draw ($(u1) + (0.18, -0.52)$) -- ($(u1) + (0.21, -0.8)$);
        \draw ($(u1) + (0.95, -0.52)$) -- ($(u1) + (1.1, -0.8)$);

        \draw ($(u2) + (-0.65, -0.52)$) -- ($(u2) + (-1, -0.8)$);
        \draw ($(u2) + (0, -0.52)$) -- ($(u2) + (-0.1, -0.8)$);
        \draw ($(u2) + (-0.65, -0.52) + (0.3, 0)$) -- ($(u2) + (-1, -0.8) + (0.45, 0)$);
        \draw ($(u2) + (0.3, -0.52)$) -- ($(u2) + (0.35, -0.8)$);
        \draw ($(u2) + (0.65, -0.52)$) -- ($(u2) + (1, -0.8)$);

        \draw ($(u3) + (-0.65, -0.52)$) -- ($(u3) + (-1, -0.8)$);
        \draw ($(u3) + (-0.65, -0.52) + (0.3, 0)$) -- ($(u3) + (-1, -0.8) + (0.45, 0)$);
        \draw ($(u3) + (0.3, -0.52)$) -- ($(u3) + (0.35, -0.8)$);
        \draw ($(u3) + (0.65, -0.52)$) -- ($(u3) + (1, -0.8)$);

        \draw ($(u4) + (-0.65, -0.52)$) -- ($(u4) + (-1, -0.8)$);
        \draw ($(u4) + (0, -0.52)$) -- ($(u4) + (-0.1, -0.8)$);
        \draw ($(u4) + (-0.65, -0.52) + (0.3, 0)$) -- ($(u4) + (-1, -0.8) + (0.45, 0)$);
        \draw ($(u4) + (0.3, -0.52)$) -- ($(u4) + (0.35, -0.8)$);
        \draw ($(u4) + (0.65, -0.52)$) -- ($(u4) + (1, -0.8)$);

        \end{scope}

        \begin{scope}[line width=0.1ex]
        \path[->] (u0) ++ (0.15, -0.52) edge[out=south, in=north] node[sloped,above] {} (u1) ;
        \path[->] (u0) ++ (0.5, -0.52) edge[out=south, in=north] node[sloped,above] {} (u2) ;
        \path[->] (u1) ++ (-0.2, -0.52) edge[out=south, in=north] node[sloped,above] {} (u3) ;
        \path[->] (u1) ++ (0.55, -0.52) edge[out=south, in=north] node[sloped,above] {} (u4) ;
        \path[->] (u3) ++ (0, -0.52) edge[out=south, in=north] node[sloped,above] {} (x) ;
        \path[->] (x) edge[out=south, in=north] node[sloped,above] {} (u5) ;
        \path[->] (u5) ++ (1.37, -0.3) edge[out=east, in=west] node[sloped,above] {} (u6);
        \end{scope}

        \begin{scope}[color=darkgray2,dashed,line width=0.2ex]
            \path[->] (c1) ++ (0.4, 0) edge[out=0, in=180] node[sloped,above] {} ( $(u0) + (-1.1, -0.2)$ ) ;
            \path[->] (c2) ++ (0.4, 0) edge[out=0, in=180] node[sloped,above] {} ( $(u1) + (-1.15, -0.2)$ ) ;
            \path[->] (c3) ++ (0.4, 0) edge[out=0, in=180] node[sloped,above] {} ( $(u3) + (-0.9, -0.2)$ ) ;
        \end{scope}
    \end{scope}
\end{tikzpicture}
}
\vspace{0.03in}
\captionof{figure}{Structure of a hash-trie with chained leaf nodes.}
\label{fig:hash-trie-overview}
\end{center}
\end{figure}
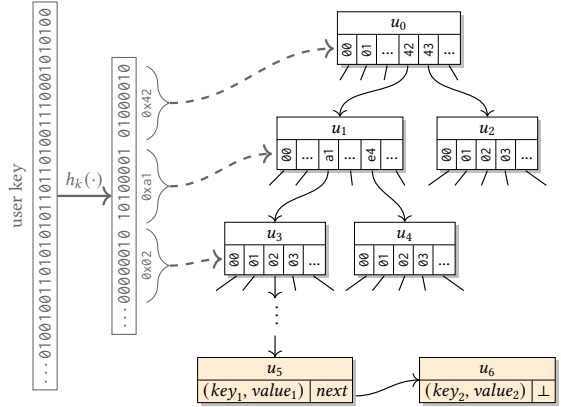

\subsection{Hash-Trie for Persistent Storage}
A \emph{trie} is a tree structure that encodes all prefix
paths of inserted key strings.
Each key is treated as a sequence of consecutive fixed-width \emph{characters}.
A trie stores paths of edges representing character sequences, collapsing
all shared prefix into a tree topology.

A \emph{hash-trie} is a trie indexed by a hash.
\DB uses a strong (well-distributed) and fast hash function (\S\ref{sec:implementation}) to
map an arbitrary key string given by the user to a 256-bit hash.
It then partitions the hash into equal-length characters.
In the trie, a \emph{tree node} consists of a character-indexed array
of pointers to its child nodes and a pointer referring back to its
parent node. We define the \emph{height} of a given node to be the
number of its ancestors.
% Thus, the root node has a height of zero.
Each leaf node maintains a linked list of user key-value pairs.
Figure~\ref{fig:hash-trie-overview} depicts
the basic structure of a hash-trie, where the user key $\key_1$ and $\key_2$ in data node $\node_5$ and $\node_6$ have the same hash prefixed by \verb|0x42a102|.
The size of the hash should be chosen so that the probability of such
collisions is small.

The insertion algorithm starts from the root node. It first visits
the child indexed by the first character of the key hash, proceeds
to the next child by the second character, and recursively walks
down the trie until the entire key sequence is consumed. Child nodes
are created as needed. At the leaf node, the algorithm adds a
\emph{data node} containing the original key-value data of the user
to the linked list.

Lookup is similar to insertion. When reaching the leaf node,
the linked list is scanned doing a full comparison using the original,
unhashed user key to locate the value.

%[Ted: remove prob. description]
%[RVR: the following is non-sensical I think.  The probability of a
%collision depends on the number of items that are being hashed, and
%is subject to the infamous birthday paradox.]
%Because the hash function is cryptographically pseudo-random,
%the probability of having a hash collision in the list is
%$2^{-256} \approx 10^{-77}$.
%
%[RVR: the following, again, depends much on the number of items in the
%trie.]
%Such possible hash collision
%will not affect correctness since all key-value pairs with the same hash key
%are preserved by the linked list at the leaf, and a complete key comparison
%is required while scanning through the list during a lookup. Since the
%probability is extremely low, one would expect on average a list of a single data node for
%any realistic storage size. However, it is possible to use shorter (say,
%128-bit, or even 64-bit) hash which could reduce the maximum tree height at the cost of
%the list scan.

There are some similarities between the hash-trie and the \bptree
data structures.
First, both are balanced, n-ary tree structures.
In practice, a typical \bptree has a branching factor of several hundreds.
As we show later in the evaluation, a good choice for the branching factor
for the hash-trie is hundreds of children per node as well.
Second, both \bptree{}s and hash-tries only store
user data at the leaves---intermediate nodes only contain metadata for indexing.

That being said, the structures have important differences.
The height of a \bptree is logarithmic in the number of keys.
When inserting data, a \bptree{} has to constantly
adjust its topology across sibling subtrees to maintain tree balance.
In a hash-trie, the path to the leaf node for a specific key is
\emph{static}: there is no reorganization of this path when
other data is inserted or deleted.  This significantly reduces the I/O
cost of maintaining the internal index structure and simplifies concurrent
access or modification to the tree, by sacrificing the support for range queries. Finally, all tree nodes of a hash-trie
have an identical size and the same storage footprint, which simplifies
the storage maintenance and reduces fragmentation. \bptree{} nodes
have a variable number of children, bounded by the branching factor.

A disadvantage of the hash-trie design is that operations always need to
visit a fixed number of tree nodes even if the data store is small.
Also, the utilization of child tables in tree nodes can be low,
potentially resulting in significant write amplication.
We next show, in two steps, how to improve upon the hash-trie structure to
build a practical, dynamically grown \emph{lazy-trie} with near-optimal
height and small storage overhead by utilizing
statistical properties of the hash function.

\begin{figure}
\begin{center}
{%
\makeatletter
\newcommand*{\Strut}[1][1em]{\vrule\@width\z@\@height#1\@depth\z@\relax}
\makeatother
\definecolor{lightgray}{HTML}{dddddd}
\definecolor{medgray}{HTML}{cccccc}
\definecolor{medgray2}{HTML}{bbbbbb}
\definecolor{darkgray}{HTML}{aaaaaa}
\definecolor{darkgray2}{HTML}{666666}
\definecolor{lightgp}{HTML}{ddddee}
\definecolor{lightorange}{HTML}{ff6929}
\definecolor{lightblue}{HTML}{246aa5}
\definecolor{lightyellow}{HTML}{ffa529}
\begin{tikzpicture}[x=1.12cm, scale=0.6, every node/.append style={transform shape}]
    \tikzstyle{data}=[inner sep=+0pt, drop shadow={shadow xshift=0.3ex,shadow yshift=-0.3ex}]
    \begin{scope}[all/.style={draw, minimum height=0.5cm, minimum width=0.5cm},line width=0.08ex]
        \setlength{\tabcolsep}{0.6ex}
        \newcommand{\tNodeChild}[1]{\rotatebox[origin=c]{90}{\footnotesize\texttt{\space#1\space}}}
        \newcommand{\mydots}{...}
        \foreach \n/\k/\x/\y/\c in {%
            1/1/-1.5/-1/27,%
            3/3/-2.2/-2/0e,%
            4/4/0/-2/5d,%
            5/5/2/-2/e9,%
            6/6/-3/-3/bb,%
            7/7/0.5/-3/21,%
            8/8/2/-3/8}
        {%
            \node[all, data, draw, fill=lightorange!20] (u\n) at (\x * 1.25, \y * 1.7) {%
                \begin{tabular}{c|c|c}\multicolumn{3}{c}{$\node_\k$}\\\hline$\nnull$& \tNodeChild{\c} & $\nnull$\end{tabular}};
        }
        \foreach \n/\k/\x/\y/\ca/\cb in {0/0/0/0/9c/b3, 2/2/1/-1/52/6f, 0a/0/4.5/0/9c/b3, 2a/2/5.5/-1/52/6f}
        {%
            \node[all, data, draw,fill=lightblue!20] (u\n) at (\x * 1.25, \y * 1.7) {
                \begin{tabular}{c|c|c|c|c}\multicolumn{5}{c}{$\node_\k$}\\\hline$\nnull$&\tNodeChild{\ca}&$\nnull$&\tNodeChild{\cb}&$\nnull$\end{tabular}};
        }
        \foreach \n/\k/\x/\y in {1/1/-3/-4,2/2/0.5/-4,3/3/2/-4,1a/1/3.5/-1,2a/2/4.2/-2,3a/3/6.2/-2}
            \node[all, data, draw,fill=lightyellow!20] (d\n) at (\x * 1.25, \y * 1.7) {\begin{tabular}{c|c}\multicolumn{2}{c}{$\dataNode_\k$}\\\hline \mydots & $\mathit{next}$\end{tabular}};

        \begin{scope}[line width=0.1ex]
        \path[->] (u0.south) ++ (-0.35, 0) edge[out=south, in=north] node[sloped,above] {} (u1) ;
        \path[->] (u0.south) ++ (0.35, 0) edge[out=south, in=north] node[sloped,above] {} (u2) ;
        \path[->] (u1) edge[out=south, in=north] node[sloped,above] {} (u3) ;
        \path[->] (u3) edge[out=south, in=north] node[sloped,above] {} (u6) ;
        \path[->] (u2.south) ++ (-0.35, 0) edge[out=south, in=north] node[sloped,above] {} (u4) ;
        \path[->] (u2.south) ++ (0.35, 0) edge[out=south, in=north] node[sloped,above] {} (u5) ;
        \path[->] (u4) edge[out=south, in=north] node[sloped,above] {} (u7) ;
        \path[->] (u5) edge[out=south, in=north] node[sloped,above] {} (u8) ;
        \path[->] (u6) edge[out=south, in=north] node[sloped,above] {} (d1) ;
        \path[->] (u7) edge[out=south, in=north] node[sloped,above] {} (d2) ;
        \path[->] (u8) edge[out=south, in=north] node[sloped,above] {} (d3) ;

        \path[->] (u0a.south) ++ (-0.35, 0) edge[out=south, in=north] node[sloped,above] {} (d1a) ;
        \path[->] (u0a.south) ++ (0.35, 0) edge[out=south, in=north] node[sloped,above] {} (u2a) ;
        \path[->] (u2a.south) ++ (-0.35, 0) edge[out=south, in=north] node[sloped,above] {} (d2a) ;
        \path[->] (u2a.south) ++ (0.35, 0) edge[out=south, in=north] node[sloped,above] {} (d3a) ;
        \end{scope}
    \end{scope}
\end{tikzpicture}
}
%\vspace{-0.20in}
\captionof{figure}{The tree structure without (left) and with (right) the path compression.}
\label{fig:lazy-trie-compress}
\end{center}
\end{figure}
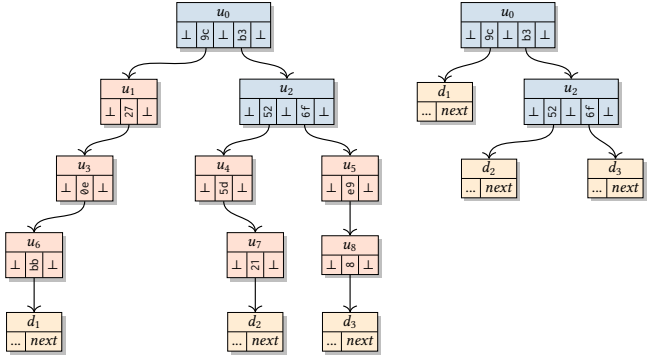

\begin{figure}
\begin{center}
\input{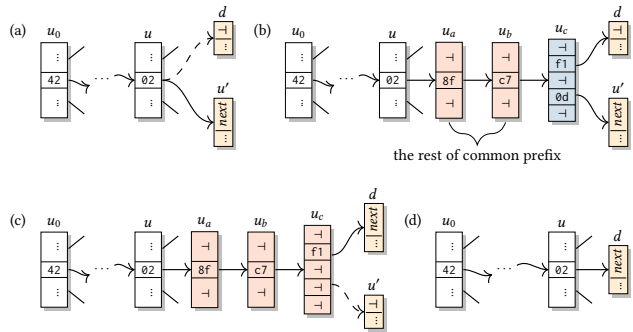}
%\vspace{-0.10in}
\captionof{figure}{The bookkeeping operations required by lazy-trie. (a)(b) show a split operation, when a new data node $\dataNode$ is inserted. (b)(c)(d) show a merge operation when the parent node $\node_c$ only has one child after the other child is removed.}
\label{fig:lazy-trie-split}
\end{center}
\end{figure}

\subsection{Path Compression}
Let the branching factor be $b$.
Then two keys have $b^{-h}$ probability of sharing the same prefix of
length~$h$.  The exponentially decreasing probability means that,
as one descends into the tree, it becomes less likely to have forks in the tree.
This led us to compress the suffix of paths in the hash-trie and only
unroll the compressed path when necessary.  Data nodes remain
at the leaves, so the collision rate remains unchanged.
Figure~\ref{fig:lazy-trie-compress} illustrates the new design.
Compared to radix trees, the difference is that we only compress
the \emph{suffixes} of paths---internal paths may still have nodes with only
one child.

Whenever inserting a new node, we lazily
create the minimum number of intermediate missing nodes just to distinguish
data nodes with different hashes. During the insertion, we first
follow the tree structure as usual.
If we reach a data node that is supposed to be a tree node, then we insert
a tree node, uncompressing the path
(see Figure~\ref{fig:lazy-trie-split}).
% Ted: see the note below
%and Algorithms~\ref{alg:lazy-trie-insert},\ref{alg:lazy-trie-split}).

The lookup algorithm stays essentially the same: \DB traverses the
lazy-trie according to the key hash until it reaches a data node,
then it scans through the data nodes.
%(Algorithm~\ref{alg:lazy-trie-lookup}).
In the delete algorithm,
when a data node is removed, a \emph{merge} operation checks whether it
is the only child of its parent node. If so, it collapses the
non-forking path repeatedly until reaching an ancestor having more
than one child.
%(see Algorithms~\ref{alg:lazy-trie-delete},\ref{alg:lazy-trie-merge}).

Unlike insertion or deletion in other tree structures, which require
complex recursive reorganization, there is at
most one non-recursive split for each insertion and one non-recursive merge for
each deletion. In a split, a simple path of intermediate nodes is created as a
chain, while, in a merge, the longest non-forking path is collapsed into a
direct parent reference. These operations only happen on a \emph{single path}
of insertion and deletion, so they do not interfere with any other siblings or
their subtrees.

%% NOTE: uncomment if #page < 12
%\begin{figure}[t]
%\small\input{code/algo-insert}
%\captionof{algorithm}{Path-compressed insertion.}\label{alg:lazy-trie-insert}
%\end{figure}
%
%\begin{figure}[t]
%\small\input{code/algo-split}
%\captionof{algorithm}{Decompress the path in a split.}\label{alg:lazy-trie-split}
%\end{figure}

%\begin{figure}[t]
%\small\input{code/algo-lookup}
%\captionof{algorithm}{\DB Lookup.}\label{alg:lazy-trie-lookup}
%\end{figure}

%\begin{figure}
%\small\input{code/algo-delete}
%\captionof{algorithm}{\DB Deletion.}\label{alg:lazy-trie-delete}
%\end{figure}
%
%\begin{figure}
%\small\input{code/algo-merge}
%\captionof{algorithm}{Merge operation for lazy-trie.}\label{alg:lazy-trie-merge}
%\end{figure}

As a result, lazy-trie dynamically adjusts the tree height depending
on how frequently the prefixes of inserted key hashes conflict. The
hash function offers uniformly distributed key hashes regardless of
the order of insertion, statistically balancing the tree with a
height of approximately
$\log_b{n}$, where $n$ is the number of keys.
Since the probability of a shared prefix decreases exponentially
with the length of the prefix, in expectation there will be
few missing intermediate nodes in a split operation.

\subsection{Sluggish Splitting}
\label{sec:lazy-trie-sluggish-splitting}

\begin{figure}[t]
\begin{center}
    \includegraphics[width=\linewidth]{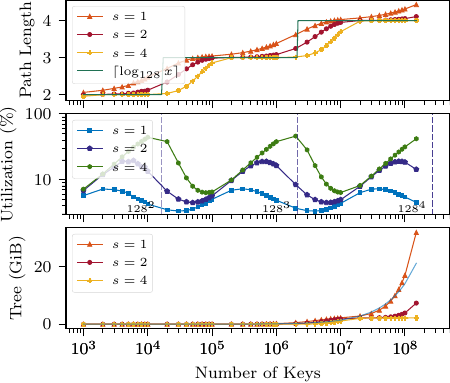}
    \captionof{figure}{
        Average path length, child table utilization, and meta-data
        storage footprint for a 128-ary lazy-trie as a function of the
        number of keys, with different sluggishness values.
        The solid blue line in the bottom graph shows the user data footprint
        for 23-byte keys and 128-byte values.}
	\label{fig:lazy-trie-sluggishness}
\end{center}
\end{figure}

The lazy-trie structure has an average path length that grows logarithmically
with the number of keys (Figure~\ref{fig:lazy-trie-sluggishness}, top, $s=1$).
However, due to the probabilistic nature, the paths for different keys can
have significantly different lengths and the height of the trie (the maximum
path length) can be significantly larger than the average path length.
Moreover, the child tables in the leaf nodes tend to be mostly empty,
leading to significant space underutilization because the majority of the
space in a node is its child table
(Figure~\ref{fig:lazy-trie-sluggishness}, middle, $s=1$).
The problem is that the lazy-trie must guarantee that keys with different
hashes cannot be in the same linked list.

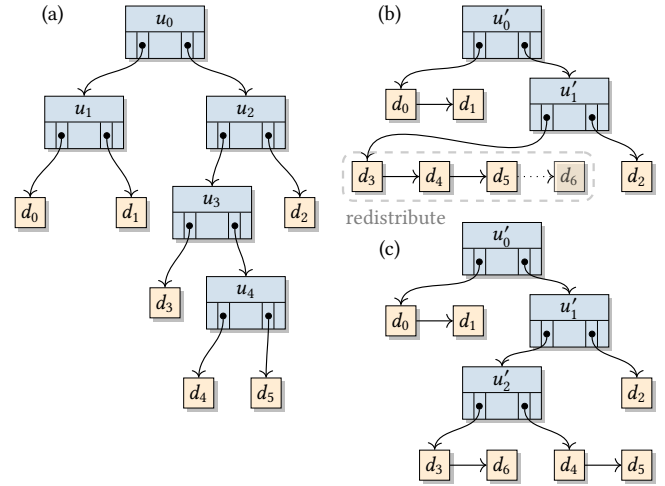
\begin{figure}[t]
\begin{center}
    {%
\makeatletter
\newcommand*{\Strut}[1][1em]{\vrule\@width\z@\@height#1\@depth\z@\relax}
\makeatother
\definecolor{lightgray}{HTML}{dddddd}
\definecolor{medgray}{HTML}{cccccc}
\definecolor{medgray2}{HTML}{bbbbbb}
\definecolor{darkgray}{HTML}{aaaaaa}
\definecolor{darkgray2}{HTML}{666666}
\definecolor{lightgp}{HTML}{ddddee}
\definecolor{lightorange}{HTML}{ff6929}
\definecolor{lightblue}{HTML}{246aa5}
\definecolor{lightyellow}{HTML}{ffa529}
\begin{tikzpicture}[x=1.12cm, scale=0.8, every node/.append style={transform shape}]
    \tikzstyle{data}=[inner sep=+0pt, drop shadow={shadow xshift=0.3ex,shadow yshift=-0.3ex}]
    \begin{scope}[all/.style={draw, minimum height=0.5cm, minimum width=0.5cm, scale=0.8},line width=0.08ex]
        \setlength{\tabcolsep}{0.6ex}
        \newcommand{\tNodeChild}[1]{\rotatebox[origin=c]{90}{\footnotesize\texttt{\space#1\space}}}
        \newcommand{\mydots}{...}
        
        \begin{scope}[every node/.style={all, data, fill=lightblue!20}]
            \node (u0a) at (0, 0) {\begin{tabular}{c|c|c|c|c}\multicolumn{5}{c}{$\node_0$}\\\hline&\tNodeChild{}&\hphantom{xx}&\tNodeChild{}&\end{tabular}};
            \node (u1a) at (-1.2, -1.5) {\begin{tabular}{c|c|c|c|c}\multicolumn{5}{c}{$\node_1$}\\\hline&\tNodeChild{}&\hphantom{xx}&\tNodeChild{}&\end{tabular}};
            \node (u2a) at (1.2, -1.5) {\begin{tabular}{c|c|c|c|c}\multicolumn{5}{c}{$\node_2$}\\\hline&\tNodeChild{}&\hphantom{xx}&\tNodeChild{}&\end{tabular}};
            \node (u3a) at (0.7, -3) {\begin{tabular}{c|c|c|c|c}\multicolumn{5}{c}{$\node_3$}\\\hline&\tNodeChild{}&\hphantom{xx}&\tNodeChild{}&\end{tabular}};
            \node (u4a) at (1.2, -4.5) {\begin{tabular}{c|c|c|c|c}\multicolumn{5}{c}{$\node_4$}\\\hline&\tNodeChild{}&\hphantom{xx}&\tNodeChild{}&\end{tabular}};

            \node (u0b) at (5, 0) {\begin{tabular}{c|c|c|c|c}\multicolumn{5}{c}{$\node'_0$}\\\hline&\tNodeChild{}&\hphantom{xx}&\tNodeChild{}&\end{tabular}};
            \node (u1b) at (6, -1.2) {\begin{tabular}{c|c|c|c|c}\multicolumn{5}{c}{$\node'_1$}\\\hline&\tNodeChild{}&\hphantom{xx}&\tNodeChild{}&\end{tabular}};

            \node (u0c) at (5, -3.6) {\begin{tabular}{c|c|c|c|c}\multicolumn{5}{c}{$\node'_0$}\\\hline&\tNodeChild{}&\hphantom{xx}&\tNodeChild{}&\end{tabular}};
            \node (u1c) at (6, -4.8) {\begin{tabular}{c|c|c|c|c}\multicolumn{5}{c}{$\node'_1$}\\\hline&\tNodeChild{}&\hphantom{xx}&\tNodeChild{}&\end{tabular}};
            \node (u2c) at (5, -6.0) {\begin{tabular}{c|c|c|c|c}\multicolumn{5}{c}{$\node'_2$}\\\hline&\tNodeChild{}&\hphantom{xx}&\tNodeChild{}&\end{tabular}};

        \end{scope}
        \begin{scope}[every node/.style={all, data, fill=lightyellow!20}]
            \node (d0a) at (-2, -3) {$d_0$};
            \node (d1a) at (-0.5, -3) {$d_1$};
            \node (d2a) at (2, -3) {$d_2$};
            \node (d3a) at (0, -4.5) {$d_3$};
            \node (d4a) at (0.5, -6) {$d_4$};
            \node (d5a) at (1.5, -6) {$d_5$};

            \node (d0b) at (3.5, -1.2) {$d_0$};
            \node (d1b) at (4.5, -1.2) {$d_1$};
            \node (d3b) at (3, -2.4) {$d_3$};
            \node (d4b) at (4, -2.4) {$d_4$};
            \node (d5b) at (5, -2.4) {$d_5$};
            \node (d2b) at (7, -2.4) {$d_2$};
            \node[opacity=0.7] (d6b) at (6, -2.4) {$d_6$};

            \node (d0c) at (3.5, -4.8) {$d_0$};
            \node (d1c) at (4.5, -4.8) {$d_1$};
            \node (d3c) at (4, -7.2) {$d_3$};
            \node (d6c) at (5, -7.2) {$d_6$};
            \node (d4c) at (6, -7.2) {$d_4$};
            \node (d2c) at (7, -6.0) {$d_2$};
            \node (d5c) at (7, -7.2) {$d_5$};

        \end{scope}
        \begin{scope}[
            line width=0.1ex,
            P/.style={postaction={
                    decorate,
                    decoration={
                        markings,
                        mark=at position 0 with {\arrow[scale=0.5]{*}},
                        mark=at position 0.99 with {\arrow[scale=1]{>}}}}}]
            \path[] (u0a) ++ (-2.5ex, -1.8ex) edge[out=south, in=north, P] node[sloped,above] {} (u1a);
            \path[] (u0a) ++ (2.5ex, -1.8ex) edge[out=south, in=north, P] node[sloped,above] {} (u2a);
            \path[] (u1a) ++ (-2.5ex, -1.8ex) edge[out=south, in=north, P] node[sloped,above] {} (d0a);
            \path[] (u1a) ++ (2.5ex, -1.8ex) edge[out=south, in=north, P] node[sloped,above] {} (d1a);
            \path[] (u2a) ++ (-2.5ex, -1.8ex) edge[out=south, in=north, P] node[sloped,above] {} (u3a);
            \path[] (u2a) ++ (2.5ex, -1.8ex) edge[out=south, in=north, P] node[sloped,above] {} (d2a);
            \path[] (u3a) ++ (-2.5ex, -1.8ex) edge[out=south, in=north, P] node[sloped,above] {} (d3a);
            \path[] (u3a) ++ (2.5ex, -1.8ex) edge[out=south, in=north, P] node[sloped,above] {} (u4a);
            \path[] (u4a) ++ (-2.5ex, -1.8ex) edge[out=south, in=north, P] node[sloped,above] {} (d4a);
            \path[] (u4a) ++ (2.5ex, -1.8ex) edge[out=south, in=north, P] node[sloped,above] {} (d5a);

            \path[] (u0b) ++ (-2.5ex, -1.8ex) edge[out=south, in=north, P] (d0b);
            \path[] (u0b) ++ (2.5ex, -1.8ex) edge[out=south, in=north, P] (u1b);
            \path[] (u1b) ++ (-2.5ex, -1.8ex) edge[out=south, in=north, P] (d3b);
            \path[] (u1b) ++ (2.5ex, -1.8ex) edge[out=south, in=north, P] (d2b);

            \path[] (u0c) ++ (-2.5ex, -1.8ex) edge[out=south, in=north, P] (d0c);
            \path[] (u0c) ++ (2.5ex, -1.8ex) edge[out=south, in=north, P] (u1c);
            \path[] (u1c) ++ (-2.5ex, -1.8ex) edge[out=south, in=north, P] (u2c);
            \path[] (u1c) ++ (2.5ex, -1.8ex) edge[out=south, in=north, P] (d2c);
            \path[] (u2c) ++ (-2.5ex, -1.8ex) edge[out=south, in=north, P] (d3c);
            \path[] (u2c) ++ (2.5ex, -1.8ex) edge[out=south, in=north, P] (d4c);
        \end{scope}
        \begin{scope}[line width=0.1ex]
            \path[->] (d0b) edge[out=east, in=west] (d1b);
            \path[->] (d3b) edge[out=east, in=west] (d4b);
            \path[->] (d4b) edge[out=east, in=west] (d5b);
            \path[->, dotted] (d5b) edge[out=east, in=west] (d6b);

            \path[->] (d0c) edge[out=east, in=west] (d1c);
            \path[->] (d3c) edge[out=east, in=west] (d6c);
            \path[->] (d4c) edge[out=east, in=west] (d5c);
        \end{scope}
        \node[draw, rounded corners, dashed, line width=0.2ex, fit=(d3b)(d4b)(d5b)(d6b), medgray] (redist) {};
        \node[anchor=north, gray, xshift=-8ex] at (redist.south) {redistribute};
        \node[xshift=-8ex, yshift=-1ex] (la) at (u0a.north west) {(a)};
        \node[xshift=-8ex, yshift=-1ex] (lb) at (u0b.north west) {(b)};
        \node[xshift=-8ex, yshift=-3ex] (lc) at (u0c.north west) {(c)};
    \end{scope}
\end{tikzpicture}
}
    %\vspace{-0.25in}
    \captionof{figure}{(a) a trie without sluggishness ($s = 1$).  (b) in a sluggish lazy-trie, an insertion ($d_6$) can lead
    to overflowing the linked list of data nodes.
    (c) for a maximum sluggishness of 3, the data nodes are redistributed.}
    \label{fig:lazy-trie-split-sluggish}
\end{center}
\end{figure}

To reduce variance in path lengths and improve utilization,
we allow keys with different hashes to share
the same linked list.  We define \emph{sluggishness} to be the
maximum number of hash values allowed in a linked list of
data nodes. The sluggishness bounds the worst-case scanning time in the linked list.
A \emph{sluggish lazy-trie} will only split the path when a linked list
overflows.  If so, the linked list will be replaced with a leaf node, and
all data nodes in the linked list will be redistributed into the child
table of the new leaf node. The redistribution is recursive and
continues until the sluggishness constraint is met.
Figure~\ref{fig:lazy-trie-split-sluggish} shows an example of the
steps in redistribution.

To demonstrate how sluggish splitting
mitigates the problems of high path length variation and low utilization,
Figure~\ref{fig:lazy-trie-sluggishness} shows
the average path length and child table utilization as the data store grows
in size for different levels of sluggisness~$s$.
The periodic change of utilization is due to the statistical growth of
the tree height. The bottom graph shows the number
of bytes used by tree nodes as a function of the number of keys.
With a sluggishness $s=4$, the storage overhead is significantly
reduced compared to no sluggish splitting ($s=1$).

We only consider sluggishness in splitting and otherwise retain the
original merge algorithm. The only downside to not % aggressively
chaining leaves back into linked lists after deletion is that the tree
metadata do not optimally shrink back after items are removed.
However, left-over nodes will still be re-used when new items are
inserted.
% Doing so simplifies maintenance and is a practical tradeoff.
We confirmed that the tree footprint does not increase over time
by running random insertion/deletion (50\% for each) experiments
with % up to
1 billion operations on an initial store of 100 million items.

\section{System Design and Implementation}
\label{sec:implementation}

In this section, we show how a lazy-trie can be used
to build \DB, a high-performance key-value store.
% from the objects used in memory
% and the syscalls we utilize to enable fast access, to how changes are made to the
% persistent storage and optimizations used for improving concurrency.
% \textbf{Programming Language.}
\DB has been fully implemented in
\textasciitilde{}8K lines of pure Rust.  Rust is a modern systems programming language that provides statically
checked memory- and thread-safety guarantees~\cite{rust,servo16,redox}.
In addition to basic constructs offered by the standard library,
it allows programmers to customize their building blocks with
different safety guarantees~\cite{rustsafety}.  For \DB, we explored
how to separate the lazy-trie algorithm from the underlying storage management.
%
% \textbf{Operating System.}
We also made use of native functionalities of Linux,
some of which may not be available on other
operating systems, such as \verb|io_submit| for kernel-based
asynchronous IO (AIO).

\subsection{Logical Spaces}
\label{sec:storage-model}

\begin{figure}[t]
\begin{center}
\begingroup
\definecolor{lightorange}{HTML}{ff6929}
\definecolor{lightblue}{HTML}{246aa5}
\definecolor{lightyellow}{HTML}{ffa529}
\definecolor{medgray}{HTML}{aaaaaa}
\makeatletter
\newcommand*{\Strut}[1][1em]{\vrule\@width\z@\@height#1\@depth\z@\relax}
\makeatother
\newcommand{\nested}[9]{%
    \pgfmathtruncatemacro{\n}{#1 - 1}
    \pgfmathtruncatemacro{\nn}{\n - 1}
    {%
        \pgfmathtruncatemacro{\x}{0}
        \pgfmathsetmacro{\xx}{\x * #5 + #2}
        \pgfmathsetmacro{\yy}{0 + #3}
        #4{\xx}{\yy}{#6p\x}
    }
    \ifthenelse{\n > 1}
    {{%
        \pgfmathtruncatemacro{\xa}{1}
        \pgfmathsetmacro{\xx}{\xa * #5 + #2}
        \pgfmathsetmacro{\yy}{0 + #3}
        #4{\xx}{\yy}{#6p\xa}
    }}{}
    \ifthenelse{\n > 2}
    {{%
        \pgfmathtruncatemacro{\xb}{2}
        \pgfmathsetmacro{\xx}{\xb * #5 + #2}
        \pgfmathsetmacro{\yy}{0 + #3}
        #4{\xx}{\yy}{#6p\xb}
    }}{}
    \pgfmathtruncatemacro{\n}{#1 - 1}
    \node[] at (\n * #5 - #9 + #2, #3) {$\cdots$};
    \pgfmathsetmacro{\xx}{\n * #5 + #8 + #2}
    \pgfmathsetmacro{\yy}{0 + #3}
    #4{\xx}{\yy}{#6p\n}
    \begin{scope}[all/.style={draw},line width=0.08ex]
        \begin{pgfonlayer}{#7}
    \node[fit=(#6p0)(#6p\n),all,data, transform shape=false] (#6) {};
        \end{pgfonlayer}
    %\node[] at (#6.north) {#6};
    \end{scope}
}

\newcommand{\region}[3]{%
    \begin{scope}[all/.style={draw, minimum height=0.5cm, minimum width=0cm},line width=0.08ex]
        \message{#3}
        \setlength{\tabcolsep}{0.6ex}
        \setlength\extrarowheight{3ex}
        \node[all, data, inner sep=+0pt] (#3) at (#1, #2)
        %{\footnotesize #3};
        {\begin{tabular}{c|c|c|c}&&&\end{tabular}};
    \end{scope}
}

\newcommand{\file}[3]{\nested{3}{#1}{#2}{\region}{0.8}{#3}{lp}{0.5}{0.1}}
\newcommand{\disk}[3]{\nested{3}{#1}{#2}{\file}{3.2}{#3}{lr}{0.6}{0.2}}
\newcommand{\writebuffer}[3]{%
    \begin{scope}[all/.style={draw, minimum height=0.1ex, minimum width=0cm},line width=0.08ex]
    \foreach \x/\c in {0/lightorange, 1/lightblue, 2/lightblue} {%
        \setlength{\tabcolsep}{0.6ex}
        \node[circle, all, data, inner sep=+2.5pt, fill=\c!60] (#3w0w\x) at (#1 + \x * 0.2, #2) {};
    }
    \node[](#3w0wx) at (#1 + 2 * 0.2, #2){};
    \begin{scope}[all/.style={draw},line width=0.08ex]
        \begin{pgfonlayer}{lr2}
        \node[rounded corners=0.5ex, fit=(#3w0w0)(#3w0wx),all,data, transform shape=false] (#3w0) {};
        \end{pgfonlayer}
    \end{scope}

    \foreach \x/\c in {0/lightorange, 1/lightyellow} {%
        \setlength{\tabcolsep}{0.6ex}
        \node[circle, all, data, inner sep=+2.5pt, fill=\c!60] (#3w1w\x) at ($(#3w0wx) + (\x * 0.2 + 0.8, 0)$){};
    }
    \node[](#3w1wx) at ($(#3w0wx) + (1 * 0.2 + 0.8, 0)$){};

    \begin{scope}[all/.style={draw},line width=0.08ex]
        \begin{pgfonlayer}{lr2}
        \node[rounded corners=0.5ex, fit=(#3w1w0)(#3w1wx),all,data, transform shape=false] (#3w1) {};
        \end{pgfonlayer}
    \end{scope}

    \foreach \x/\c in {0/lightorange, 1/lightyellow, 2/lightblue, 3/lightyellow, 4/lightorange} {%
        \setlength{\tabcolsep}{0.6ex}
        \node[circle, all, data, inner sep=+2.5pt, fill=\c!60] (#3w2w\x) at ($(#3w1wx) + (\x * 0.2 + 0.8, 0)$) {};
    }
    \node[](#3w2wx) at ($(#3w1wx) + (4 * 0.2 + 0.8, 0)$){};

    \begin{scope}[all/.style={draw},line width=0.08ex]
        \begin{pgfonlayer}{lr2}
        \node[rounded corners=0.5ex, fit=(#3w2w0)(#3w2wx),all,data, transform shape=false] (#3w2) {};
        \end{pgfonlayer}
    \end{scope}
    \end{scope}

    \node[anchor=south] (#3w0l) at ($(#3w0.north) + (0, 0.1ex)$) {Insert};
    \node[anchor=south] (#3w1l) at ($(#3w1.north) + (0, 0.1ex)$) {Delete};
    \node[anchor=south] (#3w2l) at ($(#3w2.north) + (0, 0.1ex)$) {WriteBatch};
    \node[](#3wx) at ($(20 * 0.3, #2) + (0, -2ex)$){};
    \begin{scope}[all/.style={draw},line width=0.08ex]
        \begin{pgfonlayer}{lr}
        \node[fit=(#3w0l)(#3wx),all,data, transform shape=false] (#3) {};
        \end{pgfonlayer}
    \end{scope}
}

\newcommand{\mmap}[3]{%
    \pgfmathsetmacro{\xs}{0.8}
    \pgfmathsetmacro{\xsa}{1.4}
    \pgfmathsetmacro{\xsb}{0.5}
    \pgfmathtruncatemacro{\n}{5 - 1}
    \pgfmathtruncatemacro{\nn}{\n - 1}
    {%
        \pgfmathtruncatemacro{\x}{0}
        \pgfmathsetmacro{\xx}{\x * \xs + #1}
        \pgfmathsetmacro{\yy}{0 + #2}
        %\region{\xx}{\yy}{#3p\x}
        \begin{scope}[all/.style={draw, minimum height=0.5cm, minimum width=0cm},line width=0.08ex]
            \setlength{\tabcolsep}{0.6ex}
            \setlength\extrarowheight{3ex}
            \node[all, data, inner sep=+0pt] (#3p\x) at (\xx, \yy)
            %{\footnotesize #3};
            {\begin{tabular}{c|c|c|c}&\cellcolor{black!10}&&\end{tabular}};
            \node[all, data, inner sep=+0pt, anchor=south, yshift=2ex, xshift=-0.6ex] (#3pa\x) at (#3p\x.north)
            {\begin{tabular}{c}\cellcolor{lightorange!20}\end{tabular}};
        \end{scope}
    }
    {%
        \pgfmathtruncatemacro{\xa}{1}
        \pgfmathsetmacro{\xx}{\xa * \xs + #1}
        \pgfmathsetmacro{\yy}{0 + #2}
        %\region{\xx}{\yy}{#3p\xa}
        \begin{scope}[all/.style={draw, minimum height=0.5cm, minimum width=0cm},line width=0.08ex]
            \setlength{\tabcolsep}{0.6ex}
            \setlength\extrarowheight{3ex}
            \node[all, data, inner sep=+0pt] (#3p\xa) at (\xx, \yy)
            %{\footnotesize #3};
            {\begin{tabular}{c|c|c|c}\cellcolor{black!10}&&\cellcolor{black!10}&\end{tabular}};
            \node[all, data, inner sep=+0pt, anchor=south, yshift=2ex, xshift=0.6ex] (#3pa\xa) at (#3p\xa.north)
            {\begin{tabular}{c}\cellcolor{lightyellow!20}\end{tabular}};
            \node[all, data, inner sep=+0pt, anchor=south, yshift=2ex, xshift=-1.9ex] (#3pb\xa) at (#3p\xa.north)
            {\begin{tabular}{c}\cellcolor{lightblue!20}\end{tabular}};
        \end{scope}
    }
    {%
        \pgfmathtruncatemacro{\xb}{2}
        \pgfmathsetmacro{\xx}{\xb * \xs + #1}
        \pgfmathsetmacro{\yy}{0 + #2}
        %\region{\xx}{\yy}{#3p\xb}
        \begin{scope}[all/.style={draw, minimum height=0.5cm, minimum width=0cm},line width=0.08ex]
            \setlength{\tabcolsep}{0.6ex}
            \setlength\extrarowheight{3ex}
            \node[all, data, inner sep=+0pt] (#3p\xb) at (\xx, \yy)
            %{\footnotesize #3};
            {\begin{tabular}{c|c|c|c}&\cellcolor{black!10}&&\end{tabular}};
            \node[all, data, inner sep=+0pt, anchor=south, yshift=2ex, xshift=-0.56ex] (#3pa\xb) at (#3p\xb.north)
            {\begin{tabular}{c}\cellcolor{lightblue!20}\end{tabular}};
        \end{scope}
    }
    {%
        \pgfmathtruncatemacro{\xc}{3}
        \pgfmathsetmacro{\xx}{\xc * \xs + #1}
        \pgfmathsetmacro{\yy}{0 + #2}
        %\region{\xx}{\yy}{#3p\xc}
        \begin{scope}[all/.style={draw, minimum height=0.5cm, minimum width=0cm},line width=0.08ex]
            \setlength{\tabcolsep}{0.6ex}
            \setlength\extrarowheight{3ex}
            \node[all, data, inner sep=+0pt] (#3p\xc) at (\xx, \yy)
            %{\footnotesize #3};
            {\begin{tabular}{c|c|c|c}&&&\cellcolor{black!10}\end{tabular}};
            \node[all, data, inner sep=+0pt, anchor=south, yshift=2ex, xshift=2.00ex] (#3pa\xc) at (#3p\xc.north)
            {\begin{tabular}{c}\cellcolor{lightorange!20}\end{tabular}};
        \end{scope}
    }
    %{%
    %    \pgfmathtruncatemacro{\xd}{4}
    %    \pgfmathsetmacro{\xx}{\xd * \xs + #1}
    %    \pgfmathsetmacro{\yy}{0 + #2}
    %    \region{\xx}{\yy}{#3p\xd}
    %}
    \node[] at (\n * \xs + \xsa + \xsb, #2) {$\cdots$};
    \pgfmathsetmacro{\xx}{\n * \xs + \xsb + #1}
    \pgfmathsetmacro{\yy}{0 + #2}
    \region{\xx}{\yy}{#3p\n}
    \begin{scope}[all/.style={draw},line width=0.08ex]
        \begin{pgfonlayer}{lr}
        \node[fit=(#3p0)(#3p\n)(#3pa0)(#3pa\nn),all,data, transform shape=false, dashed] (#3) {};
        \end{pgfonlayer}
    \end{scope}
}

\begin{tikzpicture}[x=1.12cm, scale=0.7, every node/.append style={transform shape}]
    \tikzstyle{data}=[drop shadow={shadow xshift=0.3ex,shadow yshift=-0.3ex}, fill=white]
    \disk{0}{0}{p0}
    \mmap{2}{2.3}{p1}
    \writebuffer{1.9}{5.5}{w0}
    \begin{scope}[all/.style={draw},line width=0.08ex]
        \node[anchor=south,rotate=90] (p1l) at (p1.west) {Mapped};
        \node[anchor=south,rotate=90] (w0l) at (w0.west) {Write Buffer};
        \begin{pgfonlayer}{bg}
        \node[fit=(p1)(w0)(p1l)(w0l),all,data, transform shape=false] (ram) {};
        \end{pgfonlayer}
    \end{scope}
    \node[anchor=north,rotate=90](raml) at (ram.east) {Memory};
    \node[anchor=north](p0l) at (p0.south) {Persistent Storage};
    \node[data,draw](wal) at ($(p0.north east) + (-10ex, 18ex)$) {Disk Thread};
    \node[data,draw,anchor=north,rotate=90,minimum width=10ex](log) at ($(p0.east) + (1ex, 0)$) {WAL};
    \begin{scope}[line width=0.1ex]
        \path[<->] (p0p0p0) edge[out=north,in=south] node[] {} (p1p0);
        \path[<->] (p0p0p2) edge[out=north,in=south] node[] {} (p1p1);
        \path[<->] (p0p1p1) edge[out=north,in=south] node[] {} (p1p2);
        \path[<->] (p0p2p0) edge[out=north,in=south] node[] {} (p1p3);
        \path[<->] (p0p2p1) edge[out=north,in=south] node[] {} (p1p4);
        \begin{scope}[densely dotted]
        \path[->] (p1p0.north) ++ (-0.6ex, 0) edge[out=north, in=south] node[] {} (p1pa0);
        \path[->] (p1p1.north) ++ (0.6ex, 0) edge[out=north, in=south] node[] {} (p1pa1);
        \path[->] (p1p1.north) ++ (-1.9ex, 0) edge[out=north, in=south] node[] {} (p1pb1);
        \path[->] (p1p2.north) ++ (-0.56ex, 0) edge[out=north, in=south] node[] {} (p1pa2);
        \path[->] (p1p3.north) ++ (2ex, 0) edge[out=north, in=south] node[] {} (p1pa3);

        \path[->] (p1pa0) edge[out=north, in=south] node[] {} (w0w0);
        \path[->] (p1pa1) edge[out=north, in=south] node[] {} (w0w1);
        \path[->] (p1pb1) edge[out=north, in=south] node[] {} (w0w0);
        \path[->] (p1pa2) edge[out=north, in=south] node[] {} (w0w0);
        \path[->] (p1pa3) edge[out=north, in=south] node[] {} (w0w1);
        \end{scope}
    \end{scope}
    \begin{scope}[T/.style={draw, rounded corners}, line width=0.2ex, color=gray]
        \path[->,T] (w0.east) -| (wal.north) node[pos=0.3, above, sloped]{Pipeline};
        \path[->,T] (wal.south) -- ($(p0.north east) + (-10ex, 0)$) node[pos=0.5, below, sloped, rotate=0]{Block AIO};
        \path[->,T] (wal.east) -| (log.east) node[pos=0.7, below, sloped, rotate=180]{Record AIO};
        \path[->,T] (wal.east) -| (log.east) node[pos=0.3, above, sloped, yshift=1ex]{Write-Ahead};
    \end{scope}
    \draw[decorate,decoration={brace,amplitude=2ex,mirror}, yshift=0ex, xshift=0ex]
        ($(p0p0.south west) + (0, -0.3)$) -- ($(p0p0.south east) + (0, -0.3)$) node [midway, xshift=-4ex,yshift=-4.8ex] (b1) {segment (file)};
    \draw[decorate,decoration={brace,amplitude=1ex,mirror}, yshift=0ex, xshift=0ex]
        ($(p0p0p2.south west) + (0, -0.8)$) -- ($(p0p0p2.south east) + (0, -0.8)$) node [midway, yshift=-3ex] (b1) {region};

\end{tikzpicture}
\endgroup

%\vspace{-0.25in}
\captionof{figure}{\DB storage hierarchy with mapped memory and write buffer. The smallest rectangle represents a page.}
\label{fig:storage-model-overview}
\end{center}
\end{figure}
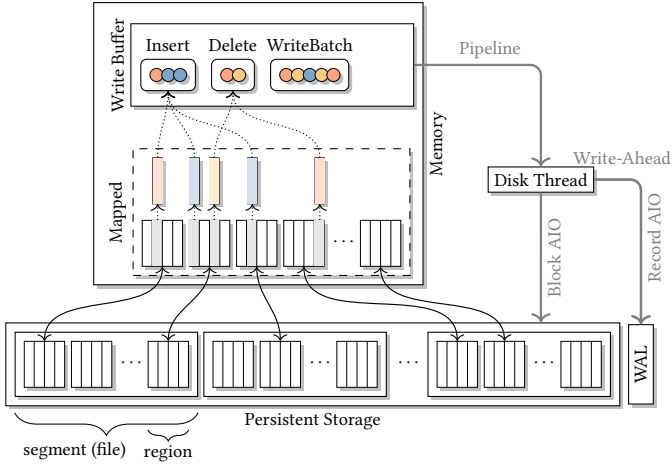

In some key-value stores, like LMDB~\cite{lmdb}, the entire store is always mapped
in memory and the maximum size has to be predetermined at its creation.
\DB supports a dynamically growing data set that may not all fit in memory.
To support this, it has \emph{logical spaces}, one for each type of objects.
A logical space is a 64-bit four layer virtual address space.
A \emph{logical address} is a 64-bit unsigned integer subdivided into four parts:
a segment number, a region number, a page number, and a page offset.
\DB associates a file with each segment.  Each region is either fully
mapped or unmapped.  When mapped, it can be accessed like ordinary memory,
that is, regions are memory-mapped that allow zero-copy read access.
Figure~\ref{fig:storage-model-overview} shows the organization of storage units with
different granularities.

Objects cannot span across regions.
Regions can only be accessed by mapping them, and thus the number of mapped
regions is effectively the cache size, blurring the boundary between the
``cache-based'' approach, used by stores like LevelDB and RocksDB, and the
``memory-based'' approach by stores like LMDB and Memcached.
Ideally, the performance is optimal when regions can remain mapped as long as
they are in active use, which is the main focus of this paper.

To map regions, we use \verb|mmap|.  While in theory we could let the
kernel keep track of dirty pages and write them back to the underlying
segment files, we found that this solution, while simple, did not
provide good performance even when \verb|madvise| is used.
The kernel ends up writing the same, actively modified pages repeatedly,
thus incurring prohibitively high write cost.
Moreover, when the bounded kernel buffer of pending writes is full, the kernel
slows down store instructions (such as x86 \texttt{mov}) made to the virtual
memory space, resulting in performance that is hard to predict.

The storage architecture of \DB is therefore hybrid.  We map the regions
in memory but keep track of our own write buffer for page writes.
Doing so also benefits write-ahead logging (\S~\ref{sec:crash-recovery}).

Logical spaces allow the lazy-trie algorithm to operate transparently
as if the entire data structure is in memory.
\DB maintains four logical spaces:
\begin{compactenum}
\item \emph{trie space}: tree nodes in use;
\item \emph{trie free list}: a stack of % indexes (pointers)
pointers to unused tree nodes;
\item \emph{data space}: data nodes in use;
\item \emph{data free list}: an array of descriptors tracking the unused portion of data space.
\end{compactenum}

\begin{figure}[t]
\begin{center}
\begingroup
\definecolor{lightorange}{HTML}{ff6929}
\definecolor{lightblue}{HTML}{246aa5}
\definecolor{lightyellow}{HTML}{ffa529}
\definecolor{medgray2}{HTML}{bbbbbb}
\definecolor{medgray}{HTML}{aaaaaa}
\newcommand{\myarraya}[6]{%
    \pgfmathsetmacro{\prev}{#1}
    \pgfmathsetmacro{\yy}{0 + #2}
    \pgfmathsetmacro{\yyy}{\yy + #5}
    \tikzstyle{data}=[preaction={transform canvas={shift={(0.4ex,-0.2ex)}},draw=medgray2,very thick}, line width=0.08ex, fill=white]
    \foreach \x/\xs/\b in {#3} {%
        \pgfmathsetmacro{\xx}{\prev}
        \pgfmathsetmacro{\tmp}{\prev + \xs}
        \global\let\prev\tmp
        \draw[draw=black, fill=white] (\xx * 1ex, \yy * 1ex) rectangle (\prev * 1ex, \yyy * 1ex) node[midway, minimum width=\xs * 1ex, minimum height=#5 * 1ex] (#4a\x) {};
        \ifthenelse{\b = 1}{%
            \draw[draw=black] (\xx * 1ex, \yy * 1ex) -- (\prev * 1ex, \yyy * 1ex) node[midway, minimum width=\xs * 1ex, minimum height=#5 * 1ex] (#4b\x) {};
        }{}
    }
    \pgfmathsetmacro{\pprev}{\prev + #6}
    \draw[draw=black] (\prev * 1ex, \yy * 1ex) -- (\pprev * 1ex, \yy * 1ex) node[midway, minimum height=#5 * 1ex] (#4ax) {};
    \draw[draw=black] (\prev * 1ex, \yyy * 1ex) -- (\pprev * 1ex, \yyy * 1ex) node[midway, minimum height=#5 * 1ex] (#4ax) {};
    \draw[draw=none] (\prev * 1ex, \yy * 1ex) rectangle (\pprev * 1ex, \yyy * 1ex) node[midway, minimum height=#5 * 1ex, minimum width=#6 * 1ex] (#4al) {$\cdots$};
    \begin{pgfonlayer}{bg}
    \draw[] (#1 * 1ex, \yy * 1ex) rectangle (\pprev * 1ex, \yyy * 1ex) node[midway, draw=white, minimum height=#5 * 1ex, minimum width=\pprev * 1ex, drop shadow={shadow xshift=0.3ex,shadow yshift=-0.3ex}, fill=white] (#4ax) {};
    \end{pgfonlayer}
}

\newcommand{\myarray}[6]{%
    \pgfmathsetmacro{\prev}{#1}
    \pgfmathsetmacro{\yy}{0 + #2}
    \pgfmathsetmacro{\yyy}{\yy + #5}
    \tikzstyle{data}=[preaction={transform canvas={shift={(0.4ex,-0.2ex)}},draw=medgray2,very thick}, line width=0.08ex, fill=white]
    \foreach \x/\xs/\c/\b in {#3} {%
        \pgfmathsetmacro{\xx}{\prev}
        \pgfmathsetmacro{\tmp}{\prev + \xs}
        \global\let\prev\tmp
        \draw[draw=black, fill=\c] (\xx * 1ex, \yy * 1ex) rectangle (\prev * 1ex, \yyy * 1ex) node[midway, minimum width=\xs * 1ex, minimum height=#5 * 1ex] (#4a\x) {};
    \ifthenelse{\b = 1}{%
        \draw[draw=black] (\xx * 1ex, \yy * 1ex) -- (\prev * 1ex, \yyy * 1ex) node[midway, minimum width=\xs * 1ex, minimum height=#5 * 1ex] (#4b\x) {};
    }{}

    }
    \pgfmathsetmacro{\pprev}{\prev + #6}
    \draw[draw=black] (\prev * 1ex, \yy * 1ex) -- (\pprev * 1ex, \yy * 1ex) node[midway, minimum height=#5 * 1ex] (#4ax) {};
    \draw[draw=black] (\prev * 1ex, \yyy * 1ex) -- (\pprev * 1ex, \yyy * 1ex) node[midway, minimum height=#5 * 1ex] (#4ax) {};
    \draw[draw=none] (\prev * 1ex, \yy * 1ex) rectangle (\pprev * 1ex, \yyy * 1ex) node[midway, minimum height=#5 * 1ex, minimum width=#6 * 1ex] (#4al) {$\cdots$};
    \begin{pgfonlayer}{bg}
    \draw[] (#1 * 1ex, \yy * 1ex) rectangle (\pprev * 1ex, \yyy * 1ex) node[midway, draw=white, minimum height=#5 * 1ex, minimum width=\pprev * 1ex, drop shadow={shadow xshift=0.3ex,shadow yshift=-0.3ex}, fill=white] (#4ax) {};
    \end{pgfonlayer}
}

\begin{tikzpicture}[x=1.12cm, scale=0.9, every node/.append style={transform shape}]
    \myarraya{0}{10}{0/2/0,1/2/0,2/2/0,3/2/0,4/2/0}{a0}{3}{10}
    \myarraya{0}{0}{0/7/0,1/6/1,2/6/0,3/6/1,4/6/1,5/6/1}{a1}{3}{10}
    \begin{scope}[myarrow/.style={decoration={markings,mark=at position 1 with {\arrow[scale=0.6,>=Latex]{>};}}, postaction={decorate}}]
        \draw[myarrow] (a0a0.south) -- ++(0, -4ex) -| (a1a1.north);
        \draw[myarrow] (a0a1.south) -- ++(0, -3ex) -| (a1a3.north);
        \draw[myarrow] (a0a2.south) -- ++(0, -1ex) -| (a1a5.north);
        \draw[myarrow] (a0a3.south) -- ++(0, -2ex) -| (a1a4.north);
    \end{scope}
    \node[anchor=east] (l1) at (a0a0.west) {trie free list};
    %\node[anchor=east] (l2) at (a1a0.west) {\begin{tabular}{r}node space or\\block data space\end{tabular}};
    \node[anchor=east] (l2) at (a1a0.west) {trie space};
    \node(l3) at (a1a0) {\footnotesize{}reserved};

    \myarray{0}{-6}{0/2/white/0,1/2/white/0,2/2/white/0,3/2/white/0,4/2/white/0}{a0}{3}{10}
    \myarray{0}{-16}{%
        0/1/lightblue!60/0,
        1/4/white/1,
        2/1/lightyellow!60/0,
        3/1/lightblue!60/0,
        4/10/white/0,
        5/1/lightyellow!60/0,
        6/1/lightblue!60/0,
        7/6/white/1,
        8/1/lightyellow!60/0,
        9/1/lightblue!60/0,
        10/3/white/0,
        11/1/lightyellow!60/0}{a1}{3}{10}
    \begin{scope}[
        myarrow/.style={decoration={markings,mark=at position 1 with {\arrow[scale=0.6,>=Latex]{>};}}, postaction={decorate}}]
        \draw[myarrow] (a0a0.south) -- ++(0, -3ex) -| (a1a1.north);
        \draw[myarrow] (a0a1.south) -- ++(0, -2ex) -| (a1a7.north);
    \end{scope}
    \begin{scope}[densely dashed,
        draw=lightblue,
        myarrow/.style={%
            decoration={markings,
            mark=at position 0.995 with {\arrow[scale=0.6,>=Latex]{>};}}, postaction={decorate}}]
        \draw[myarrow] (a1a0.north) to[out=90,in=-100,looseness=1.7] ($(a0a0.south) + (-0.5ex, 0)$);
        \draw[myarrow] (a1a6.north) to[out=90,in=-90,looseness=1.2] ($(a0a1.south) + (-0.5ex, 0)$);
    \end{scope}
    \begin{scope}[densely dashed,
        draw=lightyellow,
        myarrow/.style={%
            decoration={markings,
            mark=at position 0.9996 with {\arrow[scale=0.6,>=Latex]{>};}}, postaction={decorate}}]
        \draw[myarrow] (a1a2.south) to[out=-90,in=-70,looseness=2] (a1a0.south);
        \draw[myarrow] (a1a5.south) to[out=-90,in=-70] (a1a3.south);
        \draw[myarrow] (a1a8.south) to[out=-90,in=-70,looseness=1.5] (a1a6.south);
        \draw[myarrow] (a1a11.south) to[out=-90,in=-70,looseness=2.2] (a1a9.south);
    \end{scope}
    \node[anchor=east] (l1) at (a0a0.west) {data free list};
    \node[anchor=east] (l2) at (a1a0.west) {data space};
\end{tikzpicture}
\endgroup
%\vspace{-0.35in}
\captionof{figure}{Four logical spaces used in \DB.}
\label{fig:storage-model-space}
\end{center}
\end{figure}
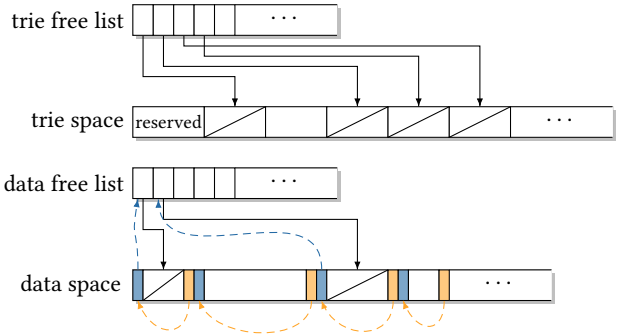

% \paragraph{Trie space.}
Tree nodes have a fixed size, and thus the trie space contains an array of
tree nodes.
The trie free list contains a stack of indexes into the tree node array.
To free a tree node, its index is pushed onto the stack.
To allocate a tree node, its index is popped from the stack.

% \paragraph{Data space.}
%Unlike most persistent stores, we use a similar layout as a standard
%heap allocator for data space allocation. This is because user data
%held in this space are always memory-mapped when accessed.
%[RVR: I don't like the prior sentence.  It's vague information that's
%better left out in my opinion.  I tried rewriting it but it still has
%little useful content.] Ted: Agree. It is subjective and could be weak (lack
%of support).
In the data space, in-use and free data nodes of different sizes
are stored continuously with a header at the front and a footer at the end.
The footer of a data node is right before the header of the next data node and
contains the size of the data node.  It supports merging of two adjacent
free data nodes (see below).
Instead of maintaining a doubly-linked list of free data nodes like the
free list in glibc's \verb|malloc|, \DB maintains a separate, unsorted
array of \emph{hole descriptors} for better locality and therefore
I/O efficiency (see Figure~\ref{fig:storage-model-space}).
Each descriptor points to a free data node, while the header of a free
data node points back to its descriptor.

\DB adopts a \emph{next-fit allocation policy}~\cite{johnstone1998memory, taocp},
by indexing into the array of hole descriptors. An index points to the last
visited descriptor and gets wrapped around when it hits the last item in the
array. The index pointer, together with the trie root pointer and other tail
pointers that indicate the boundary of logical spaces, are all stored within
the first (reserved) 4KBytes page of the trie space, before the entire tree node
array.
\subsection{Memory Abstraction and Layout}
To benefit from Rust's memory-safety and thread-safety guarantees,
we crafted our own memory-mapped object abstraction that fits in Rust's
idiom and thus utilizes its type checker.  Rust does not come natively with a type-safe
wrapper or primitive for memory-mapping.  Although nix~\cite{rustnix},
a popular Rust library,
provides related functions, they are exposed as
Rust function signatures of the original POSIX interface in C.
They are unsafe because the compiler makes normal assumptions about memory
management and variable life-times without making accommodations for
memory-mapped address space.

\subsubsection{Space and Objects}
% RVR: commented out this paragraph.  MappedMem is not used anywhere.
% The \texttt{MappedMem} trait is a general interface that describes a mapped
% chunk of memory established using \texttt{mmap}. It exposes the
% memory as a byte slice (\verb|&mut [u8]|). This interface is
% private and only used by other high-level abstractions like \texttt{MMappedSpace}.

The \texttt{MMappedSpace} struct is the core of implementing the logical space abstraction
(\S~\ref{sec:storage-model}).
It can be created from file handles and exposes safe methods.

We use an opaque struct to represent a pointer into logical space so that either it
can be reconstructed from persistent storage (like a pointer to an
existing tree node) or allocated through \texttt{MMappedSpace}.
The reason we need an abstraction for a pointer is two-fold.
First, addresses in logical spaces do not directly
correspond to virtual memory addresses even if the corresponding region is mapped.
Second, using an integer indexing into logical space would be unsafe.
Given a typed object pointer, \texttt{ObjPtr<T>}, it can dereference the pointer
to the correspondingly typed object handle, \texttt{ObjRef<T>}.
This allows manipulating the actual object typed \texttt{T} available in memory as
if there is no memory-mapping.
(It will auto-dereference to \verb|&mut T|, which is an ordinary mutable access to a
variable typed \texttt{T}.)
The object handle is accessible throughout its lifetime by pinning the affected region
in memory.

% Throughout our implementation of \DB, we make heavily use of these useful
% primitives and other similar ones to allow compilation-time checking for better
% safety.

\subsubsection{Tree Metadata}
%\begin{figure}
%\small
%\begin{minted}{rust}
%#[repr(C)]
%union ChdRaw { node: ObjPtr<Node>, data: ObjPtr<Data> }
%
%#[repr(C)]
%pub struct Node {
%	parent: ObjPtr<Node>, // parent node (for deletion)
%	height: u8,
%	pub pidx: NBranchType, // index in the parent node
%	data_mask: [u64; NBRANCH / 64], // data node bitmask
%	chd_mask: [u64; NBRANCH / 64], // validity bitmask
%	children: [ChdRaw; NBRANCH], // child table
%	// reserved field for late-initialization
%	pub rwlock: std::mem::MaybeUninit<RwLock<()>>
%}
%\end{minted}
%\captionof{figure}{Tree Node Layout.}
%\label{fig:tree-node}
%\end{figure}

A lazy-trie tree node consists predominantly of a fixed-size array of
pointers to its children.%, as shown in Figure~\ref{fig:tree-node}.
Two bitmasks
indicate the validity and type of the pointer in a particular
slot. To obtain child information, \verb|chd_mask| is first examined to determine
whether the slot contains a valid pointer.
If so, \verb|data_mask| specifies whether the pointer is to another tree node
(\verb|struct Node|) or a data node (\verb|struct Data|).
% As the most accessed component in the node structure,
Both bitmasks
are accessed using bitwise arithmetic.
On x86-64 we use inline assembly with BMI quadword instructions
such as \verb|bsfq| and \verb|bzhiq| for optimal performance.

We use the \verb|#[repr(C)]| compiler directive to ensure a C-struct layout,
guaranteeing that all memory-mapped objects can be correctly accessed even if the
Rust compiler changes the default layout of \verb|struct|.
Unfortunately we cannot take advantage of the
feature-rich Rust \verb|enum| type.
Thus, \verb|ChdRaw| (union type for each child pointer) is hidden to other parts of the \DB, which
access the trie data structure through safe enum structs and methods.

\label{sec:impl-user-data}
%\begin{figure}
%\small
%\begin{minted}[tabsize=4,obeytabs]{rust}
%#[repr(C)]
%pub struct Data {
%	lsize: u8, // the size of the linked list
%	hkey: [u8; HASH_LEN], // bytes of the hashed key
%	key_mode: KeyMode,
%	key_size: u64, // number of bytes in user key
%	val_size: u64, // number of bytes in value
%	next: ObjPtr<Data> // the next item in the list
%	// NOTE: the raw key-value bytes follow
%}
%\end{minted}
%%	dir: ObjPtr<DataDir>
%%// NOTE: both DataDir and DataSeg are page-aligned
%%pub struct DataSeg; /* a page of long user data */
%%#[repr(C)]
%%pub struct DataDir {
%%	nentry: u64,
%%	next: ObjPtr<DataDir>
%%	/* extra space until a page size */
%%}
%\captionof{figure}{Data Node Layout.}
%\label{fig:data-node}
%\end{figure}

\subsubsection{User Data}
% \paragraph{Data footprint.}
%We use a hybrid scheme for storing user data. For a short user key-value pair,
We use Google's HighwayHash~\cite{highwayhash16} to hash keys.
For each user key-value pair, we use an object of \verb|Data| struct
%(Figure~\ref{fig:data-node})
to hold the precomputed
HighwayHash of the original arbitrary-length key in \verb|hkey|,
and size information for the original key and value in \verb|key_size|
and \verb|val_size|.
The actual user key and value are placed directly after the Rust structure.
%, as long as the entire occupied space does not exceed the page size (4KBytes in Linux).
% That is, for an oversized key-value pair (most likely caused by
% having large value data, like dozens of kilobytes to megabytes),
% they will trigger an alternative approach.  In this case, such user
% data will not be entirely continuously stored as the extra content
% of \verb|Data|, but get partitioned into pages, \verb|DataSeg|. The
% pages could disperse in the block data space
% (Section~\ref{sec:block-data-space}) and they are organized as
% \verb|ObjPtr<DataSeg>| pointers in a page directory, following the
% fields in \verb|DataDir| struct. The page directory is also aligned
% to a full page and stored the same way as \verb|DataSeg|.  If the
% value is huge in size, more directories could be chained to accommodate
% all pages. This resembles a lightweight, one-level of indirection
% version of some filesystem (like ext2). While multiple levels of
% such indirections are possible, we find the single level suitable
% for the use for a key-value store, and thus it could be left as
% some future feature to support multi-levels of \verb|DataDir| and
% different bulk size for \verb|DataSeg|.
To support keys with colliding hashes and sluggish splitting,
\verb|Data| objects are
chained together into a singly-linked list using the \verb|next| pointer.
% The head of the linked list is
% the one pointed by the tree node slot and \verb|lsize| thereof keeps
% the size of the list for checking sluggishness limitation.
% \paragraph{Zero-copy read access.}
% As user key-value pairs are stored continuously in the data space,
% they could be referenced directly by the memory address allocated with
% \verb|MAP_PRIVATE|.
We use \verb|madvise|~\cite{madvise} to request that the kernel pre-fetches the memory pages
storing data objects.
%For large user data using \verb|DataSeg| and
%\verb|DataDir|, however, they are not continuously stored. Copying the pages
%into a temporary continuous space makes less sense as we already assumed long
%key-value data to be manipulated. Conversely, this is exactly the scenario
%where zero-copy could shine. Fortunately, there is \verb|mremap()| provided by Linux
%kernel, which is usually used to resize the memory-mapped space, but also
%allows changing the page table and relocating the already mapped portion of
%memory to a different base address, given \verb|MREMAP_MAYMOVE| and
%\verb|MREMAP_FIXED| as flags.  Therefore, when a large user data is
%accessed, \DB first temporarily \verb|mmap| some continuous memory that is
%not associated with any file via \verb|MAP_PRIVATE| and \verb|MAP_ANONYMOUS| as
%a placeholder, then ``scoops'' away all the data pages (\verb|DataSeg|, as they
%are all page-aligned, satisfying the requirement of the syscalls) from the
%block data space, and finally reassembles them to cover the temporary space
%using \verb|mremap|.  This will automatically release the original mapping of
%the continuous placeholder, and also leave ``holes'' on the block data
%space~\cite{mremap}. Finally, when the user is done accessing the content, \DB
%does \verb|mremap| again in the opposite direction to move back the pages to
%patch up the holes. The process is shown in Figure~\ref{fig:zero-copy-value}.

%\begin{figure}
%\begin{center}
%\input{figures/zero-copy-value}
%\captionof{figure}{Using \texttt{mremap()} to rearrange the data pages.}
%\label{fig:zero-copy-value}
%\end{center}
%\end{figure}

\subsection{Disk I/O}
As described in Section~\ref{sec:storage-model} and shown in
Figure~\ref{fig:storage-model-overview}, \DB uses a separate write buffer
to serialize all changes and % eventually
schedule them as \emph{block writes} to the physical
storage device.
Each modification made to the
internal lazy-trie data structure first writes % directly
to memory.\footnote{Due to limitations of \texttt{mmap}, we map the memory
in copy-on-write mode even though we do not require that the original contents is saved.}
The % state of the
memory thus always reflects the latest changes.
To update the underlying storage, the same write is also sent to a
\emph{disk thread} via a bounded buffer.

% \subsubsection{Block Writes}
The changes generated to the in-memory data structures
are short and frequent, which is not optimal
for secondary storage.  The disk thread therefore aggregates updates into blocks.
The Linux ext4 filesystem has the same block size as the page size, so \DB
uses 4KBytes blocks.

% \subsubsection{Asynchronous I/O}
While the lazy-trie data structure does not require log compaction such as
used in LSMs and is arguably simpler to maintain than {\bptree}s, it suffers
potentially from non-optimal locality and random writes when pointers need to be
updated.
To optimize storage updates, the \DB disk thread uses Linux native
Asynchronous I/O (AIO).
Not to be confused with the POSIX AIO standard offered by glibc~\cite{glibcaio},
Linux AIO performs concurrent writes if possible.  We access AIO through
\verb|libaio|, a thin C ABI wrapper that
is available on main-stream Linux distributions. % such as Ubuntu.
% and the other is the raw syscall interface
% provided by the kernel. Surprisingly, these two have different APIs and
% fundamentally different approaches. The glibc one is essentially an emulation
% of the given POSIX proposal using user-space threads to do writes
% asynchronously~\cite{glibcaio}.
AIO allows us to
asynchronously manage reads and writes in a non-blocking style.

Because \DB supports concurrent operations on the lazy-trie, it is possible
that multiple user threads make changes to the same page (block).
In \DB it is the disk thread's responsibility to obtain the consistent state of
a block from a file if it is not already available in its cache, rather than
copying the content from the mapped memory worked on by user threads, as there may be a data race.
The disk thread can schedule such an infrequent block read
% if a block first gets dirty by some write,
while at the same time it can schedule other writes % in the buffer
without being blocked.

\subsection{Crash Recovery}
\label{sec:crash-recovery}
\DB utilizes write-ahead logging (WAL) to achieve atomicity and durability~\cite{ARIES}.
In the disk thread I/O pipeline, disk write records are first fed into a WAL worker.
To better manage asynchronous I/O events, we implemented a library on top
of \verb|libaio| that manages I/O with \emph{futures}.
We then encapsulated the WAL logic into another library that schedules record writes via
its own I/O manager handle. The disk thread pushes a vector of records to
the WAL worker using the library, which immediately returns a future object
that gets resolved when the record write is complete.
The disk thread schedules block writes using its own I/O manager.
Likewise, when the future gets resolved upon completion,
the thread notifies the WAL worker that specific records can be pruned.

Similar to RocksDB, \DB WAL records are grouped into fixed sized chunks. Each
chunk contains a flag indicated if it is continued by the following one.
\DB operations may generate small writes to various locations in the logical space.
Encoding high-level descriptions of operations
directly as records does not work well for redos.
Instead, \DB encodes all actual low-level disk writes in each record.
To avoid write amplification,
it uses a compact format for records.
A record is the concatenation of several subrecords, each of which encodes a single
update made to a logical space.
\begin{comment}
The subrecord starts
with a 1 byte space ID, followed by 8 byte space offset. Then a 1 byte integer
$t$ is used to determine the type of the following length integer. When $0 < t <
\texttt{0xff}$, $t$ itself encodes the length of the write payload, which is sufficient
for most of the tree writes as they are short. When $t = \texttt{0xff}$, it
suggests the next 16-bit integer will be the length. Finally $t = \texttt{0x00}$
is for a full 64-bit integer that covers the maximum possible length.
\end{comment}

%\subsection{Error Handling for In-Memory Updates}
%[RVR: It's not clear what problem this section is solving.  Maybe it needs an example.]
%The lazy-trie structure is kept in logical spaces implemented by memory-mapping
%regions from files dynamically. Although crashing does not affect the
%consistency of the persisted lazy-trie thanks to WAL, for more practical use,
%\DB has to handle potential errors during the in-memory lazy-trie access. To
%safely handle this with reasonable overhead, \DB keeps the old value of the
%changed memory part during an operation only when necessary (which is
%usually short updates to tree nodes). When all in-memory operations are finished
%without error, it simply discards the old values before scheduling disk writes;
%otherwise, the in-memory changes are rolled back with the old values before the
%flow control gets back to the user.  This improves user experience as the
%in-memory state of lazy-trie is not corrupted by access failure in logical
%space.

\subsection{Concurrent Access}
The lazy-trie design does not require tree re-balancing operations, simplifying
concurrent access.  Because walks down the tree diverge exponentially fast,
gains from concurrent access can be significant.

\subsubsection{Tree-Walk Parallelism}
Locating the leaf node does not change the trie structure, while
insertion or deletion only makes changes starting from a leaf node.
We assign a reader-writer lock to each tree node to control any
access that goes through that node.
% We say that a node is
% \emph{read (write) guarded} when its read (write) lock is held.

First consider lookup operations.  For each node visited on the path to the leaf
node, a thread acquires the read lock.  Knowing that once the thread holds a read lock,
no concurrent updates can happen to the node or any node below, it is safe for the
thread to release the lock on the parent node, allowing concurrent updates to other
parts of the trie.

% NOTE: the following approach is still incorrect because an insertion could
% have ABA issue with another one.
%
%Insertions are similar to lookups in the tree walk, unless it needs the write
%access when it finally reaches the leaf node. However, one can not switch
%from a reader lock to a writer lock by simply reacquiring, because the
%node object in question could be removed by other deletion. Upgradable reader
%locks are tricky to use, and the available one in Rust can only share access
%with regular reader locks, so it is still mutually exclusive among themselves.
%Instead, we take an alternative approach by utilizing the tree topology.  Like
%lookups, the reader locks are altered with overlapping period. The difference
%is we still always keep the reader lock for the parent node during the period of
%accessing a node. The read guard at the parent guarantees the current node is
%always valid when the reader lock needs to be changed to a writer lock when the
%walk ends.

In theory, insertions could also obtain read locks similar to lookups, until
the thread needs to update a node.  At that time, the thread would have to
upgrade its read lock to a write lock.  In practice, Rust does support an upgradable
reader-writer lock, but it only allows at most one thread to hold an upgradable read
lock at a time (while other threads may concurrently hold a regular read lock).
The lock also supports a downgrade operation that converts an upgradable read lock
into a regular read lock.  We use it as follows: insertion proceeds as lookups but
obtaining upgradable read locks.  As soon as the thread determines that it will not
need to update the node, it downgrades the lock before attempting to get an
upgradable read lock on the next node down the path.

Deletion is more complex as it may remove a path that is being followed by
other concurrent threads. \DB assumes deletions are much less frequent than
other operations.  Based on this assumption, a thread performing a delete operation
obtains write locks for the entire path, ruling out some concurrent access.

\begin{figure}
\begin{center}
    \begingroup
\definecolor{lightorange}{HTML}{ff6929}
\definecolor{lightblue}{HTML}{246aa5}
\definecolor{lightyellow}{HTML}{ffa529}
\definecolor{medgray}{HTML}{aaaaaa}
\makeatletter
\newcommand*{\Strut}[1][1em]{\vrule\@width\z@\@height#1\@depth\z@\relax}
\makeatother
\newcommand{\nested}[9]{%
    \pgfmathtruncatemacro{\n}{#1 - 1}
    \pgfmathtruncatemacro{\nn}{\n - 1}
    {%
        \pgfmathtruncatemacro{\x}{0}
        \pgfmathsetmacro{\xx}{\x * #5 + #2}
        \pgfmathsetmacro{\yy}{0 + #3}
        #4{\xx}{\yy}{#6p\x}
    }
    \ifthenelse{\n > 1}
    {{%
        \pgfmathtruncatemacro{\xa}{1}
        \pgfmathsetmacro{\xx}{\xa * #5 + #2}
        \pgfmathsetmacro{\yy}{0 + #3}
        #4{\xx}{\yy}{#6p\xa}
    }}{}
    \ifthenelse{\n > 2}
    {{%
        \pgfmathtruncatemacro{\xb}{2}
        \pgfmathsetmacro{\xx}{\xb * #5 + #2}
        \pgfmathsetmacro{\yy}{0 + #3}
        #4{\xx}{\yy}{#6p\xb}
    }}{}
    \pgfmathtruncatemacro{\n}{#1 - 1}
    \node[] at (\n * #5 - #9 + #2, #3) {$\cdots$};
    \pgfmathsetmacro{\xx}{\n * #5 + #8 + #2}
    \pgfmathsetmacro{\yy}{0 + #3}
    #4{\xx}{\yy}{#6p\n}
    \begin{scope}[all/.style={draw},line width=0.08ex]
        \begin{pgfonlayer}{#7}
    \node[fit=(#6p0)(#6p\n),all,data, transform shape=false,outer sep=2] (#6) {};
        \end{pgfonlayer}
    %\node[] at (#6.north) {#6};
    \end{scope}
}

\newcommand{\tNodeChild}[1]{\rotatebox[origin=c]{90}{\footnotesize\texttt{\space#1\space}}}
\newcommand{\trienode}[3]{%
    \begin{scope}[all/.style={draw, minimum height=0.5cm, minimum width=0cm},line width=0.08ex]
        \message{#3}
        \setlength{\tabcolsep}{0.6ex}
        \node[all, data, inner sep=+0pt] (#3) at (#1, #2)
        %{\footnotesize #3};
        {\begin{tabular}{c|c}\multicolumn{2}{c}{\texttt{Node}}\\\hline
                \tNodeChild{<fields>}&\tNodeChild{0000000000}
        \end{tabular}};
    \end{scope}
}
\newcommand{\trienodemapped}[3]{%
    \begin{scope}[all/.style={draw, minimum height=0.5cm, minimum width=0cm},line width=0.08ex]
        \message{#3}
        \setlength{\tabcolsep}{0.6ex}
        \node[all, data, inner sep=+0pt] (#3) at (#1, #2)
        %{\footnotesize #3};
        {\begin{tabular}{c|c}\multicolumn{2}{c}{\texttt{Node}}\\\hline
        \tNodeChild{<fields>}&\tNodeChild{{RwLock<()>}}
        \end{tabular}};
    \end{scope}
}
\newcommand{\trienodespace}[3]{\nested{3}{#1}{#2}{\trienode}{1}{#3}{lr}{0.6}{0.2}}
\newcommand{\trienodespacemapped}[3]{\nested{3}{#1}{#2}{\trienodemapped}{1}{#3}{lr}{0.6}{0.2}}

\newcommand{\myarray}[6]{%
    \pgfmathsetmacro{\prev}{#1}
    \pgfmathsetmacro{\yy}{0 + #2}
    \pgfmathsetmacro{\yyy}{\yy + #5}
    \tikzstyle{data}=[preaction={transform canvas={shift={(0.4ex,-0.2ex)}},draw=medgray2,very thick}, line width=0.08ex, fill=white]
    \foreach \x/\xs/\b in {#3} {%
        \pgfmathsetmacro{\xx}{\prev}
        \pgfmathsetmacro{\tmp}{\prev + \xs}
        \global\let\prev\tmp
        \draw[draw=black, fill=white] (\xx * 1ex, \yy * 1ex) rectangle (\prev * 1ex, \yyy * 1ex) node[midway, minimum width=\xs * 1ex, minimum height=#5 * 1ex] (#4a\x) {\texttt{\b}};
        %\ifthenelse{\b = 1}{%
        %    \draw[draw=black] (\xx * 1ex, \yy * 1ex) -- (\prev * 1ex, \yyy * 1ex) node[midway, minimum width=\xs * 1ex, minimum height=#5 * 1ex] (#4b\x) {};
        %}{}
    }
    %\pgfmathsetmacro{\pprev}{\prev + #6}
    %\draw[draw=black] (\prev * 1ex, \yy * 1ex) -- (\pprev * 1ex, \yy * 1ex) node[midway, minimum height=#5 * 1ex] (#4ax) {};
    %\draw[draw=black] (\prev * 1ex, \yyy * 1ex) -- (\pprev * 1ex, \yyy * 1ex) node[midway, minimum height=#5 * 1ex] (#4ax) {};
    %\draw[draw=none] (\prev * 1ex, \yy * 1ex) rectangle (\pprev * 1ex, \yyy * 1ex) node[midway, minimum height=#5 * 1ex, minimum width=#6 * 1ex] (#4al) {$\cdots$};
    %\begin{pgfonlayer}{bg}
    %\draw[] (#1 * 1ex, \yy * 1ex) rectangle (\pprev * 1ex, \yyy * 1ex) node[midway, draw=white, minimum height=#5 * 1ex, minimum width=\pprev * 1ex, drop shadow={shadow xshift=0.3ex,shadow yshift=-0.3ex}, fill=white] (#4ax) {};
    %\end{pgfonlayer}
}

\begin{tikzpicture}[x=1.12cm, scale=0.85, every node/.append style={transform shape}]
    \tikzstyle{data}=[drop shadow={shadow xshift=0.3ex,shadow yshift=-0.3ex}, fill=white]
    \trienodespace{0}{0}{regn}
    \trienodespacemapped{5}{0}{regnm}
    \myarray{15}{-15}{0/2/1,1/2/1,2/2/0,3/2/0,4/2/1,5/2/0,6/4/$\cdots$,7/2/,8/2/,9/2/1}{a0}{3}{-15}
    \myarray{15}{-20}{0/2/1,1/2/0,2/2/0,3/2/0,4/2/1,5/2/0,6/4/$\cdots$,7/2/,8/2/,9/2/1}{a1}{3}{-20}
    \node [anchor=east] at (a0a0.west) {\texttt{init}};
    \node [anchor=east] at (a1a0.west) {\texttt{init\_fin}};

    \draw[decorate,decoration={brace,amplitude=1ex}, yshift=0ex, xshift=0ex]
    ($(a0a9.east) + (0.1, 0)$) -- ($(a1a9.east) + (0.1, 0)$) node [midway, xshift=10ex](rb) {Region bitmasks};

    \draw[decorate,decoration={brace,amplitude=1ex, mirror}, yshift=0ex, xshift=0ex]
    ($(a1a0.south) + (0, -0.1)$) -- ($(a1a6.south) + (0, -0.1)$) node [midway, yshift=-4ex](rb) {\texttt{std::sync::atomic::AtomicU64}};

    \draw[<->] (regn) -- (regnm);

    \begin{scope}[myarrow/.style={decoration={markings,mark=at position 0 with {\arrow[scale=0.9,>=Latex]{<};}}, postaction={decorate}}]
        \draw[myarrow] ($(regnmp0.south) + (0, -0.15)$) -- ++(0, -0.3) -| (a0a0.north);
        \draw[myarrow] ($(regnmp1.south) + (0, -0.15)$) -- ++(0, -0.4) -| (a0a1.north);
        \draw[myarrow] ($(regnmp2.south) + (0, -0.15)$) -- ++(0, -0.5) -| (a0a9.north);
    \end{scope}

    \node[anchor=south] at (regn.north) {Persistent Storage};
    \node[anchor=south] at (regnm.north) {Virtual Memory};

\end{tikzpicture}
\endgroup

    \captionof{figure}{Implementation of trie node locks.}
	\label{fig:node-lock}
\end{center}
\end{figure}
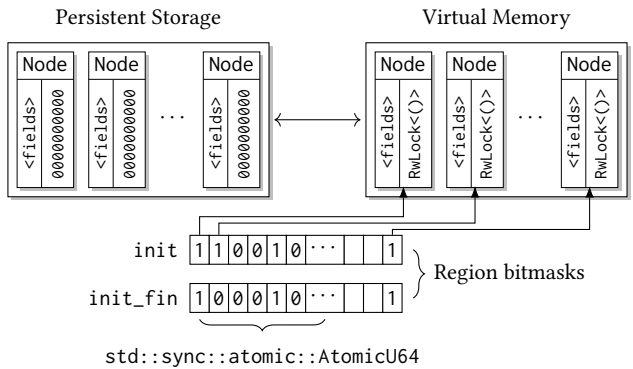

\subsubsection{Fast Node Lock}
The idiomatic way in Rust to add a lock to a type is \verb|RwLock<T>|, a wrapper
around type \verb|T|.
Because \DB trie nodes are memory-mapped, we cannot use this facility directly.
Instead, we maintain a \verb|RwLock<()>| within every trie node
%(see Figure~\ref{fig:tree-node})
, using a neglibible amount of space in the node
(which is dominated by the table of child pointers), but need a way to initialize
the locks after mapping their space from disk.

The lock objects are initialized on-the-fly as threads try to obtain
them.  We maintain two bitmaps with each region, each with one bit
per trie node.
For a region size of 16MBytes full of 256-child tree nodes, less than 2KBytes is needed for the bitmasks.
\verb|init_fin| indicates whether the corresponding lock is initialized.
\verb|init| indicates whether some thread is in the process of initializing the lock.
The bitmasks are implemented by a fixed-length array of 64-bit atomic variables so
that query and modification can be performed atomically using bit operations (Figure~\ref{fig:node-lock}).
We hide this fast node lock facility from other parts of the system behind a safe interface.
\subsubsection{Batched Writes}
Like LevelDB, \DB supports grouping several write operations (insert/delete) into a write batch. The atomicity of
a write batch is guaranteed by write-ahead logging.
Batches also need serializable isolation as they may be executed concurrently. Instead of
acquiring a global mutex lock throughout the entire batch
execution, \DB implements a simple concurrency control method that divides a write
batch execution into two phases. During the first phase, the global mutex is
held, while the batch walks down the trie by alternating read locks and stops at
the leaves where an insertion/deletion will happen. After the quick walk, the
set of nodes whose write locks are required for all operations in the batch are
recorded and deduplicated. Then the batch acquires the write locks and
releases the global lock. With the node locks held, the thread
resumes each operation in the batch and buffers all induced disk writes into a
vector (WriteBatch in Figure~\ref{fig:storage-model-overview}). In addition to node locks, a thread may also acquire a lock to update the tail
pointer of a logical space. Each tail pointer has its dedicated mutex lock.
All tail pointer locks held by the thread will only be released at the end.
Finally, the vector is submitted to the disk thread through the
multi-producer, single-consumer write buffer, after which all locks are released.
\DB optimizes for single write operations without batching, as no extra concurrency control is needed in this case.

\section{Evaluation}
\label{sec:evaluation}
%[RVR: we need some text here explaining what or goals are.]
We evaluate the implementation of \DB in order to answer the following questions:
\begin{itemize}[leftmargin=*,itemsep=1pt,topsep=1pt]
%\begin{compactitem}
    \item How does branching factor of a lazy-trie affect its performance? Which value of sluggishness should one choose? Is 256-bit practical enough for hashing? (\S\ref{sec:eval-bfactor-sluggishness}, \S\ref{sec:eval-hash-length})
    \item What is the performance for various types of \DB operations? Is it practical enough compared to other stores? (\S\ref{sec:eval-throughput})
	\item How does performance degrade when user data cannot fit entirely in memory? What is the impact of the region size? (\S\ref{sec:elastic-memory}) What is the performance of data space allocator and what is the overall storage footprint? (\S\ref{sec:fragmentation})
    \item How well does \DB perform under various kinds of realistic workloads? How well does it leverage concurrency? (\S\ref{sec:eval-ycsb})
    \item How does crash recovery in \DB compare to other approaches? (\S\ref{sec:eval-crash})
%\end{compactitem}
\end{itemize}

\subsection{Setup}
Unless otherwise noted, we use Dell R340 servers to conduct our experiments.
A server has a
hexa-core Intel Xeon E-2176G 3.70GHz CPU with 64GB DDR4 memory. For the
secondary storage medium, we use an
%% Ted: we repeatign the same experiment on a different SSD gives marginal information,
%% especially AWS is already considered
%either use Samsung SM863a V-NAND SSD with 6Gbps SATA
%interface [Ted: TBD] or
Intel Optane 905P SSD with PCIe NVMe 3.0$\times$4 interface.
All evaluations are done on Ubuntu 18.04 LTS with a dedicated ext4 filesystem for persistent files.
We allow asynchronous writes in all stores.
For most experiments, the whole data set can fit into the memory.
\begin{comment}
However, because \DB is designed with flexibility of
dynamically grown data store and region-based mapping, we are curious
about how the performance degrades as the data set grows beyond the given memory
budget. In this case, \DB uses an LRU cache to determine which region to ``swap
out'' to make space for the accessed regions.

To show our \DB is practical for use as a library,
\end{comment}

%We use LMDB 0.9.23 and RocksDB 6.1.2 as baselines and also
%compare against LevelDB 1.20
%%% Ted: uncomment the below line if we decide to include PebblesDB results
%%and PebblesDB (SOSP version)
%in some experiments.
We use FasterKV 1.8.4, LMDB 0.9.23 and RocksDB 6.1.2 as our baselines,
representing a wide spectrum of persistent key-value store designs.
In RocksDB, we enable the additional cache
feature by setting the LRU cache to 40GB, the same amount
used for \DB memory-mapped regions.
We disabled data compression to make comparison more fair.
%\DB also uses a 40GB LRU cache to determine which region to swap out
%to make space for the newly accessed regions.
%[RVR: I don't get the last sentence.]
LMDB requires specifying the maximum data store size upon
creation to preallocate the storage space---we also set it to
the same amount of memory.

Our main target for comparison is FASTER~\cite{faster2018},
as it is the key-value store that is most closely related to \DB.
Like \DB, FASTER uses key hashes for indexing and thus does not
support range queries or sorted iteration.
Despite this similarity, FASTER has a very different persistence model.
FASTER divides its memory into two parts.
One section of memory serves as a data cache for fast access, while
the other part, the \emph{HybridLog}, is used to log updates.
By default, FASTER only writes its HybridLog to disk when the
log can no longer be held in memory in its entirety.
But even when the log is flushed, a user still needs to manually invoke
the checkpoint function to make the logged changes persistent.
In practice, one has to decide how frequently one should invoke
FASTER's checkpoint function, while
other approaches persist their data continuously.

In our experiments, we made no FASTER checkpoints until the
end of each run, resulting in the minimum I/O effort that FASTER undergoes
to persist its state, but note that its in-memory index is not saved and
is lost in the event of a failure.
We made the initial hash table exactly large enough to contain the
initial number of items and used the rest of the 40GB for HybridLog.
To eliminate warmup-bias, we applied in-place updates until the memory
part of the log was saturated and made a synchronous checkpoint to flush the
leftover I/O induced by warmup operations before starting the workload.
% A checkpoint is made each time a run finishes.
We used the C++ version of FASTER that is natively available on our
Linux platform and should offer similar performance as
the C\# implementation~\cite{faster-web}.
FASTER also uses kernel-based AIO, so for fairness we set the same maximum
I/O event limit as in \DB.

For most microbenchmarks,
we use 32-byte keys and 128-byte values.
For macrobenchmarks, we evaluate \DB using YCSB~\cite{ycsb}.
YCSB is written in Java whereas \DB is in Rust and the other key-value stores
are in C or C++. To avoid overhead caused by incompatible interfacing,
we created a C
binding for \DB and invoke the APIs from a unified C++ test program dedicated to
executing the YCSB workload.
In the graphs with error bars, we run the same setup for 5 times and plot the
mean in y-axis with standard deviation bars. For other graphs, the variation
between runs is negligible.
% Such combination unifies the platform and provides a fairer and
% more realistic comparison.
For each run, we populate the data store with a
certain number of key-value items before running the test workload.
The data store is reopened when the test starts.
% and then have a cold-start of the test workload.
Finally, for multi-threaded experiments, we run a separate YCSB
workload generator for each thread in such a way that a thread only
handles keys that were preloaded or keys that were inserted by the
thread itself.

\subsection{Microbenchmarks}
\subsubsection{Branching Factor and Sluggishness}
\label{sec:eval-bfactor-sluggishness}
\begin{figure}[t]
%\ifpdfbars
\includegraphics[width=\linewidth]{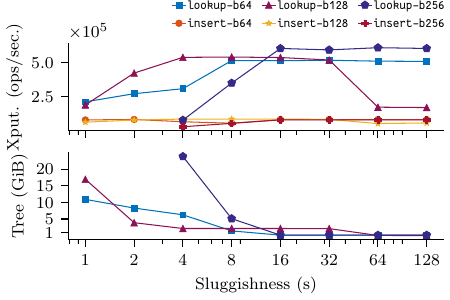}
%\else
%\includegraphics[width=\linewidth]{graphs/bfactor-sluggishness.png}
%\fi
\captionof{figure}{Lookup/insertion throughput (Xput.) and the tree footprint
with different branching factors and sluggishness.}
\label{fig:bfactor-sluggishness}
\end{figure}

There are two parameters required to instantiate a lazy-trie: the tree
branching factor and sluggishness, together determining the shape of the
tree and its statistical characteristics.
In Figure~\ref{fig:bfactor-sluggishness},
we use branching factors 64, 128, and 256.
We vary the sluggishness from 1 (no sluggishness) to 128.
For each run, a data store of 100 million items is used to perform 10
million uniformly random lookups or insertions.

As discussed in Section~\ref{sec:lazy-trie-sluggish-splitting},
we expect that a larger branching factor will make the efficacy of sluggishness
more pronounced as having more slots in the child table leads to higher
storage overhead and amplification.
Indeed, we see that for a branching factor of 256, sluggishness
significantly improves performance.
Data points for $s < 4$ are not shown in the graphs because those runs
end up with a tree footprint that exceeds the memory of our platform.
Compared to (32+128)-byte key-values, each tree node with 256 children slots
takes up around 2KBytes of space. By having more sluggishness, storage
overhead is greatly mitigated and both lookup and insertion performance
ramp up to reach or surpass those of other branching factors.
When increasing sluggishness from $4$ to $16$, the memory footprint gets cut
down from more than 23GBytes to 144MBytes.
The performance drop of
b128 with excessive sluggishness is due to the periodic change of node
utilization (Figure~\ref{fig:lazy-trie-sluggishness}, middle).
Given our store size ($10^8$ items), this leads to excessively long
linked lists.
% For optimal performance,
Thus one should choose the least sluggishness that
achieves the desired footprint. % to bound the linked list size.

When using a branching factor of 64, each tree node
only takes around 512 bytes and it requires a lower sluggishness for comparable
performance. On the other hand, because the data store can be fully cached in
memory, the choice of branching factor does not significantly affect the
maximum performance due to the low cost of memory access. To reduce the tree height
and have the best lookup performance, we use a branching factor of 256 with
sluggishness of 16 as a practical choice for all subsequent experiments.

%\vspace{-.05in}
\subsubsection{Hash Length}
\label{sec:eval-hash-length}
In the prior experiments, we used 256-bit HighwayHash to hash the keys. HighwayHash
also provides 64- and 128-bit hashing.
While generating a shorter hash is faster, it
increases the collision rate, amplifying the
cost of scanning the linked list of data nodes.
We experimented with all three hash lengths but found
no difference in performance.
We use 256-bit hashes for the remainder of the evaluation.

%\begin{figure}[t]
%\ifpdfbars
%\includegraphics[width=\linewidth]{graphs/micro-k128-nr1e8-no1e7-hlen-t1.pdf}
%\else
%\includegraphics[width=\linewidth]{graphs/micro-k128-nr1e8-no1e7-hlen-t1.png}
%\fi
%\captionof{figure}{Microbenchmark with different hash length of \DB (single-threaded, Optane NVMe).}
%\label{fig:micro-k128-nr1e8-no1e7-hlen-t1}
%\end{figure}

\subsubsection{Throughput}
\label{sec:eval-throughput}

\begin{figure}[t]
\ifpdfbars
\includegraphics[width=\linewidth]{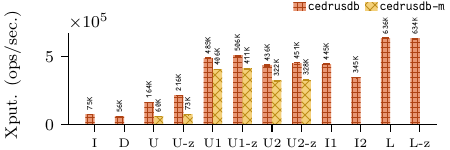}
\else
\includegraphics[width=\linewidth]{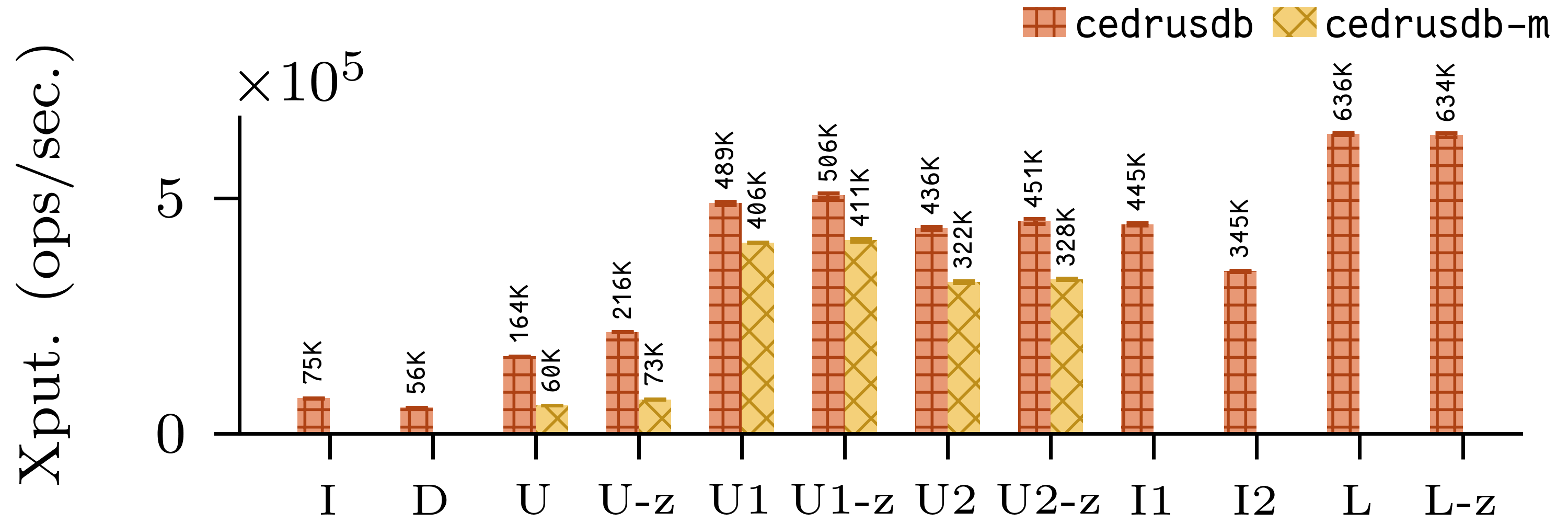}
\fi
\captionof{figure}{Microbenchmark with $10^8$ items and $5\times10^7$ operations (single-threaded).}
\label{fig:micro-k128-nr1e8-no1e7-t1}
\end{figure}
\begin{figure}[t]
\ifpdfbars
\includegraphics[width=\linewidth]{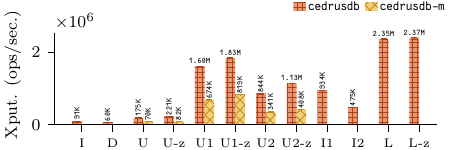}
\else
\includegraphics[width=\linewidth]{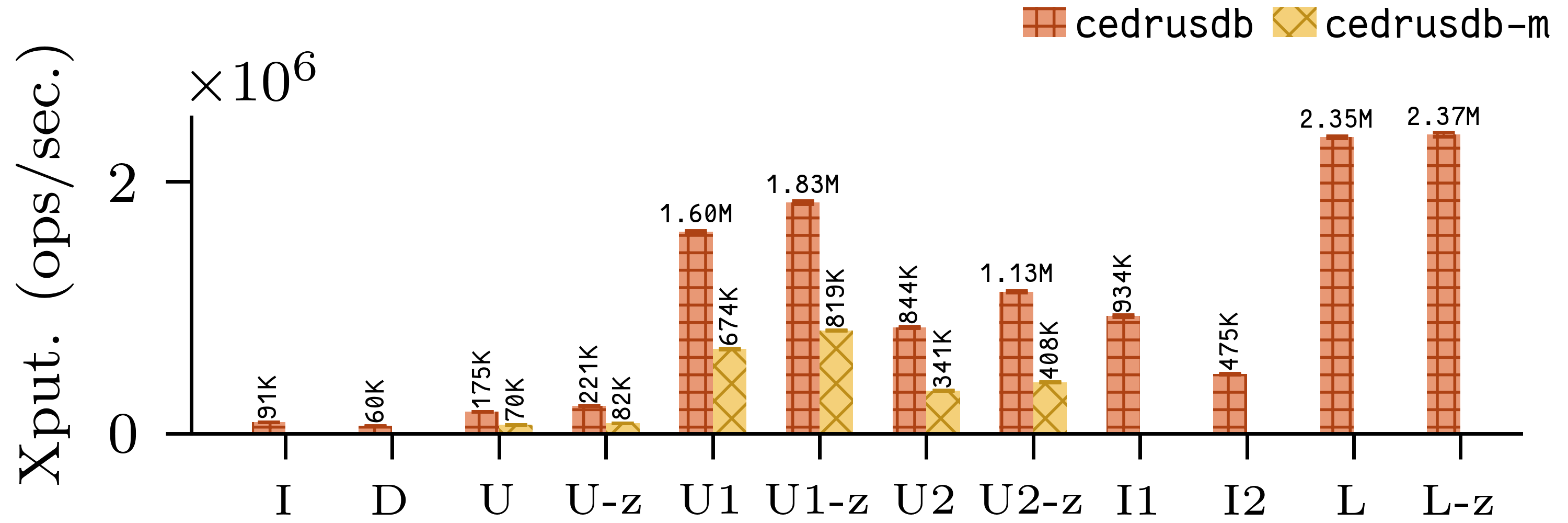}
\fi
\captionof{figure}{Concurrent microbenchmark with $10^8$ items and $5\times10^7$ operations (4 threads).}
\label{fig:micro-k128-nr1e8-no1e7-t4}
\end{figure}

We examine the performance of individual types of operations with a single client thread.
We populate the store with 100 million items and conduct 50 million reads or writes.
Figure~\ref{fig:micro-k128-nr1e8-no1e7-t1} shows that \DB performs
uniformly random (\texttt{L}) or Zipfian (\texttt{L-z}) lookups
much faster
than pure insertions (\texttt{I}), deletions (\texttt{D}) and updates (\texttt{U}).
\begin{comment}
\remove{%
The right graph of Figure~\ref{fig:micro-k128-nr1e8-no1e7-t1} shows that,
using a single thread,
LMDB outperforms \DB, especially when reads have more locality
(Zipfian distribution).
This is partly explained because \DB requires hashing and scanning in the
sluggish leaf list for each lookup.
Moreover, for Zipfian workloads, hashing in \DB eliminates locality.
However, shown on the right of Figure~\ref{fig:micro-k128-nr1e8-no1e7-t4},
\DB significantly outperforms all key-value stores for concurrent lookup
access with 4 client threads.
}

\remove{%
LSM-based key-value stores like RocksDB are optimized for write-only workloads.
In Figure~\ref{fig:micro-k128-nr1e8-no1e7-t1}, we test 10 million pure insertions (\texttt{I}), deletions (\texttt{D})
or updates (\texttt{U}) with uniformly distributed keys. Additionally, we tested
updates with a Zipfian distribution (\texttt{U-zipf}) and a very write-intensive mixed
workload with 50\% lookups + 50\% updates (\texttt{M} and \texttt{M-zipf}).
As expected, RocksDB is significantly faster than others for write-intensive operations.
}
\end{comment}
\DB has two kinds of updates: (i) \emph{in-place updates} (as used in FASTER/LMDB by default) when the new value can fit into the footprint of the current one, and (ii)
\emph{emulated updates} that remove the current values and insert the new ones.
We instrumented \DB so we could test both individually. \verb|cedrusdb|
shows in-place update performance and \verb|cedrusdb-m| shows the
performance when all updates are treated as deletions followed by insertions.
Unsurprisingly, in-place updates are fastest
since they do not alter the tree topology.
% When performed concurrently as shown in Figure~\ref{fig:micro-k128-nr1e8-no1e7-t4}, in-place update performance is close to that of lookups, particularly
% when the workload locality is better.
%
%Insertions are faster than updates because the latter engages
%the data space allocator frequently to recycle data nodes.

Although we used the fastest async channel implementation available for Rust,
we still noticed insertions are bottlenecked by our write buffer as the throughput
of insertions would have been doubled (\textasciitilde180K) had we changed the order and granularity of writes (but atomicity
would no longer be guaranteed). We observed the same bottleneck in Figure~\ref{fig:bfactor-sluggishness}.

To see the impact of writes to the overall performance when
they are mixed together with reads, we tested it with 10\% (\texttt{U1*}, \texttt{I1}) and 20\%
(\texttt{U2*}, \texttt{I2}) writes. \DB still benefits from fast reads in these workloads.

As for
concurrent access (Figure~\ref{fig:micro-k128-nr1e8-no1e7-t4}), in a 4-thread
setup, mixed writes also preserve some degree of scalability from reads.
%The current implementation of \DB uses a mutex lock for the data space
%allocator that becomes a performance bottleneck in the concurrent setup,
%whereas the allocator for the trie space is wait-free.
The performance of deletions is the lowest because they require frequent access to the data space allocator and also change the tree topology.
\begin{comment}
\remove{Nonetheless, in these experiments \DB is comparable to LMDB and it is faster than LMDB in all other workloads.}
\end{comment}

\subsubsection{Elastic Memory and Regions}
\label{sec:elastic-memory}
While \DB is optimized for the case that the entire data store fits
within a given memory budget, the region-based mapping design of \DB
allows a larger data store.
We evaluate how performance degrades as \DB runs out
of the memory budget.
We started with a baseline experiment using 100 million data items that did not
have a memory budget, so the data store could utilize all the memory.
We recorded the maximum number of regions and used that as the memory budget.
We then experimented with $0\%$--$25\%$ of additional user data to see how
performance changes.

\begin{figure}[t]
\includegraphics[width=\linewidth]{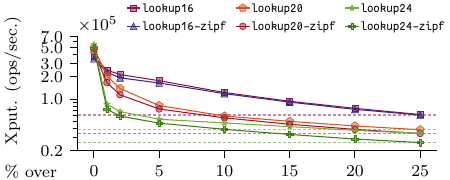}
\captionof{figure}{Read performance for different region sizes, with user data beyond the memory budget.}
\label{fig:micro-elastic-k128-nr1e8-no1e8-regn-t1}
\end{figure}

\begin{figure}[t]
\includegraphics[width=\linewidth]{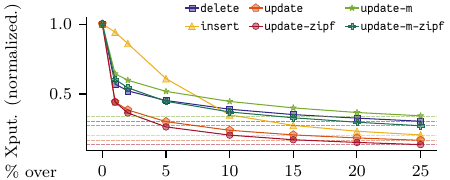}
\captionof{figure}{Normalized write performance with user data beyond the memory budget.}
\label{fig:micro-elastic-k128-nr1e8-no1e7-t1}
\end{figure}

Figure~\ref{fig:micro-elastic-k128-nr1e8-no1e8-regn-t1} shows lookup
performance for region sizes ranging from 64KBytes (\texttt{lookup16}, 16-bit)
to 16MBytes (\texttt{lookup24}, 24-bit).
The figure that shows having larger region size results in better performance when all
data fit in memory, but degrades quickly when data exceeds the memory budget.
When the whole data set fits in memory ($0\%$),
small region sizes hurt performance because there is more
overhead for managing regions and their mapping.
For large region sizes, the coarse granularity of mappings
make it more likely that cold and hot items are collocated in the same region,
resulting in increased swapping.
Figure~\ref{fig:micro-elastic-k128-nr1e8-no1e7-t1} shows the
results for write operations.
In this figure, each line is normalized to the full memory budget performance.
We see that faster operations like updates degrade more
than slower ones like deletions, due to the high penalty of swapping compared
to in-memory operations.
% Finally, because this paper focuses on when all data can fit in memory, we use
% regions of 16MBytes for all other experiments.
So while \DB will continue to operate well when running out of the given memory
budget rather than simply give up, it is important to adjust the memory budget
accordingly.

\subsubsection{Variable-Length Values and Fragmentation}
\label{sec:fragmentation}
So far (and as is common practice in many key-value store evaluations),
we used the same length for all values.  For \DB, this means that
the data space allocator only needs to take one step to find the
next-fit location that was previously freed to recycle.  When
allowing in-place updates, the allocator is not even engaged. Thus,
to throughly examine our design and effectively evaluate the
allocator, we generated a workload that mixes 128/256/1024 values
uniformly.
We first populated the store with 10 million items.  To have
each value updated many times on average, we ran 100 million
operations with the mixed workload (\texttt{M}), changing the value
of each key \textasciitilde5 times throughout the entire run.

For the next-fit allocator, scanning through the entire free list
before giving up is too expensive in practice.  Instead of making
sure to recycle a freed object whenever possible, allowing some
slack in using the free list does not cause significant fragmentation.
Therefore we limit the maximum number of steps in scanning the free
list during an allocation.

\begin{figure}[t]
\includegraphics[width=\linewidth]{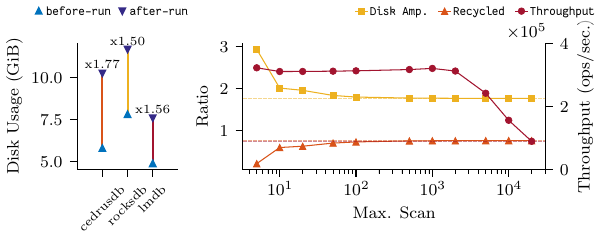}
\captionof{figure}{Data space allocator statistics and storage amplification due to fragmentation.}
\label{fig:micro-var-k128-nr1e8-no1e8-t1}
\end{figure}

In the right subgraph of Figure~\ref{fig:micro-var-k128-nr1e8-no1e8-t1},
we vary the scan step limit (\texttt{Max.~Scan}) and show the change
of the amplification factor (\texttt{Disk Amp.})~and the ratio of
reusing freed space in an allocation (\texttt{Recycled}).  The disk
amplification factor is the final disk usage divided by the initial
one. It is greater than 1 for all stores due to fragmentation.  Once
the limit exceeds 100 there is little difference compared to having
no limit.  Moreover,
as shown in the left subgraph of
Figure~\ref{fig:micro-var-k128-nr1e8-no1e8-t1},
we believe the fragmentation ratio of \DB is reasonable compared to
other key-value stores.

\subsection{Macrobenchmarks}
\label{sec:eval-ycsb}
In this section, we evaluate \DB using YCSB~\cite{ycsb} workloads
with Zipfian distributions.
We used a region size of 16 MBytes for all experiments.

\subsubsection{Single-Threaded Performance}
\begin{figure}[t]
\ifpdfbars
\includegraphics[width=\linewidth]{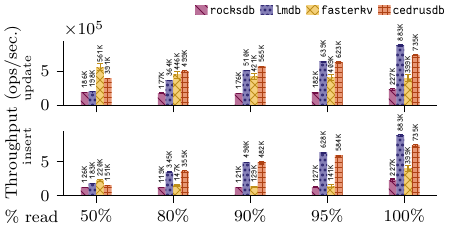}
\else
\includegraphics[width=\linewidth]{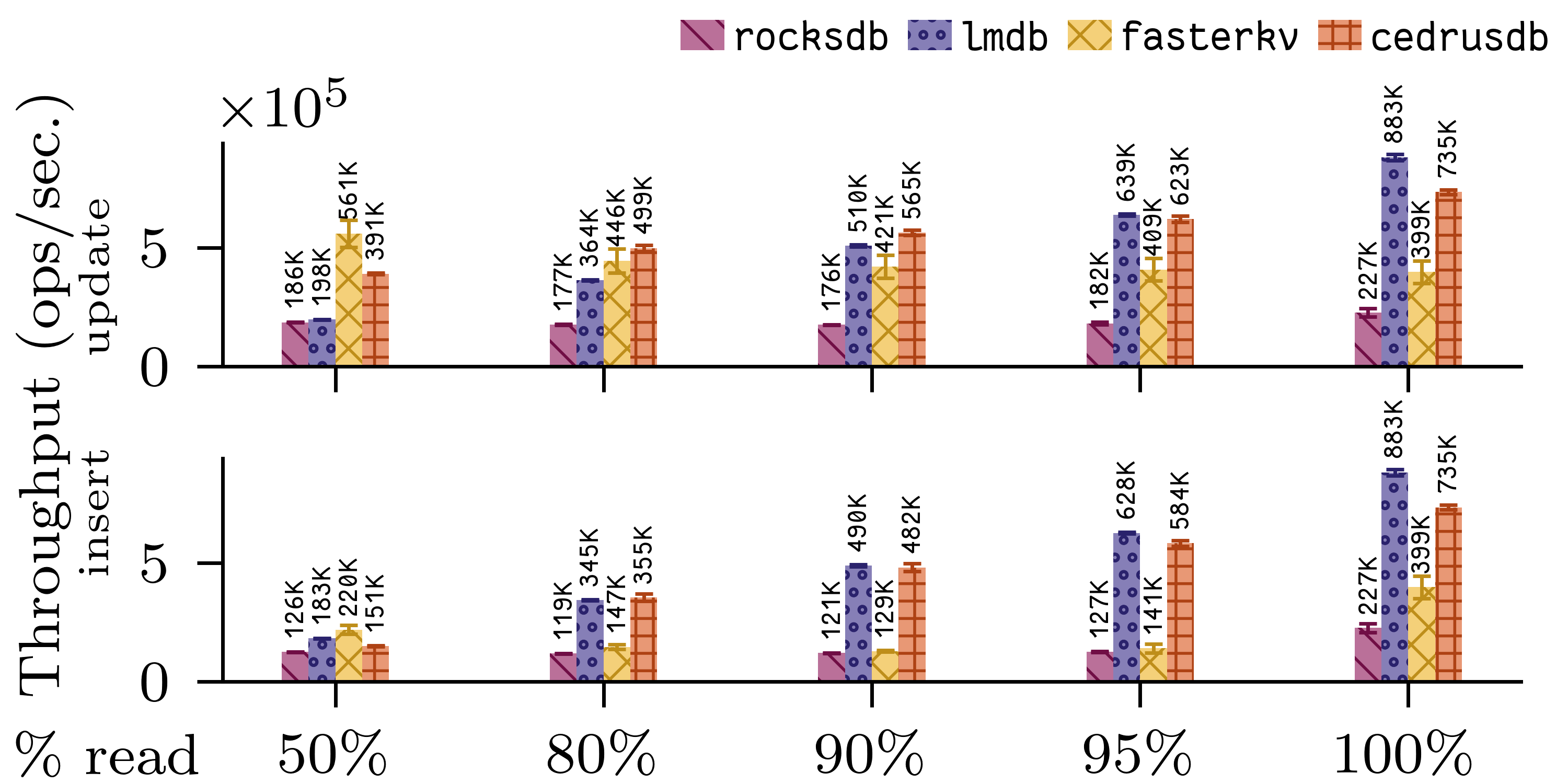}
\fi
\captionof{figure}{YCSB evaluation with $10^8$ 128-byte values and operations (single-threaded).}
\label{fig:ycsb-zipfian-k128-nr1e8-no1e8}
\end{figure}

\begin{figure}[t]
\ifpdfbars
\includegraphics[width=\linewidth]{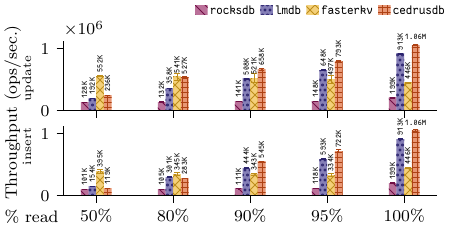}
\else
\includegraphics[width=\linewidth]{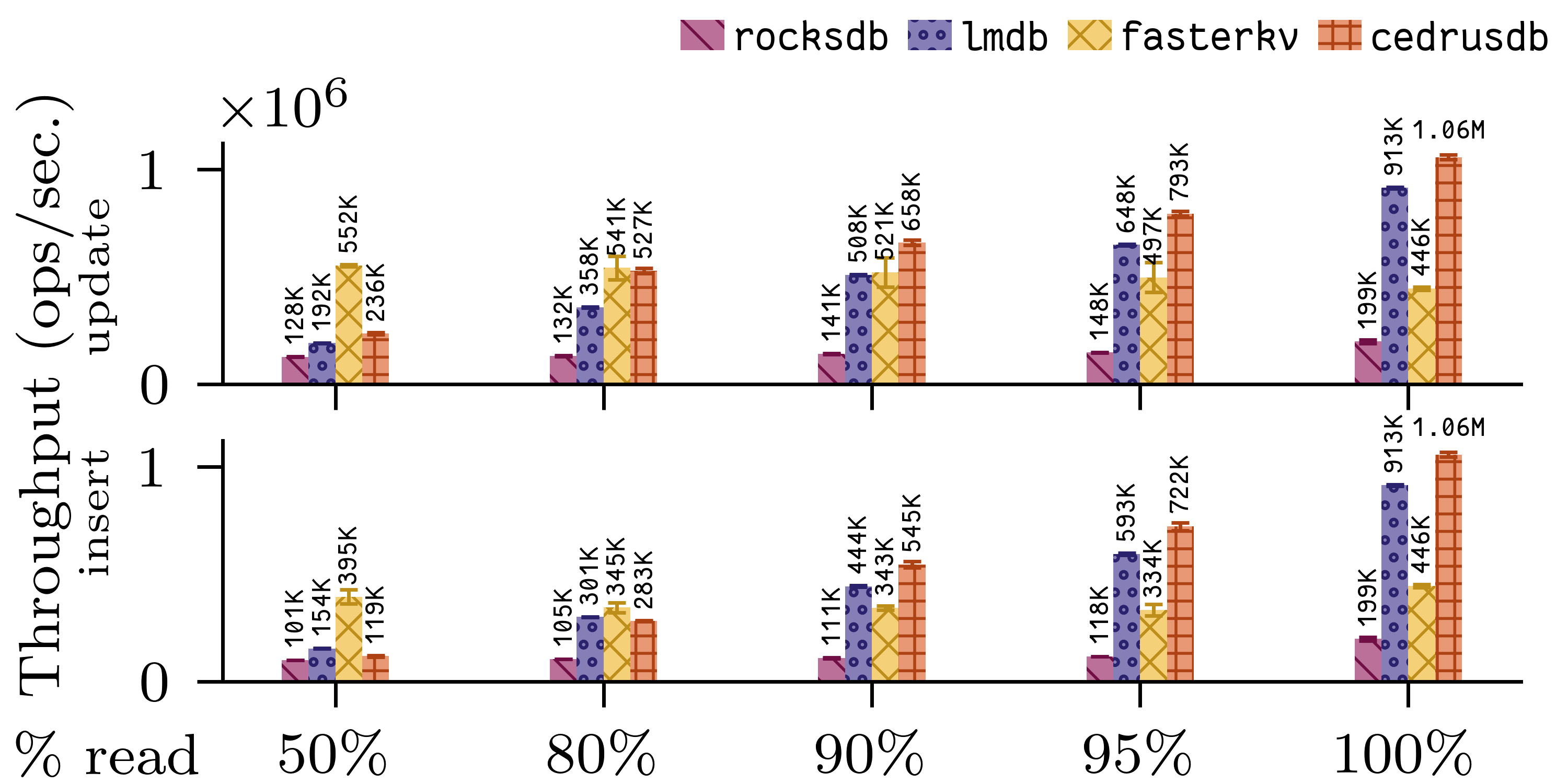}
\fi
\captionof{figure}{YCSB evaluation with $10^7$ 1024-byte values and operations (single-threaded).}
\label{fig:ycsb-zipfian-k1024-nr1e7-no1e7}
\end{figure}

Figure~\ref{fig:ycsb-zipfian-k128-nr1e8-no1e8} shows single-threaded
throughput for different read/write ratios using 128-byte values;
Figure~\ref{fig:ycsb-zipfian-k1024-nr1e7-no1e7} shows the same for 1KByte values.
The data store is populated with 100 million keys.
Two types of write operations are considered:
updating the value of an existing key and inserting a value with a
non-existing key. Although both use the same API, they trigger
different code paths (as demonstrated in our microbenchmarks).
The upper graph shows results from runs where all writes are updates
whereas the lower shows insertions;
in the rightmost bars, all operations are lookups.
%As a reminder: \verb|cedrusdb-m| simulates an in-place update by removing the
%key and then inserting the value as if it were a new key.
\DB achieves performance comparable to or better than LMDB and RocksDB in write-intensive
mixed workloads.
It outperforms others in read-intensive cases, with
the exception that LMDB is slightly faster in these cases and noticeably faster in pure lookups, when using
128-byte values.

In contrast, FASTER, which does not maintain an on-disk index,
performs better with more updates.  Changes are appended to an
in-memory log that is eventually flushed to disk when memory runs
low. In the absence of reads, this scheme is bottlenecked only by
disk bandwidth. However, if there are read operations, FASTER
performance suffers from not having an on-disk index.
FASTER is optimized for in-place updates, which get translated
into modifications inside the in-memory log. Its insertion
performance reveals more of the performance impact by writes as
shown at the bottom of the graph.

%\remove{
%This is one reason that \DB,
%unlike FASTER, does not take the hash-table, log-based approach,
%but chooses the proposed lazy-trie as the on-disk index structure,
%for more balanced performance, while still offering convenient
%persistence.
%}

To evaluate performance when values have varying sizes,
we generated two workloads. In workload A, for each
key, the value is 128, 256 or 1024 bytes chosen uniformly at random.
This illustrates a scenario in which the key-value store has items
with only a few but very different sizes.
% , whereas the dynamic range is large.
% Shorter, 128-byte items could be metadata used by the application, while the
% longer, 1024-byte items could be chunks of the actual payload.
On the other hand,
workload B considers a scenario where most values have a similar size.
The values are chosen between 128 and 256 bytes with Zipfian distribution.
Both workloads are run using 100
million mixed lookup/update operations given 10 million initial items.
\DB achieves better performance than others for read-intensive
cases in both workloads
(Figure~\ref{fig:ycsb-var-zipfian-k1024-nr1e7-no1e8}), while it degrades faster
with more updates. This is because variable-length values may not be updated
in-place and thus trigger intensive access to the data space allocator.
On the other hand,
even in the most write-intensive case the increase of disk usage
caused by fragmentation of \DB remains reasonable,
as shown at the bottom of Figure~\ref{fig:ycsb-var-zipfian-k1024-nr1e7-no1e8}.
We did not evaluate the performance of FASTER for this workload
as the C++ implementation (unlike its C\# version) does not support
variable-length values.

\begin{figure}[t]
\ifpdfbars
\includegraphics[width=\linewidth]{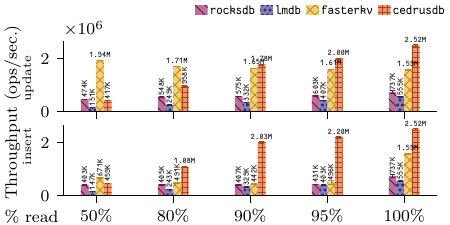}
\else
\includegraphics[width=\linewidth]{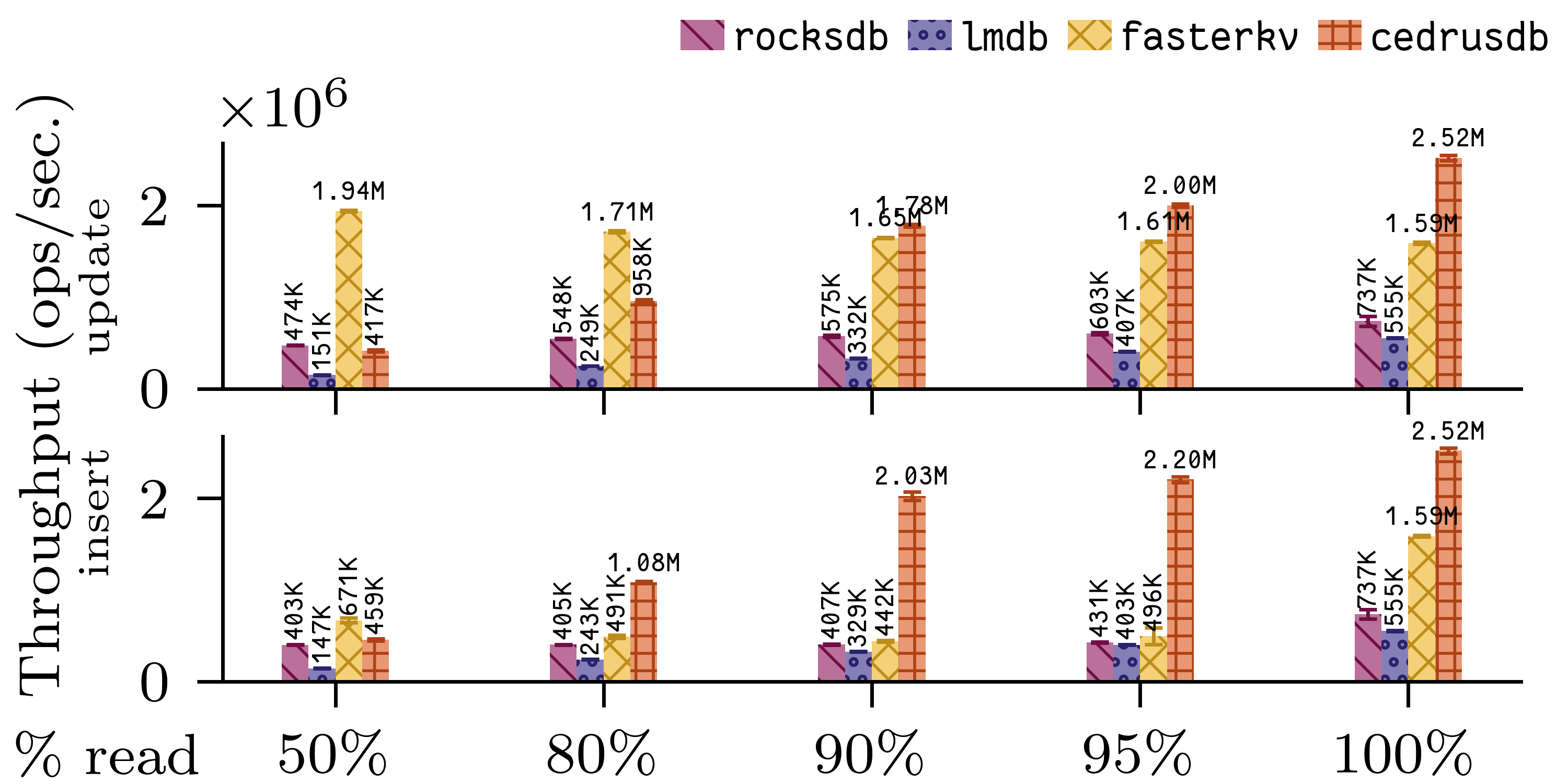}
\fi
\captionof{figure}{YCSB evaluation with $10^8$ 128-byte values and operations (4 threads).}
\label{fig:ycsb-zipfian-k128-nr1e8-no1e8-t4}
\end{figure}

\begin{figure}[t]
\ifpdfbars
\includegraphics[width=\linewidth]{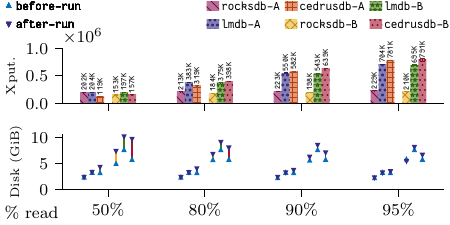}
\else
\includegraphics[width=\linewidth]{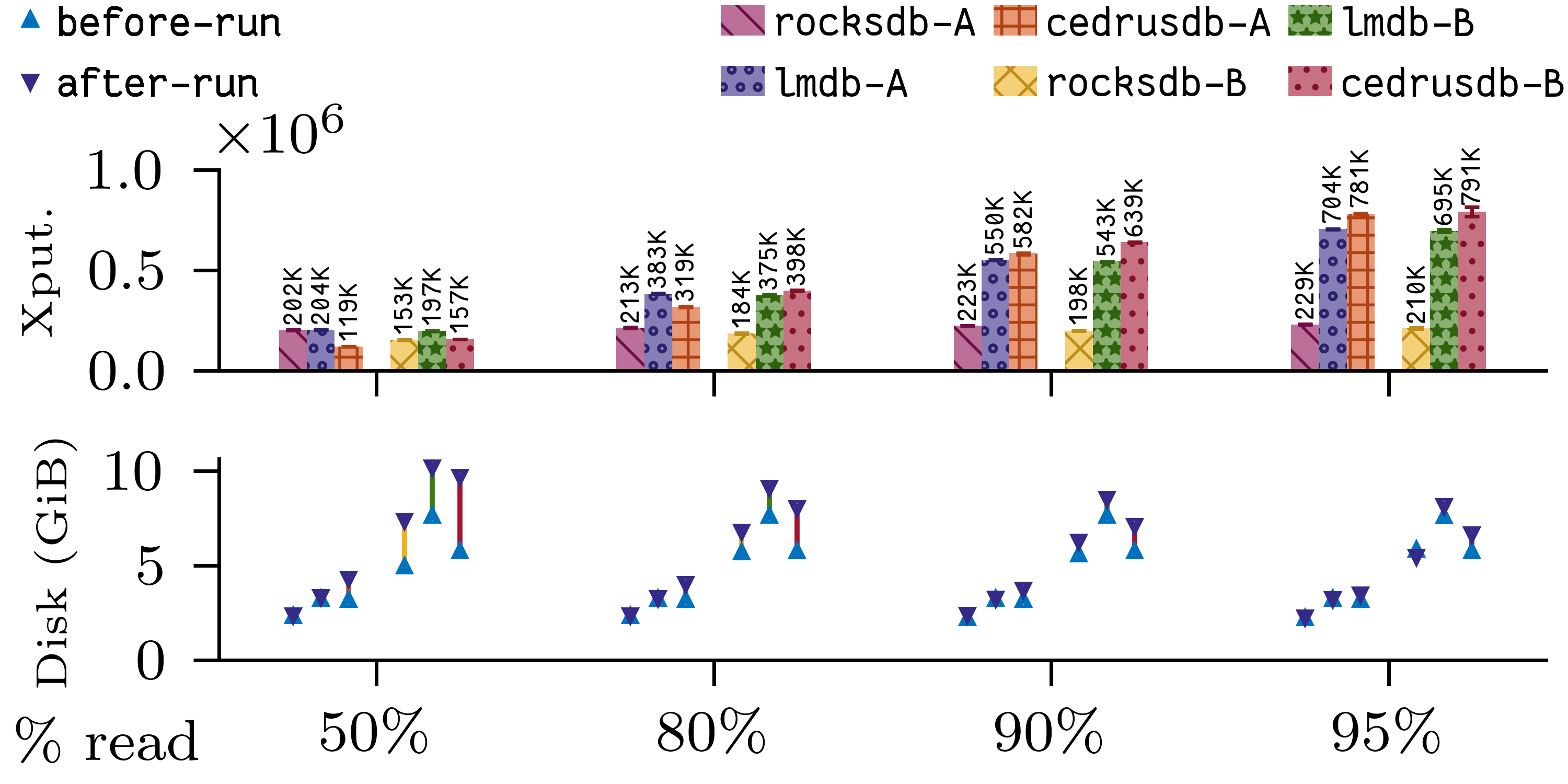}
\fi
\captionof{figure}{YCSB variable-length throughput (Xput.) with $128$/$256$/$1024$
(\texttt{*-A}) or $128$--$256$ (\texttt{*-B}) byte values (single-threaded).}
\label{fig:ycsb-var-zipfian-k1024-nr1e7-no1e8}
\end{figure}

\subsubsection{Concurrent Performance}
%\vspace{-0.05in}
\begin{comment}
As multi-core processors and multi-threaded applications become pervasive,
being able to scale well in this new scenario is important to modern key-value stores.
Thus, we evaluate the systems with concurrent access.
\end{comment}
Next we evaluate multi-threaded performance.
Both LMDB and RocksDB have specific optimizations to take advantage
of concurrency~\cite{rocksdb-doc, lmdb-doc} whereas FASTER uses atomic operations on
its in-memory hash table.
Our lazy-trie requires no reorganization when keys are inserted or
deleted, simplifying concurrent access, which could be viewed in a way as having small ``hash tables''
with some tree hierarchy.
% Compared to Figure~\ref{fig:ycsb-zipfian-k128-nr1e8-no1e8},
Figure~\ref{fig:ycsb-zipfian-k128-nr1e8-no1e8-t4} shows the aggregated
throughput when 4 client threads access the data store at the same time.
% with the same pattern as the specified workload.
\DB outperforms others in read-intensive workloads. FASTER performs
best under intensive updates, but the performance suffers when
the writes are insertions.
%[RVR: I'm still not sure if the following is interesting.]
%When in-place update is disabled,
%it still outperforms others in read-intensive workloads but gets degraded with
%more portion of updates, bottlenecked by our aforementioned data space allocator mutex lock.
%Even in this extreme case, it still performs better than LMDB.

\begin{figure}[t]
\includegraphics[width=\linewidth]{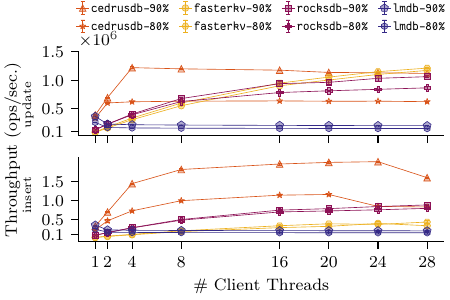}
\captionof{figure}{Concurrent YCSB evaluation with $10^8$ 128-byte values and operations (AWS NVMe SSD).}
\label{fig:ycsb-zipfian-k128-nr1e8-no1e8-aws-concurrent}
\end{figure}

To test the scalability in the number of threads, we conducted the
same experiments on an Amazon AWS \texttt{r5d.8xlarge} instance.
We used one NVMe SSD and run $80\%$--$90\%$ read and
$20\%$--$10\%$ update/insert workloads with
a varying number of client threads.\footnote{
FASTER experienced consistency issues when running with $24$--$28$ threads,
when some existing keys were reported missing. We went ahead and collected
the measurements regardless.
}
Figure~\ref{fig:ycsb-zipfian-k128-nr1e8-no1e8-aws-concurrent} shows that
\DB scales well until it hits the shared lock bottleneck at around 8
threads. \DB update performance plateaus much earlier than insertions because
Zipfian updates induce contention of frequently accessed items (hashing does not
prevent this because the same key always corresponds to the same hash), whereas
insertions create new items and benefit more from non-overlapping tree walks.
The Rust MPSC (Multi-Producer, Single-Consumer) write-buffer library limits insertion performance when there are
more than 20 threads.
Nevertheless, the read-intensive (90\% read) update throughput of \DB is still
1.05--3.07x that of RocksDB, 1.07--5.74x the throughput of LMDB, and 0.95--4.79x the throughput of FASTER.
For insert, \DB is either competitive or superior to all others.
While the performance of FASTER is inferior to that of \DB
for various workloads,
its in-memory, lock-free hash table implementation causes FASTER
to plateau much later than \DB in update workloads.
% In summary, it has demonstrated the great potential for concurrent workloads.

%\subsection{Application: LemonGraph}
%(Hongbo: working on coding and hacking)

\begin{table*}[t]
    \small
    \begin{center}
\begin{tabular}{r|c|c|c|c|c}
    & Persistence & Indexing & Recovery (sec.) & Checkpoint (sec.) & Throughput (Kops/sec.)\\
    \hline
    CedrusDB & always, disk-indexed & hashed & 0.645 & - & 391 \\
    \hline
    FASTER & manual, log-based & hashed & 28.16 & 19.38 & 446 (disk) / 1125 (volatile) \\
    \hline
    LMDB & always, disk-indexed & sorted & 0.003 & - & 198 \\
    \hline
    RocksDB & always, disk-indexed & sorted & 0.267 & - & 186 \\
    \hline
    Masstree & always, log-based & sorted & 31.8 per $10^8$ ops & $\approx 33.5^{*}$ & 1123 \\
    \hline
    KVell & always, slab-based & sorted & 91.9 & - & 410 \\
\end{tabular}
\end{center}
\caption{Key-value stores with different schemes. YCSB (50\% updates), Zipfian, 128-byte values and $10^8$ items. Throughput shows the normal operation performance where no checkpoint is made.
{*} Extrapolated value---see text.}
%\vspace{-0.20in}
%\footnotetext{Masstree could not finish its checkpoint with a data store of 100 million keys, so we measured fewer keys and interpolated this value.}}
\label{table:crash-recovery}
\end{table*}

\subsection{Crash Recovery}
\label{sec:eval-crash}
In recent years, various projects have proposed boosting
the performance of key-value stores by eschewing the on-disk index and
emphasizing a fast in-memory index.
The potential price paid is increased recovery time.
Similar to LMDB and RocksDB, \DB takes a more traditional approach by having
an up-to-date on-disk index.
In this section, we consider the performance trade-offs of the two approaches.
We used \DB, LMDB, and RocksDB as representatives of key-value stores that maintain an
up-to-date on-disk index, and
FASTER, Masstree~\cite{masstree2012}, KVell~\cite{kvell} as representatives
of the alternative approach.

We ran the following experiment with each key-value store.
After populating the store with 100 million keys,
we ran 100 million 50\% update YCSB workload operations before killing the program.
We then measured the recovery time.

As shown in Table~\ref{table:crash-recovery}, the sorted on-disk index of RocksDB comes at the expense of throughput compared to \DB whose index
with hashed keys does not support range queries. RocksDB has faster recovery time because its WAL records
high-level operations, whereas \DB logs the induced low-level writes. Nonetheless,
the recovery time for both is short and mostly dependent on the implementation and
configuration and not affected by store size or total number of writes. LMDB
uses \emph{shadow paging}, a copy-on-write technique to persist data. Without the need
for WAL, it takes neglibible time to recover.
Next we compare with FASTER, Masstree and KVell.

\vspace{-0.10in}
\paragraph{FASTER.}
As recommended by the developers~\cite{fasterdoc}, we made checkpoints with the
``fold-over'' setup, which result in lightweight, incremental checkpoints. Each
fold-over checkpoint blocked on-going operations for around 19 seconds, and recovery took 28 seconds.
The overhead of the  checkpoint is partially due to saving the in-memory hash
table, a limitation of using flat structure that is not friendly to
persistent storage.
% We also noticed a strange phenomenon that
%When running the experiment in which all writes were updates and should not lead to
%increased memory requirements, FASTER kept consuming more and more memory.
%To investigate this, we ran another 200 million updates and saw it
%caused throughput degradation to only several thousands of
%operations per second as it depleted the free memory.
For its normal performance, we used the same hash function as in \DB
%and
%limited
%the in-memory part of HybridLog to 1G, but also 
and also tried the ``volatile'' case where
there is no warm-up phase to saturate the memory part of HybridLog so there are no disk writes. In this case, we found
the performance similar to Masstree.
%Performance is significantly reduced when it keeps writing HybridLog to disk---even so,
%the store is still not persistent as no checkpoint is created.
%We did another 90\% read-intensive YCSB test to confirm that read operations cause
%the slowdown.

\vspace{-0.10in}
\paragraph{Masstree.}
Masstree~\cite{masstree2012} maintains a trie-like concatenation of \bptree{}s in memory
and logs all writes to the disk. This enables high throughput
as no on-disk index is maintained.
On the other hand, during a recovery the log has to be replayed to restore
the in-memory index.
In the experiment we performed, it took around
31.8 seconds for playing back around 50 million write operations (50\% of operations are writes).

To mitigate this and also to recycle the log storage,
Masstree supports checkpoints that serve as new initial states for replay.
We took snapshots of the store at 1 million (0.29 seconds) to 10 million (3.35 seconds) items
and confirmed linear growth of the checkpointing time.
Unfortunately, the snapshot with 100 million items was not successful with the Masstree code,
so the number in Table~\ref{table:crash-recovery} is extrapolated.

\vspace{-0.10in}
\paragraph{KVell.} Instead of using an append-only log, KVell~\cite{kvell} uses slabs to preallocate
space for items. We used its most recent GitHub code to run the experiment.
The code currently does not support YCSB keys
% as they have some common prefix
(the paper used 8-byte random integer keys for YCSB instead).
% and will otherwise crash with the message indicating this issue.
We generated
23-byte keys with random bytes using the same distribution to best approximate the
YCSB workload.
% Unlike Masstree,
KVell's recovery takes 91.9 seconds regardless of the number of operations.
This time depends only on the size of the data store.
KVell is currently unable to perform a crash recovery for a workload with a mix of
insert and delete operations.

\section{Related Work}
\label{sec:related-work}

Various prior systems have looked into better leveraging available
memory to speed up performance of key-value stores.
% Instead of using tree structures,
SILT~\cite{silt2011} has a pipeline of three data structures to
improve memory-efficiency and write-friendliness.
%Masstree~\cite{masstree2012} focuses on performance improvements for SMP
%concurrent access using a trie-like concatenation of \bptree{}s.
LMDB~\cite{lmdb} is a popular open-source key-value store that leverages a
memory-mapped, copy-on-write \bptree.
\DB's usage of memory-mapped storage is inspired by LMDB.
% For memory-mapped or purely in-memory data structures, hash tables
% require trading off storage overhead and access performance.
% A large, flat hash table reduces hash collisions but comes
% at the cost of having a large sparse storage space.
As discussed in \S\ref{sec:eval-crash}, there are also persistent
key-value stores that only keep a fast, concurrent hash table or other indexing data structures in-memory and
pipe all writes directly into append-only logs or pre-allocated slabs~\cite{masstree2012, nibble17, faster2018,
kvell}.
Such architectures suffer from significant recovery/checkpoint overhead.
\begin{comment}
Tries (aka \emph{prefix trees}) present an interesting alternative~\cite{trie}.
They have become popular in functional programming
languages~\cite{SV18, steindorfer2015, bagwell2001ideal, AreiasR14}
and are used for in-memory indexing.
ART~\cite{art2013} is an in-memory
database that indexes user keys directly with a radix tree
(a space-optimized trie).
HOT~\cite{hot18} dynamically varies the number of bits used for each tree
node to optimize the height.
Tries are also used to optimize \bptree or LSM- based key-value
stores.  SuRF~\cite{surf2018} uses a trie to optimize range queries.
\end{comment}

There has been extensive work to reduce write-amplification in LSM-based
persistent key-value stores.
RocksDB~\cite{rocksdb, rocksdb2} is a fork of LevelDB
improved by developers at Facebook. It provides more features such as
multi-threaded compaction and support for transactions.
Inspired by skip lists and based on HyperLevelDB~\cite{hyperleveldb},
PebblesDB~\cite{pebblesdb} proposes the Fragmented LSM data structure,
carefully choosing the SSTs during compaction to reduce amplification.
LSM-trie~\cite{lsmtrie2015} uses a static hash-trie merge structure
that keeps reorganizing data for more efficient compaction.
SuRF~\cite{surf2018} uses an LSM design with a trie-based filter to optimize
range queries.
Accordion~\cite{accordion2018} improves the memory
organization for LSM. Monkey~\cite{DayanAI17} reduces the lookup cost for LSM by allocating
memory to filter across different levels, minimizing the number of false
positives. Dostoevsky~\cite{DayanI18} introduces lazy-leveling to remove
superfluous merging. mLSM~\cite{mlsm} is tailored for blockchain applications and
significantly improves the performance of the Ethereum storage subsystem.

There are also recent proposals to combine LSM and \bptree designs.
Jungle~\cite{jungle2019} reduces update cost without sacrificing lookup cost in
LSM using a \bptree. SLM-DB~\cite{slmdb2019} assumes persistent memory hardware.
It uses a \bptree for indexing and stages insertions to LSM.

Existing storage data structures have evolved in response to changes in hardware.
w\bptree~\cite{CJ15} reduces transaction logging overhead for a \bptree in
non-volatile main memory.
L\bptree~\cite{LiuCW20} optimizes the index performance using 3DXPoint
persistent memory.  S3~\cite{alibaba2019} uses an in-memory skip-list index for
a customized version of RocksDB in Alibaba Cloud. RECIPE~\cite{recipe19} offers
a principled way to convert concurrent indexes on DRAM to the one on
persistent-memory with crash-consistency.

Like \DB, other systems have embraced the trie for in-memory indexes.
ART~\cite{art2013} uses a radix tree, also
compresses the non-diverging paths. HOT~\cite{hot18} uses an adaptive number of
children for each node.
Compared to these in-memory indexes, there are several major differences:
(1) \DB is designed to be persistent and has an optimized lazy-trie for an on-disk index;
% for example, lazy-trie tree nodes have a fixed number of slots in their child table to
% improve persistent storage performance.
(2) To ensure near-optimal tree height like a \bptree, instead of directly using key strings
to index the trie, lazy-trie uses fixed-length hashes, having different statistical properties;
(3) In addition to path compression, lazy-trie employs
sluggish splitting to reduce variance in tree height, making the tree storage
footprint practical and even lower than that for a \bptree.

%(Ted: talk about Hash-Table/in-memory index + Log approach here carefully)

\section{Conclusion}
\label{sec:conclusion}
%\vspace{-.09in}
This paper explored the idea of an in-memory index that is
also storage-friendly, allowing both fast in-memory access and fast
failure recovery.
We designed the lazy-trie data structure and implemented \DB to this end.
\DB represents a new trade-off between fast access and fast recovery.
Potential future work directions include further optimizing performance, for
example by removing bottlenecks in shared locks and the Rust write buffer.
Furthermore, to increase applicability, we are also interested in designing
a sorted data structure that could support range queries.
\bibliographystyle{acm}
\interlinepenalty=10000
\bibliography{\jobname}

\begin{thebibliography}{10}

\bibitem{madvise}
madvise(2) - {Linux} manual page.
\newblock \url{http://man7.org/linux/man-pages/man2/madvise.2.html}.
\newblock Accessed: 2020-04-15.

\bibitem{jungle2019}
{\sc Ahn, J., Qader, M.~A., Kang, W., Nguyen, H., Zhang, G., and
  Ben{-}Romdhane, S.}
\newblock Jungle: Towards dynamically adjustable key-value store by combining
  {LSM}-tree and copy-on-write {B+}-tree.
\newblock In {\em 11th {USENIX} Workshop on Hot Topics in Storage and File
  Systems, HotStorage 2019, Renton, WA, USA, July 8-9, 2019\/} (2019), D.~Peek
  and G.~Yadgar, Eds., {USENIX} Association.

\bibitem{highwayhash16}
{\sc Alakuijala, J., Cox, B., and Wassenberg, J.}
\newblock Fast keyed hash/pseudo-random function using {SIMD} multiply and
  permute, 2016.

\bibitem{servo16}
{\sc Anderson, B., Bergstrom, L., Goregaokar, M., Matthews, J., McAllister, K.,
  Moffitt, J., and Sapin, S.}
\newblock Engineering the servo web browser engine using rust.
\newblock In {\em Proceedings of the 38th International Conference on Software
  Engineering, {ICSE} 2016, Austin, TX, USA, May 14-22, 2016 - Companion
  Volume\/} (2016), L.~K. Dillon, W.~Visser, and L.~Williams, Eds., {ACM},
  pp.~81--89.

\bibitem{BadamPPP09}
{\sc Badam, A., Park, K., Pai, V.~S., and Peterson, L.~L.}
\newblock Hashcache: Cache storage for the next billion.
\newblock In {\em Proceedings of the 6th {USENIX} Symposium on Networked
  Systems Design and Implementation, {NSDI} 2009, April 22-24, 2009, Boston,
  MA, {USA}\/} (2009), J.~Rexford and E.~G. Sirer, Eds., {USENIX} Association,
  pp.~123--136.

\bibitem{hot18}
{\sc Binna, R., Zangerle, E., Pichl, M., Specht, G., and Leis, V.}
\newblock {HOT:} {A} height optimized trie index for main-memory database
  systems.
\newblock In {\em Proceedings of the 2018 International Conference on
  Management of Data, {SIGMOD} Conference 2018, Houston, TX, USA, June 10-15,
  2018\/} (2018), G.~Das, C.~M. Jermaine, and P.~A. Bernstein, Eds., {ACM},
  pp.~521--534.

\bibitem{accordion2018}
{\sc Bortnikov, E., Braginsky, A., Hillel, E., Keidar, I., and Sheffi, G.}
\newblock Accordion: Better memory organization for {LSM} key-value stores.
\newblock {\em {PVLDB} 11}, 12 (2018), 1863--1875.

\bibitem{rocksdb2}
{\sc Cao, Z., Dong, S., Vemuri, S., and Du, D.~H.}
\newblock Characterizing, modeling, and benchmarking {RocksDB} key-value
  workloads at {Facebook}.
\newblock In {\em 18th {USENIX} Conference on File and Storage Technologies
  ({FAST} 20)\/} (Santa Clara, CA, Feb. 2020), {USENIX} Association,
  pp.~209--223.

\bibitem{faster2018}
{\sc Chandramouli, B., Prasaad, G., Kossmann, D., Levandoski, J.~J., Hunter,
  J., and Barnett, M.}
\newblock {FASTER}: {A} concurrent key-value store with in-place updates.
\newblock In {\em Proceedings of the 2018 International Conference on
  Management of Data, {SIGMOD} Conference 2018, Houston, TX, USA, June 10-15,
  2018\/} (2018), G.~Das, C.~M. Jermaine, and P.~A. Bernstein, Eds., {ACM},
  pp.~275--290.

\bibitem{CJ15}
{\sc Chen, S., and Jin, Q.}
\newblock Persistent {B+}-trees in non-volatile main memory.
\newblock {\em {PVLDB} 8}, 7 (2015), 786--797.

\bibitem{ycsb}
{\sc Cooper, B.~F., Silberstein, A., Tam, E., Ramakrishnan, R., and Sears, R.}
\newblock Benchmarking cloud serving systems with {YCSB}.
\newblock In {\em Proceedings of the 1st {ACM} Symposium on Cloud Computing,
  SoCC 2010, Indianapolis, Indiana, USA, June 10-11, 2010\/} (2010), J.~M.
  Hellerstein, S.~Chaudhuri, and M.~Rosenblum, Eds., {ACM}, pp.~143--154.

\bibitem{lmdb-doc}
{\sc Corporation, S.}
\newblock {LMDB: Lightning Memory-Mapped Database Manager}.
\newblock \url{http://www.lmdb.tech/doc/}.
\newblock Accessed: 2020-05-16.

\bibitem{lmdb}
{\sc Corporation, S.}
\newblock Symas lightning memory-mapped database.
\newblock \url{https://symas.com/lmdb/}.
\newblock Accessed: 2020-04-15.

\bibitem{DayanAI17}
{\sc Dayan, N., Athanassoulis, M., and Idreos, S.}
\newblock Monkey: Optimal navigable key-value store.
\newblock In {\em Proceedings of the 2017 {ACM} International Conference on
  Management of Data, {SIGMOD} Conference 2017, Chicago, IL, USA, May 14-19,
  2017\/} (2017), S.~Salihoglu, W.~Zhou, R.~Chirkova, J.~Yang, and D.~Suciu,
  Eds., {ACM}, pp.~79--94.

\bibitem{DayanI18}
{\sc Dayan, N., and Idreos, S.}
\newblock Dostoevsky: Better space-time trade-offs for {LSM}-tree based
  key-value stores via adaptive removal of superfluous merging.
\newblock In {\em Proceedings of the 2018 International Conference on
  Management of Data, {SIGMOD} Conference 2018, Houston, TX, USA, June 10-15,
  2018\/} (2018), G.~Das, C.~M. Jermaine, and P.~A. Bernstein, Eds., {ACM},
  pp.~505--520.

\bibitem{dynamo07}
{\sc DeCandia, G., Hastorun, D., Jampani, M., Kakulapati, G., Lakshman, A.,
  Pilchin, A., Sivasubramanian, S., Vosshall, P., and Vogels, W.}
\newblock Dynamo: {Amazon}'s highly available key-value store.
\newblock In {\em Proceedings of the 21st {ACM} Symposium on Operating Systems
  Principles 2007, {SOSP} 2007, Stevenson, Washington, USA, October 14-17,
  2007\/} (2007), T.~C. Bressoud and M.~F. Kaashoek, Eds., {ACM}, pp.~205--220.

\bibitem{redox}
{\sc Developers, R.~O.}
\newblock Redox - your next(gen) {OS}.
\newblock \url{https://www.redox-os.org/}.
\newblock Accessed: 2020-04-15.

\bibitem{rustsafety}
{\sc Developers, T. R.~P.}
\newblock The {Rustonomicon}: the dark arts of advanced and unsafe {Rust}
  programming.
\newblock \url{https://doc.rust-lang.org/stable/nomicon/}.
\newblock Accessed: 2020-04-15.

\bibitem{rocksdb}
{\sc Dong, S., Callaghan, M., Galanis, L., Borthakur, D., Savor, T., and Strum,
  M.}
\newblock Optimizing space amplification in {RocksDB}.
\newblock In {\em {CIDR} 2017, 8th Biennial Conference on Innovative Data
  Systems Research, Chaminade, CA, USA, January 8-11, 2017, Online
  Proceedings\/} (2017), www.cidrdb.org.

\bibitem{hyperleveldb}
{\sc Escriva, R.}
\newblock Inside {HyperLevelDB}.
\newblock \url{https://hackingdistributed.com/2013/06/17/hyperleveldb/}.
\newblock Accessed: 2020-04-20.

\bibitem{rocksdb-doc}
{\sc Facebook}.
\newblock Memtable - facebook/rocksdb wiki.
\newblock
  \url{https://github.com/facebook/rocksdb/wiki/MemTable#concurrent-insert}.
\newblock Accessed: 2020-05-16.

\bibitem{snappydb}
{\sc Hachicha, N.}
\newblock {SnappyDB}: a fast and lightweight key/value database library for
  {Android}.
\newblock \url{https://www.snappydb.com/}.
\newblock Accessed: 2020-04-15.

\bibitem{johnstone1998memory}
{\sc Johnstone, M.~S., and Wilson, P.~R.}
\newblock The memory fragmentation problem: Solved?
\newblock {\em ACM Sigplan Notices 34}, 3 (1998), 26--36.

\bibitem{slmdb2019}
{\sc Kaiyrakhmet, O., Lee, S., Nam, B., Noh, S.~H., and Choi, Y.}
\newblock {SLM-DB:} single-level key-value store with persistent memory.
\newblock In {\em 17th {USENIX} Conference on File and Storage Technologies,
  {FAST} 2019, Boston, MA, February 25-28, 2019\/} (2019), A.~Merchant and
  H.~Weatherspoon, Eds., {USENIX} Association, pp.~191--205.

\bibitem{taocp}
{\sc Knuth, D.~E.}
\newblock {\em The Art of Computer Programming, Volume 1: Fundamental
  Algorithms}, vol.~1.
\newblock Pearson Education, 1997.

\bibitem{fastercloud}
{\sc Kulkarn, C., Chandramouli, B., and Stutsman, R.}
\newblock Achieving high throughput and elasticity in a larger-than-memory
  store.
\newblock In {\em Proc. VLDB Endow. Volume 14, Issue 8, 2021 (to appear)}.

\bibitem{recipe19}
{\sc Lee, S.~K., Mohan, J., Kashyap, S., Kim, T., and Chidambaram, V.}
\newblock Recipe: converting concurrent {DRAM} indexes to persistent-memory
  indexes.
\newblock In {\em Proceedings of the 27th {ACM} Symposium on Operating Systems
  Principles, {SOSP} 2019, Huntsville, ON, Canada, October 27-30, 2019\/}
  (2019), T.~Brecht and C.~Williamson, Eds., {ACM}, pp.~462--477.

\bibitem{art2013}
{\sc Leis, V., Kemper, A., and Neumann, T.}
\newblock The adaptive radix tree: Artful indexing for main-memory databases.
\newblock In {\em 29th {IEEE} International Conference on Data Engineering,
  {ICDE} 2013, Brisbane, Australia, April 8-12, 2013\/} (2013), C.~S. Jensen,
  C.~M. Jermaine, and X.~Zhou, Eds., {IEEE} Computer Society, pp.~38--49.

\bibitem{kvell}
{\sc Lepers, B., Balmau, O., Gupta, K., and Zwaenepoel, W.}
\newblock {KVell}: the design and implementation of a fast persistent key-value
  store.
\newblock In {\em Proceedings of the 27th {ACM} Symposium on Operating Systems
  Principles, {SOSP} 2019, Huntsville, ON, Canada, October 27-30, 2019\/}
  (2019), T.~Brecht and C.~Williamson, Eds., {ACM}, pp.~447--461.

\bibitem{silt2011}
{\sc Lim, H., Fan, B., Andersen, D.~G., and Kaminsky, M.}
\newblock {SILT:} a memory-efficient, high-performance key-value store.
\newblock In {\em Proceedings of the 23rd {ACM} Symposium on Operating Systems
  Principles 2011, {SOSP} 2011, Cascais, Portugal, October 23-26, 2011\/}
  (2011), T.~Wobber and P.~Druschel, Eds., {ACM}, pp.~1--13.

\bibitem{voldemort}
{\sc LinkedIn, M.}
\newblock Project {Voldemort} - a distributed database.
\newblock \url{https://www.project-voldemort.com/voldemort}.
\newblock Accessed: 2020-12-10.

\bibitem{LiuCW20}
{\sc Liu, J., Chen, S., and Wang, L.}
\newblock {LB+}-trees: Optimizing persistent index performance on {3D XPoint}
  memory.
\newblock {\em {PVLDB} 13}, 7 (2020), 1078--1090.

\bibitem{orientdb}
{\sc Ltd, O.}
\newblock {OrientDB}.
\newblock \url{https://www.orientdb.org/}.
\newblock Accessed: 2020-12-10.

\bibitem{masstree2012}
{\sc Mao, Y., Kohler, E., and Morris, R.~T.}
\newblock Cache craftiness for fast multicore key-value storage.
\newblock In {\em European Conference on Computer Systems, Proceedings of the
  Seventh EuroSys Conference 2012, EuroSys '12, Bern, Switzerland, April 10-13,
  2012\/} (2012), P.~Felber, F.~Bellosa, and H.~Bos, Eds., {ACM}, pp.~183--196.

\bibitem{MarmolGA16}
{\sc M{\'{a}}rmol, L., Guerra, J., and Aguilera, M.~K.}
\newblock Non-volatile memory through customized key-value stores.
\newblock In {\em 8th {USENIX} Workshop on Hot Topics in Storage and File
  Systems, HotStorage 2016, Denver, CO, USA, June 20-21, 2016\/} (2016),
  N.~Agrawal and S.~H. Noh, Eds., {USENIX} Association.

\bibitem{nibble17}
{\sc Merritt, A., Gavrilovska, A., Chen, Y., and Milojicic, D.~S.}
\newblock Concurrent log-structured memory for many-core key-value stores.
\newblock {\em {PVLDB} 11}, 4 (2017), 458--471.

\bibitem{faster-web}
{\sc Microsoft}.
\newblock {FASTER - Key Features}.
\newblock \url{https://microsoft.github.io/FASTER/}.
\newblock Accessed: 2021-04-25.

\bibitem{fasterdoc}
{\sc Microsoft}.
\newblock {FasterKV} basics - {FASTER}.
\newblock
  \url{https://microsoft.github.io/FASTER/docs/fasterkv-basics/#checkpointing-and-recovery}.
\newblock Accessed: 2020-12-07.

\bibitem{ARIES}
{\sc Mohan, C., Haderle, D., Lindsay, B., Pirahesh, H., and Schwarz, P.}
\newblock Aries: A transaction recovery method supporting fine-granularity
  locking and partial rollbacks using write-ahead logging.
\newblock {\em ACM Transactions on Database Systems 17}, 1 (Mar. 1992),
  94–--162.

\bibitem{radixtree}
{\sc Morrison, D.~R.}
\newblock {PATRICIA} - practical algorithm to retrieve information coded in
  alphanumeric.
\newblock {\em J. {ACM} 15}, 4 (1968), 514--534.

\bibitem{mlsurvey}
{\sc Nguyen, G., Dlugolinsky, S., Bob{\'{a}}k, M., Tran, V.~D., Garc{\'{\i}}a,
  {\'{A}}.~L., Heredia, I., Mal{\'{\i}}k, P., and Hluch{\'{y}}, L.}
\newblock Machine learning and deep learning frameworks and libraries for
  large-scale data mining: a survey.
\newblock {\em Artif. Intell. Rev. 52}, 1 (2019), 77--124.

\bibitem{rustnix}
{\sc nix-rust Project~Developers, T.}
\newblock nix --- crates.io: Rust package registry.
\newblock \url{https://crates.io/crates/nix}.
\newblock Accessed: 2020-04-15.

\bibitem{rangequery}
{\sc Pirzadeh, P., Tatemura, J., Po, O., and Hacig{\"{u}}m{\"{u}}s, H.}
\newblock Performance evaluation of range queries in key value stores.
\newblock {\em J. Grid Comput. 10}, 1 (2012), 109--132.

\bibitem{glibcaio}
{\sc Project, T.~G.}
\newblock Asynchronous {I/O} (the {GNU C} library).
\newblock
  \url{https://www.gnu.org/software/libc/manual/html_node/Asynchronous-I_002fO.html}.
\newblock Accessed: 2020-04-15.

\bibitem{pebblesdb}
{\sc Raju, P., Kadekodi, R., Chidambaram, V., and Abraham, I.}
\newblock {PebblesDB}: Building key-value stores using fragmented
  log-structured merge trees.
\newblock In {\em Proceedings of the 26th Symposium on Operating Systems
  Principles, Shanghai, China, October 28-31, 2017\/} (2017), {ACM},
  pp.~497--514.

\bibitem{mlsm}
{\sc Raju, P., Ponnapalli, S., Kaminsky, E., Oved, G., Keener, Z., Chidambaram,
  V., and Abraham, I.}
\newblock {mLSM}: Making authenticated storage faster in {Ethereum}.
\newblock In {\em 10th {USENIX} Workshop on Hot Topics in Storage and File
  Systems, HotStorage 2018, Boston, MA, USA, July 9-10, 2018\/} (2018), A.~Goel
  and N.~Talagala, Eds., {USENIX} Association.

\bibitem{ShenCJS16}
{\sc Shen, Z., Chen, F., Jia, Y., and Shao, Z.}
\newblock Optimizing flash-based key-value cache systems.
\newblock In {\em 8th {USENIX} Workshop on Hot Topics in Storage and File
  Systems, HotStorage 2016, Denver, CO, USA, June 20-21, 2016\/} (2016),
  N.~Agrawal and S.~H. Noh, Eds., {USENIX} Association.

\bibitem{rust}
{\sc Team, T.~R.}
\newblock Rust programming language.
\newblock \url{https://www.rust-lang.org/}.
\newblock Accessed: 2020-05-26.

\bibitem{WangDLXZCCOR18}
{\sc Wang, S., Dinh, T. T.~A., Lin, Q., Xie, Z., Zhang, M., Cai, Q., Chen, G.,
  Ooi, B.~C., and Ruan, P.}
\newblock Forkbase: An efficient storage engine for blockchain and forkable
  applications.
\newblock {\em {PVLDB} 11}, 10 (2018), 1137--1150.

\bibitem{lsmtrie2015}
{\sc Wu, X., Xu, Y., Shao, Z., and Jiang, S.}
\newblock {LSM}-trie: An {LSM}-tree-based ultra-large key-value store for small
  data items.
\newblock In {\em 2015 {USENIX} Annual Technical Conference, {USENIX} {ATC}
  '15, July 8-10, Santa Clara, CA, {USA}\/} (2015), S.~Lu and E.~Riedel, Eds.,
  {USENIX} Association, pp.~71--82.

\bibitem{surf2018}
{\sc Zhang, H., Lim, H., Leis, V., Andersen, D.~G., Kaminsky, M., Keeton, K.,
  and Pavlo, A.}
\newblock Surf: Practical range query filtering with fast succinct tries.
\newblock In {\em Proceedings of the 2018 International Conference on
  Management of Data, {SIGMOD} Conference 2018, Houston, TX, USA, June 10-15,
  2018\/} (2018), G.~Das, C.~M. Jermaine, and P.~A. Bernstein, Eds., {ACM},
  pp.~323--336.

\bibitem{alibaba2019}
{\sc Zhang, J., Wu, S., Tan, Z., Chen, G., Cheng, Z., Cao, W., Gao, Y., and
  Feng, X.}
\newblock {S3:} {A} scalable in-memory skip-list index for key-value store.
\newblock {\em {PVLDB} 12}, 12 (2019), 2183--2194.

\end{thebibliography}

%\lstset{
%basicstyle=\small\ttfamily,
%columns=flexible,
%breaklines=true
%}
\newpage
\ifdefined\submission
\begin{appendices}
\section{Changes Since Last Submission}
We have made two majors changes to the paper since our OSDI'21 submission:
\begin{enumerate}[leftmargin=*,topsep=1pt]
    \item We re-positioned \DB as a direct competitor to unsorted, persistent
        key-value stores like FASTER as it provides similar functionalities
        with similar limitations. Our motivation is also improved: we rethought
        the recent trend of keeping an in-memory (storage-unfriendly but fast
        for access) index and logging all high-level operations to the disk ---
        while this approach usually provides better performance and I/O utilization,
        it defers the
        overhead of on-disk index maintenance to the recovery of the volatile store state. While such a different
        persistence approach may be suitable for some use cases (e.g. system logging), it is
        desirable that a design have the benefits of both worlds:
        \begin{itemize}
            \item The persistence model is the same as on-disk index
                solutions. No checkpoints or playbacks of operations are
                required for persistence.
            \item A data structure that is good for fast access, with performance
                at least as good as the log-based approach.
        \end{itemize}
        To this end, lazy-trie is a data structure that is memory-efficient but also storage-friendly. \DB
        is a practical implementation based upon this data structure.
    \item We realized that FASTER is our closest competitor. We carefully designed and conducted experiments comparing to FASTER. From the results, we confirmed
        the drawbacks and trade-offs of log-based approaches, which try to balance the performance between normal
        operations and persisting the in-memory state: pure
        writes are very fast while mixed workloads degrade its performance.
        \DB outperforms FASTER in many scenarios and demonstrates
        a promising capability of concurrent access,
        a key benefit of in-memory indexes.
\end{enumerate}
\section{Reviews and Rebuttal from OSDI 2021}
\subsection{Final Comment from Reviewer A}
Thank you for submitting to OSDI'21. The PC had discussed this paper intensively and have read your response carefully. Reviewers like the idea of lazy trie, but all feel that the paper is not full ready due to insufficient experimental evaluation to compare with previous works, and also include operations such as range queries where the proposed technique falls short.  It would be useful for reviewers to understand how general and applicable the proposed solution is.  We encourage the authors addresses these concerns and submit the work to future conferences.

\subsection{Reviews}
\subsubsection{Review \#150A}
\begin{itemize}[leftmargin=*,itemsep=1pt,topsep=1pt]
\item\textbf{Novelty.} 3. New contribution
\item\textbf{Experimental methodology.} 4. Good
\item\textbf{Writing quality.} 4. Well-written
\item\textbf{Overall merit.} 4. Weak accept  (OK paper, but I'm not enthusiastic)
\item\textbf{Paper summary.}
This paper proposed a new data structure called "lazy-trie" for key-value store, especially for cases where most data can fit into memory.  Based on this data structure, it then built CedrusDB, a persistent key-value store which can better leverage concurrent processing than past approaches.

\item\textbf{Strengths.}

+  Up front about the proposed approach limitations and trade-offs

+  The paper is easy to read 

+  The experimental evaluation is relatively solid.

\item\textbf{Significant weaknesses.}

- The approach cannot well support range queries, not sure how much would it limit the applicability of the proposed approach.

- One of the major benefits over previous work is concurrent processing.  It would be  nicer to show deeper insights what enabled such benefits.  Can the same technique apply to previous approaches?

\item\textbf{Comments for author.}
This paper is easy to read.  The authors did a great job presenting the motivation, the technical approach as well as the limitations up front instead of trying to hide or downplay the limitations.   The experimental evaluation is relatively convincing.

I just have a few minor concerns/questions:

(1) How bad is the limitation related to range query support?  While the paper cited a few past work with similar limitations, I was hoping that the paper can provide some real application examples or use case scenario that do not need range queries.  Is this the trade-off that real world applications are willing to tolerate?

(2) CedrusDB can support datasets that are too large to fit into main memory by dividing the space into segments and map to memory segments by segments.  How about applications with operations that span multiple segments?  Would it affect performance?

(3) In order to balance the  "lazy-trie", it introduces the idea of "sluggish lazy-trie".  What did you lose on this?  Can it be determined dynamically only when a large path variance is detected? 

Overall, it seems a solid paper with good evaluation.

\item\textbf{Questions for authors' response.}
Questions above.
\end{itemize}

\subsubsection{Review \#150B}
\begin{itemize}[leftmargin=*,itemsep=1pt,topsep=1pt]
\item\textbf{Novelty.} 3. New contribution
\item\textbf{Experimental methodology.} 2. Poor
\item\textbf{Writing quality.} 3. Adequate
\item\textbf{Overall merit.} 3. Weak reject (This paper should be rejected, but I'll not fight strongly)

\item\textbf{Paper summary.}
CedrusDB is a key-value store for point queries and
insertions/deletions/updates. It has four main design elements:

1. An in-memory hash-trie. Keys are hashed to 256 bits and stored in a
trie.  Prefix compression is not done. However suffixes are collapsed:
unique suffixes are stored as a single node rather than a unary
subtree. "Data nodes" (leaves) with the same parent are stored as a
linked list up to some threshold list length (called "sluggishness")
and then assembled into the trie structure.

2. Lazy flushing of hash-trie updates. Write-ahead logging is used for
atomicity.  Udates to the trie are batched and flushed lazily, after
which log records are pruned

3. "Logical spaces": essentially an on-disk allocator using
regions. Each region is either fully mapped into memory (using mmap)
or not mapped at all. Trie nodes are mapped to a fixed-size allocator
while data nodes are mapped to a variable-sized allocator using a
next-fit allocation policy.

4. Multi-core concurrency control is provided by node-level
locking. CedrusDB uses hand-over-hand locking going down the tree from
the root. Lookups take read locks; insertions take upgradable read
locks which are upgraded to write locks if needed; deletions
write-lock the entire path to the root.

CedrusDB is implemented in Rust on Linux.

Performance is compared to LevelDB, RocksDB, and LMDB using YCSB with
Zipfian access distribution, and 50--100\% reads.  CedrusDB and LMDB
generally outperform the other two, with CedrusDB being slightly
better on a smaller tree with larger values (1024 byte) and LMDB
slightly better on a larger tree with smaller values (128 bytes). The
highest performance reported for CedrusDB is 2.5 M lookups/s (100%
read) with 4 threads. Scalability beyond 4 threads is not shown.

Crash recovery is compared with RocksDB, MassTree, KVell, and FASTER.
On a store with 100 M items, 50\% Zipfian updates, and 128-byte values,
CedrusDB recovers in 1.85s with a performance of 389 Kops/s.By
comparison Masstree recovers in around 30 s with a checkpoint interval
of 33.5 s, but has 3x higher throughput.

Sensitivity comparisons with CedrusDB alone show that some
"sluggishness" improves both performance and memory consumption, and
that performance drops drastically when the data size increases even
modestly beyond the available memory size.

\item\textbf{Strengths.}

+ Interesting data structure

+ Allocation strategy considered

+ Recovery implemented and evaluated

\item\textbf{Significant weaknesses.}

- No range queries

- Performance much worse than Masstree

- No performance comparison to hash tables or B+-trees

- Poor scalability especially with deletes.

\item\textbf{Comments for author.}
Thank you for your paper. There seem to be four main design elements
to CedrusDB: the hash-trie, the lazy flushing of dirty data (plus
WAL), the "logical spaces" aka allocator, and the locking (latching)
protocol for shared-memory concurrency control. It is nice to see a
complete design and implementation that includes all four, and I also
liked the fact that recovery is both implemented and evaluated. It was
also nice to see small values and non-uniform access evaluated as
these are often neglected.

The main claim of the paper is improved performance, or an improved
performance/recovery tradeoff. Here the paper falls short. In
particular:

1. The comparison is on point queries/updates only, against
RocksDB/LMDB/Masstree, all of which support range queries. This is not
apples-to-apples. As CedrusDB does not support range queries, there
needs to be a comparison with a state of the art hash table.
According to the Masstree Eurosys paper, hash tables are at least 2.5x
faster than Masstree itself. In Table 1 CedrusDB has 3x \emph{worse}
throughput than Masstree, so presumably it would be 7.5x (or more)
relative to a hash table.

2. The main benefit of CedrusDB seems to be that the WAL can be pruned
by lazy flushing of dirty data thus reducing recovery time. But this
kind of asynchronous checkpointing is an old idea and widely used. It
is also not tied to a specific data structure and could also be used
with Masstree, B+-tree, hash table, etc. E.g. in the evaluation
showing Masstree with 10x higher recovery time and 3x higher
throughput: what would happen with more aggressive/more efficient
checkpointing in Masstree?

3. Some design features perform poorly or are not evaluated. E.g. why
is scalability only shown up to 4 threads given that key-value stores
are typically run on machine with many more hyperthreads? Why is the
very poor concurrency on deletes (and the resulting poor performance)
acceptable? Is there any point in supporting out-of-core operation
given that huge performance degradation (many factors for just 25\%
out-of-core)?

The locking/latching protocol and allocator are interesting and mildly
novel in the context of the hash-trie data structure but there is no
evaluation to show whether they are better than alternative designs.

At a higher level it is also worth remembering that the "problem" of
persistence for in-DRAM data has a simple solution that works with any
data structure with zero foreground cost for checkpointing: a Li-ion
battery and a cheap SSD that can be used to save the memory state only
when there is a power failure.

Detailed comments/questions:

The data structure is somewhat similar to burst tries and Masstree but
is clearly not identical. I expected the paper to more clearly explain
the rationale and the pros/cons of the design choices (such as using a
"sluggish" linked list).

There are repeated claims that the hash-trie is "simpler" than a
B+-tree due to the lack of recursive split/merge. But no evidence is
provided for this and it is a very subjective claim. The hash-trie can
also have recursive balancing operations as a corner case. The locking
protocol even as described does not sound simple and making deletes
concurrent would make it even more complex.

Is the locking protocol from Section 3.5 one of the novel
contributions?  If so it needs to be explained in more detail. In DB
terminology, this kind of locking is often referred to as "latching"
as it provides single-key atomicity and data structure integrity but
not transactions/multi-key atomicity. How does the CedrusDB strategy
relate to DB latching strategies as described in
\url{https://www.hpl.hp.com/techreports/2010/HPL-2010-9.pdf}? Is it provably
deadlock-free?

The paper says that deletion sacrifices "some" concurrency: by keeping
the root node write-locked in my view it sacrifices \emph{all}
concurrency. This is quite clear from Figures 10 11 for the "D"
workload (and for cedrusdb-m) where the performance at 4 threads is
almost the same (sometimes worse) than 1 thread. This suggests the
system will not scale well when the workload contains deletions.

How does CedrusDB handle concurrent updates to a page that is being
written to disk? I.e. locking, make a copy first, other ...

How is paging-in of clean pages done: i.e. bulk-loaded when region
is mapped, demand-paged by CedrusDB, demand-paged by OS, other, ...

The paper is full of details about the Rust implementation: what
primitives Rust supports, etc.  Mostly they do not seem relevant to
the design or the performance. If they are please make it clear
how. Otherwise I would consider omitting or shortening these sections.
\end{itemize}

\subsubsection{Review \#150C}
\begin{itemize}[leftmargin=*,itemsep=1pt,topsep=1pt]
\item\textbf{Novelty.} 2. Incremental improvement
\item\textbf{Experimental methodology.} 3. Average
\item\textbf{Writing quality.} 4. Well-written
\item\textbf{Overall merit.} 3. Weak reject (This paper should be rejected, but I'll not fight strongly)
\item\textbf{Paper summary.}
This paper presents CedrusDB--a KV-store for systems that can store most (but not necessarily all) of their data in memory. It replaces traditional B+ tree and LSM tree structures with a new trie-based one. The user keys are hashed, and the hashed keys are split into byte segments that represent characters in the trie. Paths in the trie are compressed to prevent the allocation and traversal of mostly-empty internal nodes. The trie leaves point to a linked list of user key-value pairs, and this list may contain pairs belonging to several leaves. This way paths can remain compressed and their splits delayed.

Hashing the user keys ensures their uniform distribution, which results in a balanced trie, but it prevents support for range queries on the user keys. Thus, CedrusDB trades this feature for improved update efficiency, concurrent access, and flexible space overhead.  Additional optimizations include mapping of file segments in memory, the management of the free tried nodes and user data space to minimize fragmentation, and the use of upgradable reader-writer locks.

CedrusDB is implemented in Rust and compared in the experiments to LMDB (B+ tree) and LevelDB/RocksDB (LSM-tree) on microbenchmarks and YCSB. CedrusDB outperforms each of the other KV-stores in different scenarios, but not in all of them (for single-threaded experiments). It is superior in multi-threaded use cases.

\item\textbf{Strengths.}
The motivation is strong: a KV-store that is designed for cases where most of the data fit in memory, but not necessarily all of it.

The design of the trie is clever, and the reduced update complexity clearly pays off when it comes to concurrent accesses. 

The paper is very well written and many aspects of the design are explained with helpful illustrations and examples. 

The evaluation covers all performance aspects of CedrusDB, as well as a comprehensive analysis of its parameters, and scenarios where its limitations are demonstrated and explained.

\item\textbf{Significant weaknesses.}
The evaluation does not give the full picture of the tradeoffs in CedrusDB. The design of Lazy Trie trades range queries for more efficient updates, but it is only compared to KV-stores that do support range queries (based on B+ and LSM trees). It should also be compared to hash-based structures that do not support range queries, as they are expected to be more efficient and smaller than the tree-based structures.

\item\textbf{Comments for author.}
I enjoyed reading this paper and think the optimizations are neat. The description of the design was easy to follow (in most parts), and the design decisions are well-motivated and explained. I appreciate the careful construction of the experiments with the goal of covering all the interesting aspects of this design, and comparing to several relevant baseline. There are certainly many relevant papers and research prototypes out there, and it is not possible to compare a new design to all the previous (relevant) ones.

My main concern is that, despite the extensive evaluation, CedrusDB is discussed and evaluated only in the context of tree-based KV-stores. This overlooks an entire section of the design space: tree-based structures dedicate storage and update overhead to provide not only lookups but also ordered scans (range queries) of the keys. However, hash-based structures provide fast lookups and updates but do not support such scans. 

CedrusDB does not support scans, which is a valid design decision, but it still incur the fundamental overheads of the trees: it stores and maintains the internal nodes and leaves. This is of course also a valid design decision, and a major contribution of the paper is the path compression that minimizes these overheads. However, to show the true value of this design, it must be compared to some hash-based design that avoids these overheads altogether. 

The related work section in the WiscKey paper (FAST '16) discusses hash-based KV-stores that could be used for this comparison (and should certainly be discussed). In a related note, in Sections 4.4 and 5 you claim that hash-based KV-stores have inefficient crash-recovery. Why are these design aspects connected? 

To summarize this point, even if the hash-based KV-stores are inferior when it comes to recovery, they present an important part of the design space, because of their reduced overhead. To establish the benefit of CedrusDB, we need to see how it compares to these structures in terms of performance and space overhead, and not only in terms of recovery.

The description of the disk I/O and crash recovery (Sections 3.3 and 3.4) is not as clear as the other sections. Perhaps a figure describing these processes (or at least one of them) would help. After reading these subsections back and forth, it is still not entirely clear to me how the integrity of the tree is ensured. I am confused because we may have multiple threads updating the tree structure, on one hand, and the disk thread aggregating writes in the background, on the other hand. How can we be sure that this aggregation does not result in updates only partially reaching the disk? I suppose this is taken care of by the WAL worker, I am not sure how.

In Section 4.2.4, you only compare CedrusDB with different regions, but the context is important: how does RocksDB perform in these settings? RocksDB is designed for this scenario, where we don't have all the data in memory, so this comparison is important.

Why is LMDB excluded from the crash recovery comparisons? It supposedly maintains its disk-resident data up-to-date, which slows down its writes, but its recovery time should be minimal.

Minor comments and nits:

On page 2 you describe an advantage of hash-trie as having fixed-sized nodes, in contrast to B+ trees with a variable number of children. However, just like in a hash-tire, the size of the B+ tree nodes is constant, and not all ``children'' slots are necessarily populated.

The analysis and presentation of Figure 4 are very helpful. It would be good to include LMDB and RocksDB in the figure for comparison.

Page 5: ``Each descriptor points to a free data node, while the header of a free data node points back to its descriptor'' --- from Figure 7 and the description of the allocation policy, I wonder if you meant that the header points to the \emph{next} free descriptor.

Page 6: ``it is the disk thread's responsibility to obtain the consistent state of a block from a file if it is not already available in its cache, rather than copying the content from memory as there may be a data race'' - I don't understand why copying content from memory is an alternative if the data is not cached. Are you referring to reads or to writes here?  

Figure 8 is never referenced from the text.

Page 7: ``CedrusDB optimizes for single write operations without batching, as no extra concurrency control is needed in this case'' - it is clear why a single write is simpler, but please explain what you mean by ``CedrusDB optimizes''. Optimizes how?

All the bar-graphs are painfully small (albeit well-formed). Please increase their size, and preferably rotate the labels so that they are horizontal.

Page 9 left bottom: ``The figure that show'' $\rightarrow$ figure shows that.

Page 10 left: ``updating the value of an existing key and to insert'' $\rightarrow$ and inserting

Section 4.3.1, Figure 16 vs. Figure 15: why does the value size have such an effect on LMDB?

\item\textbf{Questions for authors' response.}

1. Can you justify comparing your approach only to tree based KV-stores? Perhaps there is a fundamental limitation of hash-based structures that eludes me.

2. Please clarify how the disk thread maintains the integrity of the tree when it is updated concurrently from several threads.
\end{itemize}

\subsubsection{Review \#150D}
\begin{itemize}[leftmargin=*,itemsep=1pt,topsep=1pt]
\item\textbf{Novelty.} 3. New contribution
\item\textbf{Experimental methodology.} 2. Poor
\item\textbf{Writing quality.} 4. Well-written
\item\textbf{Overall merit.} 2. Reject (This paper should be rejected, I'll argue against it)
\item\textbf{Paper summary.}
This paper introduces a new unsorted index, named lazy trie, for a persistent key-value store and shows its Rust implementation, called CedrusDB. The key idea is to create a trie using a hash of a key. Compared to other popular indexes, like B+-tree and LSM tree, the complexity of a trie is only dependent on the key length, not the number of keys, so it is beneficial to handle many key-value pairs. The paper proposes a ``sluggish split/merge'', which is controlled by a sluggish factor, to reduce the space overhead. The authors implement the lazy trie using Rust. CedrusDB is the Rust implementation of the lazy trie. CedrusDB relies on mmap() for the lazy trie access and uses WAL
logging for crash consistency and read-write lock for concurrency. The
evaluation results show the performance and space usage comparison mostly with RocksDB (LSM tree) and LMDB (B+tree). CedrusDB shows better performance than RocksDB for read-intensive workloads when the (almost) entire data set is cached to DRAM.

\item\textbf{Strengths.}

- The idea of lazy trie and sluggish split/merge is novel and interesting.

- Designing and implementing a high-performance key-value store is an important problem as a key-value store becomes a critical component in many applications.

- The paper is mostly well written and easy to follow.

\item\textbf{Significant weaknesses.}

- While the key of the lazy trie is path compression, the paper does not provide an in-depth comparison with other state-of-the-art path compression techniques in trie indexes, such as ART and HOT.

- The evaluation does not show an in-depth comparison against the most relevant key-value stores, including KVell (SSD-optimized hybrid B-tree), FASTER (hybrid hash table), the traditional disk-based extensible hash table.

- CedrusDB shows poor performance when a workload is write-intensive or when the entire data is not cached. Hence it is questionable if CedrusDB can handle a large volume of data.

- The memory (disk space) management seems to be ad-hoc (see detail comments).

\item\textbf{Comments for author.}
Thank you for submitting the paper! I enjoy reading it. I like the idea of lazy trie and sluggish split/merge. The paper is well written and easy to follow. I wanted to like it more, but I think the idea is still premature, and the paper is not ready to be published. I hope the following comments are helpful to improve the algorithm and the paper for the next version.

Design

1. I like the idea of lazy trie and sluggish split/merge, which is simple but effective for path compression in a trie structure.

  However, I am not convinced if the lazy trie is better than the path compression techniques in recently proposed trie structures, such as ART [27] and HOT [6]. While ART and HOT are designed for in-memory index, I think that applying/customizing ART and HOT for CedrusDB will be straightforward. Similarly, I don't see any reason that lazy trie cannot be used for in-memory index. Unfortunately, I could not find any concrete reason why lazy trie would be better than ART and HOT from the description in the paper. 

  The paper should present in-depth comparison between lazy trie and the recent path compression techniques, including ART and HOT in design and evaluation. Will CedrusDB be better than merely using ART-ROWEX
or HOT-ROWEX on mmapped-region in terms of performance and space consumption?

2. The current read-write lock-based concurrency control in CerdrusDB is OK (not incorrect) but not great. I think there are many better concurrency protocols, which can be easily applicable to CedrusDB, than the simple read-write lock. I think ROWEX (Read-Optimized Write EXclusion) [R1] would be a good candidate. Also, the lock-free search and lock-based structural modification in FAST \& FAST [R2] would be a good candidate.

   [R1] The ART of Practical Synchronization, DaMon 2016

   [R2] Endurable Transient Inconsistency in Byte-Addressable Persistent
   B+-Tree, FAST 2018

3. The space management seems ad-hoc to me. Why using an "unsorted array of hole descriptor" achieve better locality and IO efficiency? In my understanding, CedrusDB does not maintain a size-classified list, like a buddy list in a memory allocator. Is the proposed unsorted free list better than the standard buddy structure design or free block tree design (e.g., [R3])?

   [R3] Memory Management Techniques for Large-Scale Persistent-Main-Memory Systems, VLDB 2017

 Also, it is not clear how to manage the DRAM usage. When to map and munmap regions? How to decide which region should be munapped (evicted)?

4. The description on Disk IO (3.3) is confusing. In my understanding, the region is mmaped with a private option, and a disk thread read the clean block from disk (if not cached) and applies the WAL log to create the up-to-date block image and then issues an AIO write. Is my understanding correct? If this is the case, the page once dirtied, cannot be released from the memory even after the disk block is written, so the entire regions should take a physical memory in write-intensive workloads. Is this correct?

5. One related idea to lazy trie is HTree used in ext3/4 file system. HTree indexes a hashed file name in a directory. Both HTree and lazy trie use a hashed key on a sorted index.

Evaluation

6. The paper should show extensive experimental comparison with the state-of-the-art key-value stores such as FASTER (hybrid hash table) and KVell (optimized hybrid B-tree for fast SSD). Also, the extensible hash table designed for disk should be included in the evaluation.

 As Table 1 shows, the performance of CedrusDB is worse than KVell even though CedrusDB (presumably) uses much more DRAM than KVell, and it does not support a range query.

7. After reading 4.2.4, I became skeptical that CedrusDB can handle a large volume of data. It seems that the current CedrusDB design/implementation considers only when the entire data set is cached.

  For example, is CedrusDB able to handle 2TB of data (which is just one SSD capacity) with 64GB DRAM in what performance? Is it expected that CedrusDB can provide better performance than KVell, FASTER, or extensible hash table?

8. The paper needs to show the execution time breakdown. It will demonstrate whether CedrusDB can maximize the IO bandwidth without making kernel or concurrency control a bottleneck. In particular, it will be useful to co-related the IO usage (max IOPS) to the achieved performance.

9. Fig 17 and Fig 19 show that the scalability of CedrusDB is not very good. The performance starts saturated only at 4 to 8 cores. I wish to see more extensive evaluation and analysis on scalability, which is essential today.

\item\textbf{Questions for authors' response.}
Please address the comments 1, 2, 3, 6, 7, and 8 above.
\end{itemize}

\subsubsection{Review \#150E}
\begin{itemize}[leftmargin=*,itemsep=1pt,topsep=1pt]
\item\textbf{Novelty.} 3. New contribution
\item\textbf{Experimental methodology.} 3. Average
\item\textbf{Writing quality.} 4. Well-written
\item\textbf{Overall merit.} 3. Weak reject (This paper should be rejected, but I'll not fight strongly)

\item\textbf{Paper summary.}
The paper proposes a data structure called lazy-trie for indexing data on storage. Based on this data structure, the paper builds a persistent key-value store called CedrusDB. The lazy-trie structure has similarities with B+-tree but its simpler design allows for higher concurrency. The main reason for the simpler design is that unlike B+trees, tries do not need to require complex reorganization for splitting and merging nodes on updates. However, tries can lead to unbalanced depth. CedrusDB reduces this problem by using key hashes to index into the trie, which ensures that the tree depth is static, but this means that CedrusDB cannot support range queries.

The lazy-trie structure is a variant of a standard trie that uses two compression techniques. The first borrows from radix trees to compress paths and the second uses a linked list to compress lower levels of the tree with few keys.

CedrusDB is primarily designed for data sets that fit in memory. The paper shows that CedrusDB provides much higher performance compared to LMDB, a B+-tree key-value store that is designed for data sets that fit entirely in memory. However, CedrusDB has much lower performance compared to RocksDB for write-intensive workloads.

CedrusDB does not provide any performance comparison with key-value stores using in-memory indexes, such as KVell, but shows that their recovery time is much higher than CedrusDB.

\item\textbf{Strengths.}
Problem space is interesting.

The basic design is simple and easy to follow.

The paper is clear about the trade-offs inherent in their design.

Paper is well written.

\item\textbf{Significant weaknesses.}
The type of applications that will benefit appears to be limited.

Comparison with other, higher performing, B+-tree key-value stores is missing.

Lack of range queries is a significant limitation.

\item\textbf{Comments for author.}
The use of tries for indexing in persistent stores is an interesting idea. The design of CedrusDB is well thought and the paper shows the benefits of the design. I thought that sluggish splitting is a relatively simple idea that provides good performance benefits.

However, I think the use cases for CedrusDB are somewhat limited. It primarily benefits read dominated workloads, performing much worse for writes. It doesn't support range scans, and I believe it is not competitive with key-value stores that use in-memory indexes (although this is not evaluated). It will help to clarify what realistic workloads and applications will benefit with CedrusDB.

My main concerns are related to the evaluation.

The paper compares against LMDB but I am not sure whether LMDB is specifically designed for concurrent operation. Its overall performance goes down when moving from 1 to 4 threads in all the experiments shown in the paper. The paper needs to compare against a store that provides better support for multi-core scaling (Wiredtiger?) or argue why LMDB is the best alternative.

I would have also liked to see a performance comparison with KVell to understand the design tradeoffs.

Finally, SplinterDB (ATC 2020) is a BeTree-based design that significantly reduces compaction over LSM stores and generally provides better performance than KVell. I am not sure whether the SplinterDB source code is available, but it seems to be the most relevant for comparison.

Another issue with the evaluation is that the only macrobenchmark used in the paper (YCSB) is not run with default workloads, making it hard to compare against other papers. Why not run the various YCSB A..F workloads that CedrusDB supports? Also, the default record size in YCSB is 1000 bytes, not 128 or 1024. The paper also doesn't show multi-threaded performance for the 1024 size.

Your design makes several choices that should be evaluated individually:

- CedrusDB implements its own page replacement but doesn't show the benefits of doing so.

- CedrusDB uses hashing, which limits range queries. How much benefit does hash-trie provide over a trie?

- You could also compare with using full radix tree design where compression can occur at any point in the path.

writing:

Is a "child table" the pointer array?

Section 2.3 introduces S but defines sluggishness later.

The reason for using segments and regions was not clear to me. Why not use a separate file per region?

Are the blue arrows in Figure 7 correct?

"In CedrusDB it is the disk thread's responsibility to obtain the consistent state of a block from a file if it is not already available in its cache, rather than copying the content from memory as there may be a data race."

who is copying memory, from where? or is the text saying that user threads don't do the copying from disk, but delegate it to the disk thread?

The text mentions that deletions are hard to handle without locking the entire path. You may be able to use tombstones and delay deletion.

The section on batched writes is a bit confusing because I initially thought this section was referring to aggregating writes into blocks for writing. If I understand correctly, you mean batching writes in a transactional manner. Also, it appears that the purpose of the global lock is to avoid deadlocks and is only held for batched operations. It will help to clarify the design.

The numbers on top of the plots in Figure 11 are hard to read (some of them are too close to the error bar).
\end{itemize}

\subsubsection{Review \#150F}
\begin{itemize}[leftmargin=*,itemsep=1pt,topsep=1pt]
\item\textbf{Novelty.} 2. Incremental improvement
\item\textbf{Experimental methodology.} 3. Average
\item\textbf{Writing quality.} 3. Adequate
\item\textbf{Overall merit.} 3. Weak reject (This paper should be rejected, but I'll not fight strongly)
\item\textbf{Paper summary.}
CedrusDB is a persistent key-value store that indexes all data in memory for fast access while using a SSD for persistence. It does not handle range queries. It utilizes the new lazy trie data structure for its index. The lazy trie is sluggish and hash based. User keys are hashed to 256 bit keys that are indexed in the trie. The trie lazily inserts tree nodes only when necessary to differentiate data nodes underneath it. The trie sluggishly allows multiple data nodes with different hashes to be combined in a single linked list up to a fixed sluggishness factor (e.g., 4 or 16). Laziness and sluggishness keep the height of the trie low for faster lookups and less metadata overhead.

Compared to RocksDB it trades off no range queries and storing all data in memory to get considerably higher read and write throughput. Compared to LMDB it trades off slightly lower single threaded performance to get much higher multi-threaded performance. Compared to MassTree it trades off much lower throughput to get much shorter recovery times.

\item\textbf{Strengths.}

+ Sluggish lazy hash trie is a new data structure

+ CedrusDB wins at multi-threaded performance for data that fits in memory when very fast recovery is necessary

+ CedrusDB prototype seems more robust than other academic prototypes in evaluation

\item\textbf{Significant weaknesses.}

- CedrusDB only wins in a very specific scenarios

- Target deployment of all data in DRAM backed by SSD needs better motivation

\item\textbf{Comments for author.}
Thank you for submitting your work to OSDI 2021. I liked reading about your new data structure.

My biggest concerns about the paper are how realistic the target deployment scenario are and how exactly it fits in with related work.

The target deployment scenario of all data in DRAM backed by SSD should have better motivation. We can now address very large amount of memory, but memory is still very very expensive compared to SSDs per GB. I believe that DRAM is still much more expensive that NVM like Optane PM per GB. Are there target deployment scenarios where all data in DRAM+SSD is the best choice of hardware?

There has been quite a lot of work on indexing data structures and persistent key value stores. The paper includes mostly pair-wise comparisons between lazy-tries and other datastructures, and between CedrusDB and other KV stores. It would be very helpful to provide an overall comparison with all relevant data structures and lazy-tries.  Similarly it would be very helpful to provide an overall comparison with all relevant persistent key-value stores and CedrusDB. What are the biggest difference and trade offs between B+ trees, LSM, B-tree variants, trie variants, and lazy tries? Similarly, what are the biggest differences and tradeoffs between CedrusDB and the other persistent key-value stores?

The evaluation compares to many competing systems. These competitors should be included in each graph. For instance, MassTree should be included on all of the graphs in 4.2 and 4.3. It is not fair to only include it in the experiments where CedrusDB will beat it.

MassTree recovery in 30 seconds seems fast enough for many deployments. How slow will MassTree be for realistic deployments? When is the faster recovery of CedrusDB worth the lack of range queries and lower throughput?

The evaluation setup needs a few more details. Why does disabling data compression result in a fairer comparison? What is the parameter for the Zipf workloads?
\end{itemize}

\subsection{Response by the Author}

We thank the reviewers for their helpful feedback. Below we address the most major concerns only.

\paragraph{Limitation and applications} Lack of range query support is indeed a limitation of this work. However, many applications only need ``map-like'' persistent storage: user apps that store configurations/metadata and web browsers (like Chrome) that keeps users' local data for websites. Blockchain infrastructures can also benefit from this as their data objects are massive in number and usually already hashed (unsorted). A major blockchain company plans to use CedrusDB in its runtime (and allegedly a prototype already works).

\paragraph{Comparisons to other works} More than one reviewer expressed concern about the fairness of comparing CedrusDB to tree-based solutions. The unfortunate reality of ``embedded KV stores'' (libraries that support pure local storage, used as part of a larger system) is that there aren't many practical and robust open-source libraries available. In Section 4.4, we compared CedrusDB to those with different models and guarantees, as our best effort \emph{not} to downplay any of them, but to clarify the following:

1a) We are indeed closer to the hash-based design such as FASTER as for our \emph{data structure}.

1b) But our storage system/model falls entirely in the many "tree-based categories" (embedded, not log-based).

2) Unlike us, prior hash-based works are typically coupled with a log-base approach. We speculate that this is because making a hash structure indexed directly on disk is non-trivial, and we believe this is the major contribution of our work. But it makes an apple-to-apple comparison difficult.

3) Log/slab-based approaches are indeed much faster in failure-free cases, but they have downsides for recovery/persistence in various ways, including that checkpoint overhead increases as the data set grows in size, and it is hard to pick a proper frequency for making such checkpoints. Code bases provided by prior work such as KVell and FASTER also have issues in normal use (crashed in some simple operations, which prevented us to do very basic evaluations, and the unique persistence model of FASTER makes the comparison even harder).

\paragraph{Relationship to ART and HOT} both ART and HOT are \emph{in-memory} indexes whereas CedrusDB's index is on disk, therefore tricks such as sluggishness are required in order to ensure a small storage footprint. The fundamental three differences in the data structure are summarized in Section 5.

\paragraph{Concurrency} CedrusDB utilizes a simple concurrency control (locking) mechanism to ensure consistency as described in 3.5. As a brief response to how it works together with the disk thread: changes are first made to the mapped memory directly, guarded by locking, during which writes are collected. Then, right before releasing all locks at the end of the (single operation)/batch, all low-level space writes are bundled into a record and pushed to the disk thread, which schedules WAL and the subsequent async writes to the disk to reflect the change. The key facts are:

1) We optimize the case for single-write (unbatched) operations so they just need to grab read/write locks accordingly, at a node's granularity.

2) We support ``transactional'' writes that atomically operate on multiple keys, like LevelDB's WriteBatch. It also makes it easy to support RMW patterns and we do support it in the library.

3) We agree that the concurrency control here is not perfect, but it is simple to implement on this data structure, while providing more concurrency than a single mutex scheme (as used in other popular embedded key-value store libraries).
\end{appendices}
\fi

%%%%%%%%%%%%%%%%%%%%%%%%%%%%%%%%%%%%%%%%%%%%%%%%%%%%%%%%%%%%%%%%%%%%%%%%%%%%%%%%
\end{document}